\documentclass[twocolumn,showpacs,preprintnumbers,amsmath,amssymb,superscriptaddress,nofootinbib,english]{revtex4-1}
\pdfoutput=1
\usepackage{times,amsmath,amsfonts,amssymb,epstopdf}
\usepackage{graphicx}
\usepackage{dcolumn}
\usepackage{bm}
\usepackage{epsfig}
\usepackage{graphicx}
\usepackage{hyperref}
\usepackage[usenames]{color}
\usepackage{url}
\usepackage[normalem]{ulem}
\usepackage[T1]{fontenc}

\usepackage[usenames]{color}

\def\be{\begin{equation}}
\def\ee{\end{equation}}
\def\ben{\begin{eqnarray}}
\def\een{\end{eqnarray}}
\def\ba{\begin{array}}
\def\ea{\end{array}}

\newcommand{\bq}{\begin{eqnarray}}
\newcommand{\eq}{\end{eqnarray}}
\newcommand{\bes}{\begin{subequations}}
\newcommand{\ees}{\end{subequations}}

\begin{document}
\newcommand{\half}{{\textstyle\frac{1}{2}}}
\allowdisplaybreaks[3]
\def\triangledown{\nabla}
\def\grad3{\hat{\nabla}}
\def\a{\alpha}
\def\b{\beta}
\def\g{\gamma}\def\G{\Gamma}
\def\d{\delta}\def\D{\Delta}
\def\ep{\epsilon}
\def\et{\eta}
\def\z{\zeta}
\def\t{\theta}\def\T{\Theta}
\def\l{\lambda}\def\L{\Lambda}
\def\m{\mu}
\def\f{\phi}\def\F{\Phi}
\def\n{\nu}
\def\r{\rho}
\def\s{\sigma}\def\S{\Sigma}
\def\ta{\tau}
\def\x{\chi}
\def\o{\omega}\def\O{\Omega}
\def\k{\kappa}
\def\pa {\partial}
\def\ov{\over}
\def\br{\\}
\def\ud{\underline}

\def\nucubic{{\nu} \rm{Cubic}}
\def\nuquartic{{\nu} \rm{Quartic}}
\def\nuquintic{{\nu} \rm{Quintic}}
\def\nulcdm{{\nu} \Lambda\rm{CDM}}

\newcommand\lsim{\mathrel{\rlap{\lower4pt\hbox{\hskip1pt$\sim$}}
    \raise1pt\hbox{$<$}}}
\newcommand\gsim{\mathrel{\rlap{\lower4pt\hbox{\hskip1pt$\sim$}}
    \raise1pt\hbox{$>$}}}
\newcommand\esim{\mathrel{\rlap{\raise2pt\hbox{\hskip0pt$\sim$}}
    \lower1pt\hbox{$-$}}}
\newcommand{\dpar}[2]{\frac{\partial #1}{\partial #2}}
\newcommand{\sdp}[2]{\frac{\partial ^2 #1}{\partial #2 ^2}}
\newcommand{\dtot}[2]{\frac{d #1}{d #2}}
\newcommand{\sdt}[2]{\frac{d ^2 #1}{d #2 ^2}}    

\title{The observational status of Galileon gravity after Planck}

\author{Alexandre Barreira}
\email[Electronic address: ]{a.m.r.barreira@durham.ac.uk}
\affiliation{Institute for Computational Cosmology, Department of Physics, Durham University, Durham DH1 3LE, U.K.}
\affiliation{Institute for Particle Physics Phenomenology, Department of Physics, Durham University, Durham DH1 3LE, U.K.}

\author{Baojiu Li}
\affiliation{Institute for Computational Cosmology, Department of Physics, Durham University, Durham DH1 3LE, U.K.}

\author{Carlton M. Baugh}
\affiliation{Institute for Computational Cosmology, Department of Physics, Durham University, Durham DH1 3LE, U.K.}

\author{Silvia Pascoli}
\affiliation{Institute for Particle Physics Phenomenology, Department of Physics, Durham University, Durham DH1 3LE, U.K.}

\preprint{IPPP/14/51DCPT/14/102}

\begin{abstract}
We use the latest CMB data from Planck, together with BAO measurements, to constrain the full parameter space of Galileon gravity. We constrain separately the three main branches of the theory known as the Cubic, Quartic and Quintic models, and find that all yield a very good fit to these data. Unlike in $\Lambda{\rm CDM}$, the Galileon model constraints are compatible with local determinations of the Hubble parameter and predict nonzero neutrino masses at over $5\sigma$ significance. We also identify that the low-$l$ part of the CMB lensing spectrum may be able to distinguish between $\Lambda{\rm CDM}$ and Galileon models. In the Cubic model, the lensing potential deepens at late times on sub-horizon scales, which is at odds with the current observational suggestion of a positive ISW effect. Compared to $\Lambda$CDM, the Quartic and Quintic models predict less ISW power in the low-$l$ region of the CMB temperature spectrum, and as such are slightly preferred by the Planck data. We illustrate that residual local modifications to gravity in the Quartic and Quintic models may render the Cubic model as the only branch of Galileon gravity that passes Solar System tests.
\end{abstract} 
\maketitle

\section{Introduction}

The idea of modifying Einstein's theory of General Relativity (GR) has attracted substantial attention recently in cosmological studies (see \cite{Clifton:2011jh} for a review). Despite the overwhelming success of GR in passing a vast number of experimental tests \cite{Will:2014kxa}, these only probe the laws of gravity on length scales that are much smaller than those relevant for cosmology. As a result, it remains unclear whether or not the success of GR can be extended to larger scales, which leaves room for alternative scenarios. This provides the main motivation for looking at modified gravity models: the study of alternative theories should result in a better understanding of the different observational signatures one may expect, and hence, help to design future observational missions that aim to test gravity on cosmological scales. Further motivation to study modified gravity comes from the fact that these theories may also offer an explanation for the current accelerated expansion of the Universe. In the context of GR, such an acceleration can only be explained by postulating the existence of a "dark energy" fluid that accelerates the expansion due to its negative pressure. The energy of the vacuum behaves as a cosmological constant ($\Lambda$). This represents the simplest explanation for dark energy. However, there is a huge discrepancy between the value that particle physics theories predict for $\Lambda$, and the value inferred from cosmological observations. Modified gravity models address this by attempting to explain the acceleration without invoking any "dark energy". 

Here, we focus on the Galileon model of modified gravity first proposed by Ref.~\cite{PhysRevD.79.064036}. The action of this model is made up of five Lagrangian densities for a scalar field $\varphi$ (cf.~Eq.~\ref{Galileon action}). These constitute the most general set of terms that, in four-dimensional Minkowski space, (i) are invariant under the Galilean shift transformation $\partial_\mu\varphi \rightarrow \partial_\mu\varphi + b_\mu$ (where $b_\mu$ is a constant four-vector); and (ii) lead to equations of motion that remain at second-order in field derivatives. The latter property makes this model a subset of the general Horndeskii theory \cite{Horndeski:1974wa}, protecting it against the presence of Ostrogradski ghosts (see Ref.~\cite{Woodard:2006nt} for a discussion). In Refs.~\cite{PhysRevD.79.084003, Deffayet:2009mn} it was shown that the original action of Ref.~\cite{PhysRevD.79.064036} needs to be augmented with certain couplings to curvature tensors, if one requires the field equations to remain second-order in dynamical spacetimes like the Friedmann-Robertson-Walker (FRW). This breaks the Galileon symmetry, but helps to keep the theory free from pathologies.

The action of the Galileon model contains nonlinear covariant derivative self-couplings of $\varphi$, which induce couplings between partial derivatives of the Galileon and metric fields. This is a process known as {\it kinetic gravity braiding} \cite{Deffayet:2010qz, Babichev:2012re}, and it is why the Galileon model is called a modified gravity theory. These derivative interactions can cause the Universe to accelerate, but also introduce extra terms that modify the gravitational force law. The latter are proportional to the spatial gradients of $\varphi$ and are usually referred to as "fifth force" terms. A vital observational requirement is that the amplitude of these terms must be suppressed in regions such as the Solar System, where GR has been proven an extremely successful theory \cite{Will:2014kxa}. Interestingly, the same nonlinear nature of the derivative couplings that give rise to the fifth force is also what allows for these modifications to be suppressed near massive bodies like the Sun. Far away from matter sources, the equation of motion in $\varphi$ can be linearized, resulting in a Poisson-like equation that gives rise to a sizeable spatial gradient of $\varphi$, and hence, to nonvanishing fifth force terms. However, near massive bodies, where the density is higher, the nonlinear terms can no longer be neglected. This effectively leads to a suppression of the spatial gradient of $\varphi$, in such a way that the magnitude of the fifth force becomes negligible compared to normal gravity. This "screening" effect is known as the Vainshtein mechanism \cite{Vainshtein1972393, Babichev:2013usa, Koyama:2013paa}.

The Galileon model we study here has been the subject of many recent papers \footnote{See, e.g., Refs.~\cite{Gannouji:2010au, Chow:2009fm, PhysRevD.82.103015, Neveu:2014vua, DeFelice:2010nf, kimura1, kimura2} for studies of models motivated in similar ways.}. References~\cite{DeFelice:2010pv, Nesseris:2010pc} explored the phenomenology of the Galileon model at the background level, and derived observational constraints using data from type Ia Supernovae (SNIa), Baryonic Acoustic Oscillations (BAO) and the distance priors encoded in the peak positions of the Cosmic Microwave Background (CMB). By working under the quasi-static approximation,  Refs.~\cite{DeFelice:2010as, Appleby:2011aa, Appleby:2012ba, Okada:2012mn, Neveu:2013mfa} have placed constraints using data for the linear growth rate of matter fluctuations. The first predictions for the CMB temperature, CMB lensing and  linear matter power spectra in Galileon cosmologies were obtained in Ref.~\cite{Barreira:2012kk}, and later used to perform the first thorough exploration of the cosmological parameter space in these models using the full information contained in the CMB temperature spectrum \cite{Barreira:2013jma}. Reference \cite{Bartolo:2013ws} studied the bispectrum of matter density perturbations. The Galileon model predictions for the clustering of matter on small (nonlinear) scales have been studied by performing N-body simulations of structure formation \cite{Barreira:2013eea, Li:2013tda} and by making use of the spherical collapse model in the Excursion Set formalism \cite{Barreira:2013xea}. An analytical halo model for Galileon gravity was developed by Ref.~\cite{Barreira:2014zza}, and Ref.~\cite{Hellwing:2014nma} investigated the potentially powerful role that low order statistics in the galaxy velocity field can play in distinguishing Galileon gravity from GR. All of these studies have identified some observational tensions that seem to question the viability of the Galileon model.

The above studies have assumed neutrino particles to be massless. However, the detection of neutrino flavour oscillations in atmospheric, solar and reactor experiments demonstrates that the sum of the three neutrinos masses is non-zero, $\Sigma m_\nu > 0$. These experiments measure $m_2^2 - m_1^2 \approx \left(7.5 \pm 0.19\right)\times 10^{-5}\ {\rm eV}^2$ ($1\sigma$) and $|m_2^2 - m_3^2| \approx \left(2.427 \pm 0.007\right)\times 10^{-3}\ {\rm eV}^2$ ($1\sigma$) \cite{GonzalezGarcia:2012sz}, where $m_1, m_2, m_3$ are the masses of the three neutrino eigenstates. Assuming a normal mass ordering ($m_1 < m_2 < m_3$), the data imply $\Sigma m_\nu > 0.06\ {\rm eV}$, whereas for an inverted mass ordering ($m_3 < m_2 < m_1$), $\Sigma m_\nu > 0.1\ {\rm eV}$. Currently, the upper bound on the summed neutrino masses from terrestrial experiments is $\Sigma m_\nu < 6.6\ {\rm eV}$ \cite{Kraus:2004zw, Aseev:2011dq}. Such large values of $\Sigma m_\nu$ can affect substantially a number of different cosmological observables. As a result, it seems reasonable to require that consistent cosmological constraints treat $\Sigma m_\nu$ as a free parameter. This is of particular interest in modified gravity models, where some degeneracies may arise \cite{He:2013qha, Motohashi:2012wc, Baldi:2013iza}. In the context of Galileon gravity, the impact of massive neutrinos has been explored recently in Ref.~\cite{Barreira:2014ija}. In particular, it has been shown that the modifications introduced by massive neutrinos at the background and linear perturbation levels can effectively eliminate observational tensions that were thought to rule out this theory of gravity.

Here, we update the Galileon constraints presented in Ref.~\cite{Barreira:2013jma} to include the Planck satellite data \cite{Ade:2013zuv}. In particular, for the first time, we make use of the CMB lensing power spectrum as measured by Planck \cite{Ade:2013tyw} in the constraints. We also analyze the impact of considering the possibility of massive neutrinos by extending the analysis of Ref.~\cite{Barreira:2014ija}, who focused on the so-called Cubic Galileon model. Our ultimate goal is to provide a general overview of the current observational status of the Galileon model, and to identify the types of observables and future experiments that have the greatest potential to further constrain or rule out this model.

The layout of this paper is as follows. In Sec.~\ref{sec:model} we present the action of the model and display the relevant background and linearly perturbed equations. Section \ref{sec:methodology} is devoted to the methodology adopted in our constraints. In particular, we describe the full cosmological parameter space, as well as the datasets used in the constraints. In Sec.~\ref{sec:pastconstraints} we review briefly some of the results obtained by previous Galileon constraint studies. The cosmological constraints for the so-called Cubic sector of the Galileon model are discussed in Sec.~\ref{sec:resultscubic}. In this section, we address explicitly the important role played by massive neutrinos. We also analyze the resulting best-fitting cosmologies, and show how these can be further constrained by future data. Section \ref{sec:results45} focuses on the constraints for the so-called Quartic and Quintic sectors of Galileon gravity. We discuss the implementation of the Vainshtein mechanism in these sectors of the model, and how such an implementation can be used to place tight constraints on the model parameters. Finally, we summarize in Sec.~\ref{sec:summary}.

We assume the metric convention $(+,-,-,-)$ and work in units in which the speed of light $c = 1$. Greek indices run over $0,1,2,3$ and $8\pi G=\kappa=M^{-2}_{\rm Pl}$, where $G$ is Newton's constant and $M_{\rm Pl}$ is the reduced Planck mass.

\section{The Galileon model}\label{sec:model}

\label{The model}

The action of the covariant Galileon model (in the absence of explicit couplings to matter) is given by

\bq\label{Galileon action}
&& S = \int {\rm d}^4x\sqrt{-g} \left[ \frac{R}{16\pi G} - \frac{1}{2}\sum_{i=1}^5c_i\mathcal{L}_i - \mathcal{L}_m\right],
\eq
with
\bq\label{L's}
\mathcal{L}_1 &=& M^3\varphi, \nonumber \\
\mathcal{L}_2 &=& \nabla_\mu\varphi\nabla^\mu\varphi,  \nonumber \\
\mathcal{L}_3 &=& \frac{2}{M^3}\Box\varphi\nabla_\mu\varphi\nabla^\mu\varphi, \nonumber \\
\mathcal{L}_4 &=& \frac{1}{M^6}\nabla_\mu\varphi\nabla^\mu\varphi\Big[ 2(\Box\varphi)^2 - 2(\nabla_\mu\nabla_\nu\varphi)(\nabla^\mu\nabla^\nu\varphi) \nonumber \\
&& -R\nabla_\mu\varphi\nabla^\mu\varphi/2\Big], \nonumber \\
\mathcal{L}_5 &=&  \frac{1}{M^9}\nabla_\mu\varphi\nabla^\mu\varphi\Big[ (\Box\varphi)^3 - 3(\Box\varphi)(\nabla_\mu\nabla_\nu\varphi)(\nabla^\mu\nabla^\nu\varphi) \nonumber \\
&& + 2(\nabla_\mu\nabla^\nu\varphi)(\nabla_\nu\nabla^\rho\varphi)(\nabla_\rho\nabla^\mu\varphi) \nonumber \\
&& -6 (\nabla_\mu\varphi)(\nabla^\mu\nabla^\nu\varphi)(\nabla^\rho\varphi)G_{\nu\rho}\Big],
\eq
where $R$ is the Ricci curvature scalar, $g$ is the determinant of the metric $g_{\mu\nu}$, and $M^3\equiv M_{\rm Pl}H_0^2$ with $H_0$ being the present-day Hubble expansion rate. The five terms in the Lagrangian density are fixed by the Galilean invariance in a flat spacetime, $\partial_\mu\varphi \rightarrow \partial_\mu\varphi + b_\mu$, and $c_{1-5}$ are dimensionless constants. The explicit couplings to the Ricci scalar $R$ and the Einstein tensor $G_{\mu\nu}$ in $\mathcal{L}_4$ and $\mathcal{L}_5$ break the Galilean symmetry, but are necessary to limit the equations of motion to second-order in field derivatives (and hence free from Ostrogradski ghosts) in spacetimes such as FRW \cite{PhysRevD.79.084003}. We set the potential term to zero ($c_1 = 0$), as we are only interested in cases where cosmic acceleration is driven by the kinetic terms of the field. The Einstein equations and the Galileon field equation of motion can be found as Eqs.~(A1--A7) of Ref.~\cite{Barreira:2012kk}.

Throughout the paper we shall refer to three sectors of the Galileon model. We dub Cubic and Quartic Galileon the models made up by $\left\{\mathcal{L}_2, \mathcal{L}_3\right\}$ and $\left\{\mathcal{L}_2, \mathcal{L}_3, \mathcal{L}_4\right\}$, respectively. We shall use the name Quintic Galileon when referring to the most general model, i.e., $\left\{\mathcal{L}_2, \mathcal{L}_3, \mathcal{L}_4, \mathcal{L}_5\right\}$.

\subsection{Linearly perturbed equations}

The linearly perturbed gauge-invariant equations in Galileon gravity have the same structure as in GR, but with extra contributions to the energy-momentum tensor, $T_{\mu\nu}$, from the Galileon field.  Here, for brevity, we simply list these contributions and lay out also the main premises of the method of $3+1$ decomposition used in the derivation. For more details we refer interested readers to  Ref.~\cite{Barreira:2012kk}.

In the $3+1$ decomposition approach, the main idea is to make spacetime splits of the physical quantities w.r.t. an observer's four-velocity, $u^\mu$ \cite{PhysRevD.40.1804, Challinor:1998xk, Challinor:1999xz, Challinor:2000as}. This is achieved via the projection tensor, $h_{\mu\nu} = g_{\mu\nu} - u_\mu u_\nu$, which projects covariant tensors onto 3-dimensional hyperspaces that are orthogonal to $u^\mu$. For instance, the covariant spatial derivative $\hat{\nabla}$ of a generic tensor field $A^{\beta...\gamma}_{\sigma...\lambda}$ is defined as
\bq
\hat{\nabla}^\alpha A^{\beta\cdot\cdot\cdot\gamma}_{\sigma\cdot\cdot\cdot\lambda} \equiv h^{\alpha}_{\mu}h^{\beta}_{\nu}\cdot\cdot\cdot h^{\gamma}_{\kappa}h^{\rho}_{\sigma}\cdot\cdot\cdot h^{\eta}_{\lambda}\nabla^\mu A^{\nu\cdot\cdot\cdot\kappa}_{\rho\cdot\cdot\cdot\eta}.
\eq

The energy-momentum tensor and the covariant derivative of $u^{\mu}$ can be decomposed, respectively, as 

\bq
\label{Tuv} T_{\mu\nu} &=& \pi_{\mu\nu} + 2q_{(\mu}u_{\nu)} + \rho u_\mu u_\nu - ph_{\mu\nu},\\
\nabla_\mu u_\nu &=& \sigma_{\mu\nu} + \varpi_{\mu\nu} + \frac{1}{3}\theta h_{\mu\nu} + u_\mu A_\nu,
\eq
in which $\pi_{\mu\nu}$ is the projected symmetric and trace-free (PSTF) anisotropic stress, $q_\mu$ is the heat flux vector, $\rho$ is the energy density, $p$ is the isotropic pressure, $\sigma_{\mu\nu}$ is the PSTF shear tensor, $\varpi_{\mu\nu} = \hat{\nabla}_{[\mu}u_{\nu]}$ is the vorticity, $\theta = \nabla^\alpha u_\alpha = 3\dot{a}/a = 3H$ is the expansion scalar (where $a$ is the mean expansion scale factor and $H$ is the Hubble rate) and $A_\mu = \dot{u}_\mu$ is the observer's four-acceleration. A dot denotes derivatives w.r.t.~physical time, which can be expressed in terms of covariant derivatives as $\dot{\varphi} = u^\alpha\nabla_\alpha\varphi$. Square brackets and parentheses indicate antisymmetrization and symmetrization, respectively. Note that $u^\alpha u_\alpha = 1$, in accordance with the metric signature adopted.

As mentioned above, the Galileon field contributes to the Einstein equations via additional terms to the components of the total energy-momentum tensor $T_{\mu\nu}$:
\bq
\label{physical quantities 1} \rho &=& \rho^f + \rho^\varphi,\\
\label{physical quantities 2} p &=& p^f + p^\varphi, \\ 
\label{physical quantities 3} q_\mu &=& q_\mu^f + q_\mu^\varphi, \\
\label{physical quantities 4} \pi_{\mu\nu} &=& \pi_{\mu\nu}^f + \pi_{\mu\nu}^\varphi,
\eq
where the superscripts $^\varphi$ and $^f$ indicate, respectively, the Galileon terms and the terms corresponding to the rest of the matter fluid. The latter includes photons, neutrinos, baryonic and cold dark matter. Up to first order in perturbed quantities, the Galileon terms are given by

\begin{widetext}
\bq\label{perturbed1}
\rho^\varphi &\doteq& c_2\left[ \frac{1}{2}\dot{\varphi}^2\right] + \frac{c_3}{M^3} \left[ 2\dot{\varphi}^3\theta + 2\dot{\varphi}^2\hat{\Box}\varphi\right]  +\frac{c_4}{M^6}\left[ \frac{5}{2}\dot{\varphi}^4\theta^2 + 4\dot{\varphi}^3\theta\hat{\Box}\varphi + \frac{3}{4}\dot{\varphi}^4\hat{R}\right]     \nonumber \\
&& + \frac{c_5}{M^9}\left[\frac{7}{9}\dot{\varphi}^5\theta^3 + \frac{5}{3}\dot{\varphi}^4\theta^2\hat{\Box}\varphi +\frac{1}{2}\dot{\varphi}^5\theta\hat{R} \right], \\
\label{perturbed2}
p^\varphi &\doteq& c_2\left[ \frac{1}{2}\dot{\varphi}^2\right] + \frac{c_3}{M^3} \left[ -2\ddot{\varphi}\dot{\varphi}^2\right] \nonumber \\
&&  + \frac{c_4}{M^6}\left[ -4\ddot{\varphi}\dot{\varphi}^3\theta - \dot{\varphi}^4\dot{\theta} - \frac{1}{2}\dot{\varphi}^4\theta^2 - 4\ddot{\varphi}\dot{\varphi}^2\hat{\Box}\varphi - \frac{4}{9}\dot{\varphi}^3\theta\hat{\Box}\varphi + \dot{\varphi}^4\hat{\triangledown}\cdot A + \frac{1}{12}\dot{\varphi}^4\hat{R}\right] \nonumber \\
&& + \frac{c_5}{M^9}\left[-\frac{5}{3}\ddot{\varphi}\dot{\varphi}^4\theta^2 - \frac{2}{3}\dot{\varphi}^5\dot{\theta}\theta - \frac{2}{9}\dot{\varphi}^5\theta^3 - \frac{2}{9}\dot{\varphi}^4\theta^2\hat{\Box}\varphi - \frac{8}{3}\ddot{\varphi}\dot{\varphi}^3\theta\hat{\Box}\varphi - \frac{1}{2}\ddot{\varphi}\dot{\varphi}^4\hat{R} - \frac{2}{3}\dot{\varphi}^4\dot{\theta}\hat{\Box}\varphi + \frac{2}{3}\dot{\varphi}^5\theta\hat{\triangledown} \cdot A\right],\\
\label{perturbed3}
q_\mu^\varphi &\doteq& c_2\left[ \dot{\varphi}\hat{\triangledown}_\mu\varphi \right] + \frac{c_3}{M^3} \left[ 2\dot{\varphi}^2\theta\hat{\triangledown}_\mu\varphi - 2\dot{\varphi}^2\hat{\triangledown}_\mu\dot{\varphi}\right] \nonumber \\
&&  + \frac{c_4}{M^6}\left[ -4\dot{\varphi}^3\theta\hat{\triangledown}_\mu\dot{\varphi} + 2\dot{\varphi}^3\theta^2\hat{\triangledown}_\mu\varphi - \dot{\varphi}^4\hat{\triangledown}_\mu\theta + \frac{3}{2}\dot{\varphi}^4\hat{\triangledown}^\alpha\sigma_{\mu\alpha} + \frac{3}{2}\dot{\varphi}^4\hat{\triangledown}^\alpha{\varpi}_{\mu\alpha} \right] \nonumber \\
&& + \frac{c_5}{M^9}\left[ - \frac{5}{3}\dot{\varphi}^4\theta^2\hat{\triangledown}_\mu\dot{\varphi} + \frac{5}{9}\dot{\varphi}^4\theta^3\hat{\triangledown}_\mu\varphi - \frac{2}{3}\dot{\varphi}^5\theta\hat{\triangledown}_\mu\theta + \dot{\varphi}^5\theta\hat{\triangledown}^\alpha\sigma_{\mu\alpha} + \dot{\varphi}^5\theta\hat{\triangledown}^\alpha{\varpi}_{\mu\alpha} \right],\\
\label{perturbed4}
\pi_{\mu\nu}^\varphi &\doteq& \frac{c_4}{M^6}\left[  -\dot{\varphi}^4 \left( \dot{\sigma}_{\mu\nu} - \hat{\triangledown}_{\langle\mu}A_{\nu\rangle} - \mathcal{E}_{\mu\nu}\right) - \left( 6\ddot{\varphi}\dot{\varphi}^2 + \frac{2}{3}\dot{\varphi}^3\theta\right)\hat{\triangledown}_{\langle\mu}\hat{\triangledown}_{\nu\rangle}\varphi - \left(6\ddot{\varphi}\dot{\varphi}^3 + \frac{4}{3}\dot{\varphi}^4\theta\right)\sigma_{\mu\nu}\right] \nonumber \\
&& + \frac{c_5}{M^9}\left[ -\left(\dot{\varphi}^5\dot{\theta} +\dot{\varphi}^5\theta^2 + 6\ddot{\varphi}\dot{\varphi}^4\theta\right)\sigma_{\mu\nu} -  \left(\dot{\varphi}^5\theta + 3\ddot{\varphi}\dot{\varphi}^4\right)\dot{\sigma}_{\mu\nu} -  \left( 4\ddot{\varphi}\dot{\varphi}^3\theta + \dot{\varphi}^4\dot{\theta} + \frac{1}{3}\dot{\varphi}^4\theta^2 \right)\hat{\triangledown}_{\langle\mu}\hat{\triangledown}_{\nu\rangle}\varphi \right. \nonumber \\
&& \left. \ \ \ \ \ \ \ \ \ \ \ \ \ + \left(\dot{\varphi}^5\theta + 3\ddot{\varphi}\dot{\varphi}^4 \right)\hat{\triangledown}_{\langle\mu}A_{\nu\rangle} - 6\ddot{\varphi}\dot{\varphi}^4\mathcal{E}_{\mu\nu} \right],
\eq
\end{widetext}
in which $\hat{\Box}\equiv\hat{\nabla}^\mu\hat{\nabla}_\mu$. In Eq.~(\ref{perturbed4}), $\hat{\nabla}_{\langle\mu}\hat{\nabla}_{\nu\rangle}\varphi$ and $\mathcal{E}_{\mu\nu} = u^{\alpha}u^{\beta}\mathcal{W}_{\mu\alpha\nu\beta}$ are PSTF rank-2 tensors that live in the 3-dimensional hypersurface orthogonal to $u^\mu$ (i.e.~$u^\mu\hat{\nabla}_{\langle\mu}\hat{\nabla}_{\nu\rangle}\varphi  = u^\mu\mathcal{E}_{\mu\nu} = 0$), where $\mathcal{W}_{\mu\alpha\nu\beta}$ is the Weyl tensor. Angular brackets indicate trace-free quantities. 

The linearly perturbed Galileon field equation of motion is given by
\begin{widetext}
\begin{eqnarray}
\label{perturbed EoM}
0 &\doteq& c_2\left[\ddot{\varphi} + \hat{\Box}\varphi + \dot{\varphi}\theta \right] + \frac{c_3}{M^3} \left[ 4\ddot{\varphi}\dot{\varphi}\theta  + \frac{8}{3}\dot{\varphi}\theta\hat{\Box}\varphi + 4\ddot{\varphi}\hat{\Box}\varphi + 2\dot{\varphi}^2\theta^2 + 2\dot{\varphi}^2\dot{\theta} - 2\dot{\varphi}^2 \hat{\nabla} \cdot A\right] \nonumber \\
&&  + \frac{c_4}{M^6}\left[  6\ddot{\varphi}\dot{\varphi}^2\theta^2 + 4\dot{\varphi}^3\dot{\theta}\theta + 2\dot{\varphi}^3\theta^3 + 8\ddot{\varphi}\dot{\varphi}\theta\hat{\Box}\varphi + \frac{26}{9}\dot{\varphi}^2\theta^2\hat{\Box}\varphi - 4\dot{\varphi}^3\theta\hat{\nabla}\cdot A + 4\dot{\varphi}^2\dot{\theta}\hat{\Box}\varphi + 3\ddot{\varphi}\dot{\varphi}^2\hat{R} + \frac{1}{3}\dot{\varphi}^3\theta\hat{R}\right] \nonumber \\
&& + \frac{c_5}{M^9}\left[ \frac{5}{9}\dot{\varphi}^4\theta^4   +\frac{20}{9}\ddot{\varphi}\dot{\varphi}^3\theta^3  +\frac{5}{3}\dot{\varphi}^4\dot{\theta}\theta^2 +\frac{8}{9}\dot{\varphi}^3\theta^3\hat{\Box}\varphi +\frac{1}{2}\dot{\varphi}^4\dot{\theta}\hat{R} \right. \nonumber \\
&&\ \ \ \ \ \ \ \ \ \ \ \ \ \ \ \ \ \  \left. +\frac{1}{6}\dot{\varphi}^4\theta^2\hat{R} - \frac{5}{3}\dot{\varphi}^4\theta^2\hat{\nabla} \cdot A  + 4\ddot{\varphi}\dot{\varphi}^2\theta^2\hat{\Box}\varphi + \frac{8}{3}\dot{\varphi}^3\dot{\theta}\theta\hat{\Box}\varphi + 2\ddot{\varphi}\dot{\varphi}^3\theta\hat{R} \right].
\end{eqnarray}
\end{widetext}

In the above equations, $\varphi = \bar{\varphi} + \delta\varphi$, where $\bar{\varphi}$ and $\delta\varphi$ denote, respectively, the background averaged (an overbar indicates background quantities) and the perturbation values of the Galileon field. Note that if a spatial derivative acts on $\varphi$, then it acts only on the perturbed part of the Galileon field, e.g. $\hat{\Box}\varphi = \hat{\Box}\delta\varphi$. On the other hand, $\dot{\varphi} = \dot{\bar{\varphi}} + \dot{\delta\varphi}$.

\subsection{Background evolution}

In order to solve the perturbation equations, one first needs to know the time evolution of the background quantities. This can be achieved by solving one of the Friedmann equations:

\bq\label{background1}
&&\frac{\theta^2}{3} = \kappa {\bar{\rho}}, \\
\label{background2}
&&\dot{\theta} + \frac{1}{3}\theta^2 + \frac{\kappa}{2}({\bar{\rho}} + 3{\bar{p}}) = 0,
\eq
and the equation of motion of the background Galileon field:
\bq\label{background EoM}
0 &=& c_2\left[\ddot{\varphi} + \dot{\varphi}\theta \right] + \frac{c_3}{M^3} \left[ 4\ddot{\varphi}\dot{\varphi}\theta + 2\dot{\varphi}^2\theta^2 + 2\dot{\varphi}^2\dot{\theta}\right] \nonumber \\
&&  + \frac{c_4}{M^6}\left[  6\ddot{\varphi}\dot{\varphi}^2\theta^2 + 4\dot{\varphi}^3\dot{\theta}\theta + 2\dot{\varphi}^3\theta^3 \right] \nonumber \\
&& + \frac{c_5}{M^9}\left[ \frac{5}{9}\dot{\varphi}^4\theta^4   +\frac{20}{9}\ddot{\varphi}\dot{\varphi}^3\theta^3  +\frac{5}{3}\dot{\varphi}^4\dot{\theta}\theta^2 \right],
\eq
where we have dropped the overbars on $\varphi$ to lighten the notation. In Eqs.~(\ref{background1}) and (\ref{background2}), the Galileon contribution to $\bar{\rho}$ and $\bar{p}$ is obtained by extracting the zeroth-order part (non-hatted terms) of Eqs.~(\ref{perturbed1}) and (\ref{perturbed2}), respectively:

\bq\label{density-background}
\bar{\rho}_\varphi &=& c_2\left[ \frac{1}{2}\dot{\varphi}^2\right] + \frac{c_3}{M^3} \left[ 2\dot{\varphi}^3\theta\right]  +\frac{c_4}{M^6}\left[ \frac{5}{2}\dot{\varphi}^4\theta^2\right]     \nonumber \\
&& + \frac{c_5}{M^9}\left[\frac{7}{9}\dot{\varphi}^5\theta^3\right],
\eq

\bq\label{pressure-background}
\bar{p}_\varphi &=& c_2\left[ \frac{1}{2}\dot{\varphi}^2\right] + \frac{c_3}{M^3} \left[ -2\ddot{\varphi}\dot{\varphi}^2\right] \nonumber \\
&& + \frac{c_4}{M^6}\left[ -4\ddot{\varphi}\dot{\varphi}^3\theta - \dot{\varphi}^4\dot{\theta} - \frac{1}{2}\dot{\varphi}^4\theta^2\right] \nonumber \\
&& + \frac{c_5}{M^9}\left[-\frac{5}{3}\ddot{\varphi}\dot{\varphi}^4\theta^2 - \frac{2}{3}\dot{\varphi}^5\dot{\theta}\theta - \frac{2}{9}\dot{\varphi}^5\theta^3\right].
\eq 

In general, the background evolution is obtained by solving Eqs.~(\ref{background1} - \ref{background EoM}) numerically. Reference \cite{DeFelice:2010pv} did this and showed that different initial conditions of the background Galileon field give rise to different time evolution that eventually merge into a common trajectory called the {\it tracker solution}. Moreover, Refs.~\cite{Barreira:2012kk} and \cite{Barreira:2013jma} have shown that it is a requirement that the tracker solution is reached well before the onset of the accelerated expansion era so that a reasonable fit to the CMB data can be achieved (we reillustrate this in Appendix \ref{ap:tracker}). Since before the accelerated era the Galileon field makes a subdominant contribution to the dynamics of the expansion, this justifies the use of the tracker solution at all cosmological epochs. The advantage of assuming the tracker is twofold. First, it allows one to derive analytical formulae for the background evolution (just like, e.g., $\Lambda$CDM models), which greatly simplifies and speeds up the numerical calculations;  second, it also allows us to reduce the number of free parameters by one, which is helpful when exploring the high-dimensional parameter space of the model.

The tracker evolution is characterized by

\bq\label{eq:tracker}
H\dot{\bar{\varphi}} = {\rm constant} \equiv \xi H_0^2,
\eq
where $\xi$ is a dimensionless constant. Multiplying Eq.~(\ref{background1}) by $H^2$, eliminating $\dot{\bar{\varphi}}$ with Eq.~(\ref{eq:tracker}) and dividing the resulting equation by $H^4_0$, one obtains

\bq\label{eq:friedmann2}
E^4 &=& \left(\Omega_{r0}a^{-4} + \Omega_{m0}a^{-3} + \Omega_{\nu0}\frac{\bar{\rho}_{\nu}(a)}{\bar{\rho}_{\nu0}} \right)E^2 \nonumber \\
&+& \frac{1}{6}c_2\xi^2 + 2c_3\xi^3 + \frac{15}{2}c_4\xi^4 + 7c_5\xi^5,
\eq
in which $E\equiv H/H_0$, $\Omega_{i0} = \bar{\rho}_{i0}/\rho_{\rm{c0}}$, where $\rho_{\rm{c0}} = 3H_0^2/\kappa$ is the critical energy density today. The subscript $_r$ refers to radiation, $_m$ to baryonic and cold dark matter and $_\nu$ refers to neutrinos. At the present day,  Eq.~(\ref{eq:friedmann2}) gives

\bq\label{eq:constraint1}
\Omega_{\varphi0} \equiv 1-\Omega_{r0} - \Omega_{m0} - \Omega_{\nu0} \nonumber \\
= \frac{1}{6}c_2\xi^2 + 2c_3\xi^3 + \frac{15}{2}c_4\xi^4 + 7c_5\xi^5,
\eq
where we have assumed spatial flatness. This equation can be regarded as a constraint equation for one of the Galileon parameters, i.e., one of the parameters can be fixed by the condition that the Universe is spatially flat. The assumption that the field follows the tracker allows us to fix one more Galileon parameter. This second constraint equation can be obtained by plugging Eq.~(\ref{eq:tracker}) into Eq.~(\ref{background EoM}), which yields

\bq\label{eq:constraint2}
c_2\xi^2 + 6c_3\xi^3 + 18c_4\xi^4 + 15c_5\xi^5 = 0. 
\eq
Equation (\ref{eq:friedmann2}) is a second-order algebraic equation for $E(a)$, whose solution reads

\bq\label{eq:tracker_H}
&&E(a)^2 = \frac{1}{2}\left[\left(\Omega_{r0}a^{-4} + \Omega_{m0}a^{-3} + \Omega_{\nu0}\frac{\bar{\rho}_{\nu}(a)}{\bar{\rho}_{\nu0}}\right) \right. \nonumber \\
&&  \left. + \sqrt{\left(\Omega_{r0}a^{-4} + \Omega_{m0}a^{-3} + \Omega_{\nu0}\frac{\bar{\rho}_{\nu}(a)}{\bar{\rho}_{\nu0}}\right)^2 + 4\Omega_{\varphi0}}\right].
\eq
Finally, using Eq.~(\ref{eq:tracker}) we have
\bq\label{eq:tracker_galileon}
\dot{\bar{\varphi}} &=& \xi H_0/E(a).
\eq
These last two equations completely specify the background evolution in the Galileon model. Note that, on the tracker, $H(a)$ depends exclusively on the value of the $\Omega_{i0}$ and not on the Galileon parameters. As a result, observational probes such as BAO or SNIa, which are sensitive only to background quantities, cannot constrain the values of the $c_i$.

\section{Methodology}\label{sec:methodology}

To obtain our results, we have modified the publicly available {\tt CAMB} \cite{camb_notes} and {\tt COSMOMC} \cite{Lewis:2002ah} codes to follow Galileon cosmologies. Our new versions of these codes have been subjected to a number of tests. For instance, we have checked that the background evolution computed in {\tt CAMB} agrees with the background evolution computed in an independent code written in {\tt Python} (although for most of this paper we shall use the analytical expressions of Eqs.~(\ref{eq:tracker_H}) and (\ref{eq:tracker_galileon})). At the level of the perturbed equations, we have also checked that the energy and momentum conservation equations are satisfied. The latter equations do not directly enter the calculations. As a result, the fact that they hold automatically after solving for the other perturbation equations serves as an independent and nontrivial check of the code solutions.

We constrain the parameter space by running Monte Carlo Markov Chains (MCMC). For each model and data combination we have run eight chains in parallel. The {\tt CosmoMC} code checks the convergence of a given set of chains by using the Gelman and Rubin statistic, $R_{\rm G\&R}$ = "variance of the chain means"/"mean of the chain variances". Our chains are only stopped if $R_{\rm G\&R} - 1 < 0.02$. When estimating the likelihood surface from the chain samples, we discard the first third of the chains to eliminate the points sampled during the "burn-in" period.

\subsection{Datasets}

In our constraints, we consider three data combinations.  The first dataset comprises the Planck data for the CMB temperature anisotropy angular power spectrum \cite{Ade:2013kta, Ade:2013zuv}. These include its low-$l$ and high-$l$ temperature components, as well as the cross-correlation of the temperature map with the WMAP9 polarization data \cite{Hinshaw:2012fq}. For this piece of the likelihood, we also vary the nuisance parameters that are used to model foregrounds, and instrumental and beam uncertainties. We denote this dataset by {\it P}. We call our second dataset {\it PL}, which adds to {\it P} the data for the power spectrum of the lensing potential (reconstructed from the CMB), also given by the Planck satellite \cite{Ade:2013tyw}. On smaller angular scales, the CMB lensing power spectrum can be affected by nonlinearities. However, given the current level of precision of the data, such nonlinear corrections can be ignored and one can assume that linear perturbation theory holds. The final dataset, denoted by {\it PLB}, also includes the BAO measurements obtained from the 6df \cite{Beutler:2011hx}, SDDS DR7 \cite{Padmanabhan:2012hf}, BOSS DR9 \cite{Anderson:2012sa} and WiggleZ \cite{Blake:2011en} galaxy redshift surveys.

\subsubsection{Leaving out data sensitive to the nonlinear growth of structure}

In general, modified gravity models introduce nontrivial gravitational interactions, which can leave clear imprints on the way structures form in the Universe. Hence, measurements of the growth rate from redshift space distortions (RSD), the amplitude and shape of the galaxy power spectrum, cluster abundance, galaxy shear, etc., can in principle be used to place further constraints on models like Galileon gravity. However, the use of these data requires a proper understanding and modelling of structure formation on scales where nonlinear effects play an important role. These effects can only be accurately estimated via N-body simulations, which can be particularly challenging (and expensive) to run for large numbers of modified gravity cosmologies.

The first N-body simulations of the Galileon model were presented in Refs.~\cite{Barreira:2013eea, Li:2013tda}, which do not include the effects of massive neutrinos. As we shall see, one of the main results of this paper relates to the impact that massive neutrinos have on the viability of the Galileon model. This justifies investigations of the interplay between the modifications to gravity in this model and massive neutrinos on small length scales, which is left for future work. For the time being, we limit ourselves to the constraints one can derive using only linear perturbation theory. Future nonlinear studies of structure formation should then focus on the model parameters that best fit the {\it P}, {\it PL} and {\it PLB} datasets.

\subsection{Parameter space}

In addition to the Galileon model parameters
\bq\label{eq:gali-params}
\left\{ c_2, c_3, c_4, c_5, \xi\right\},
\eq
we also constrain the cosmological parameters
\bq\label{eq:cosmo-params}
\left\{\Omega_{b0}h^2, \Omega_{c0}h^2, \theta_{\rm MC}, \tau, n_s, A_s, \Sigma m_{\nu}\right\},
\eq
which are, respectively, the physical energy density of baryons, the physical energy density of cold dark matter, a {\tt CosmoMC} parameter related to the angular acoustic scale of the CMB, the optical depth to reionization,  the scalar spectral index of the primordial power spectrum, the amplitude of the primordial power spectrum at a pivot scale $k = 0.05\ \rm{Mpc}^{-1}$ and the summed mass of the three active neutrino species.

The value of the Hubble expansion rate today, $H_0 = 100 h\ {\rm km/s/Mpc}$, is a derived parameter (this differs from Ref.~\cite{Barreira:2013jma}). For a given point in parameter space, the {\tt CosmoMC} code determines, by trial-and-error, the value of $H_0$ that reproduces the sampled value of $\theta_{\rm MC}$. $\theta_{\rm MC}$ is much less correlated with the other parameters than $H_0$, which speeds up the convergence of the chains, despite of the additional trial-and-error calculations. Parameters such as $\Omega_{\varphi0}$ and the {\it rms} linear matter fluctuations at $8\ {\rm Mpc}/h$, $\sigma_8$, are also derived parameters. We always fix the number of relativistic neutrinos $N_{\rm eff} = 3.046$, the baryonic mass fraction in helium $Y_{\rm P} = 0.24$ and the running of the scalar spectral index ${\rm d}n_s/{\rm dln}k = 0$. We also set to zero the amplitude of the tensor perturbations, and its tensor spectral index, as we are only interested in scalar perturbations. We briefly investigate the impact of the tensor fluctuations in Appendix \ref{ap:tensor}.

Part of the goal of this paper is to demonstrate the impact that the parameter $\Sigma m_\nu$ has on the goodness of fit of Galileon models. To this end, we shall consider two variations of the model: one for which the number of massive neutrinos $N_{\rm massive} = 0$ and $\Sigma m_\nu = 0$, and another for which $N_{\rm massive} = 3$ and $\Sigma m_\nu$ is a free parameter. We shall refer to the first class of models as "base" Galileon models, and shall denote the second class with a prefix $\nu$, e.g., $\nucubic$ Galileon. For comparison purposes, we also consider a $\nulcdm$ model.

\subsubsection{Scaling degeneracy of the Galileon parameters}\label{subsub:scaling}

We have previously demonstrated that, if one allows all of the Galileon parameters to vary, then it becomes impossible to constrain the Galileon sector of the parameter space \cite{Barreira:2013jma}. This is because of a scaling degeneracy between the Galileon parameters \cite{DeFelice:2010pv}: all of the equations of the Galileon model remain invariant under the following transformations

\bq\label{eq:scaling}
&& c_2 \longrightarrow c_2/B^2, \nonumber \\
&& c_3 \longrightarrow c_3/B^3, \nonumber \\
&& c_4 \longrightarrow c_4/B^4, \nonumber \\
&& c_5 \longrightarrow c_5/B^5, \nonumber \\
&& \varphi \longrightarrow \varphi B,
\eq 
where $B$ is any constant. Figure 1 of Ref.~\cite{Barreira:2013jma} shows that, when all nonzero Galileon parameters are varying, then the chains develop narrow and infinitely long regions along which the likelihood does not change. To circumvent this, in Ref.~\cite{Barreira:2013jma} we employed the practical solution of fixing the value of $c_3$ and using it as a pivot parameter to construct the scaling-invariant quantities $\left\{c_2/c_3^{2/3}, c_4/c_3^{4/3}, c_5/c_3^{5/3}, \varphi c_3^{1/3}\right\}$, which could then be properly constrained by the data. Here, we follow a similar approach, but fix the value of $c_2$ instead. We have run a set of chains in which all of the Galileon parameters were allowed to vary. We found that the parameters $c_3$, $c_4$ and $c_5$ can cross zero, whereas $c_2$ is always negative. This shows that $c_2$ is the best Lagrangian parameter to fix. For instance, by fixing $c_3 > 0$ we would be discarding {\it a priori} the potentially viable regions that have $c_3 < 0$. In Ref.~\cite{Barreira:2013jma}, by performing the same test we have found that $c_3$ was also unlikely to cross zero, and hence could be used as the fixed parameter. The reason behind this difference is related to the different way the background evolution is solved. We discuss this further in subsection \ref{subsec:quintic-chains}. The magnitude of $c_2$ is not critical, as one can always rescale the resulting constraints to any value of $c_2$ (with the same sign) by using Eqs.~(\ref{eq:scaling}). For simplicity, we fix $c_2 = -1$. This way, the $\mathcal{L}_2$ term in Eq.~(\ref{Galileon action}) becomes the standard scalar kinetic term, but with the opposite sign.

The scaling degeneracy then further reduces the dimensionality of the Galileon sector of the parameter space by one. This, together with the fact that we can use Eqs.~(\ref{eq:constraint1} - \ref{eq:constraint2}) to fix two other Galileon parameters, sets the dimensionality of the Galileon sector of the parameter space to $5 - 3 = 2$.

\subsection{Theoretical constraints from the absence of ghost and Laplace instabilities}

When sampling a given point of the parameter space, our version of {\tt CosmoMC} first checks if the point is stable against the existence of ghost or Laplace instabilities of the scalar perturbations. The conditions to avoid these theoretical pathologies were derived and presented in Refs.~\cite{DeFelice:2010pv, DeFelice:2010nf, Appleby:2011aa}. We have derived our own expressions and find agreement with the result of Ref.~\cite{Appleby:2011aa}. These stability conditions, although they apply to the evolution of the perturbations, depend only on background quantities, and hence can be evaluated rather quickly. In the code, if a given point develops one of these types of instabilities, then that point is rejected automatically, without proceeding to the perturbation equations or to the evaluation of observational likelihoods. Without this initial check, these points would still be rejected as the instabilities drastically affect the evolution of the gravitational potential, and hence, lead to very poor fits to the CMB data. However, this step helps to speed up the overall performance of the code, and also avoids the numerical difficulties associated with the instabilities.

Although we consider only scalar perturbations, we note that in principle one could impose similar conditions for the avoidance of ghost and Laplace instabilities of the tensor perturbations \cite{DeFelice:2010nf, Neveu:2013mfa}. We also do not impose any conditions that ensure that the perturbation of the Galileon field does not propagate superluminally, as these cases may not necessarily lead to pathologies \cite{Babichev:2007dw}. We do not rule out cases for which $\bar{\rho}_{\varphi} < 0$ and let the data decide their viability instead.

\section{Overview of previous observational constraints}\label{sec:pastconstraints}

In this section, we summarize the constraints on the Galileon model of Eq.~(\ref{Galileon action}) that were obtained in previous work. Our goal is simply to provide a general overview of the current status of Galileon constraints and not to present a thorough review. For further details, we refer the interested reader to the cited literature and references therein.

The first observational constraints on the Galileon model were derived in Ref.~\cite{Nesseris:2010pc}, by using only data sensitive to the background dynamics. The authors allow for non-flat spatial geometries of the Universe and find, in particular, that although the tracker solution can provide a good fit to the individual datasets (which include data from SNIa, BAO and CMB distance priors), there is some tension when one combines these observational probes.

References \cite{Appleby:2012ba, Okada:2012mn} attempted to use measurements of the growth rate of structure to constrain the Galileon model. These two papers concluded that the model has difficulties in fitting the background and the growth rate data simultaneously. However, Ref.~\cite{Neveu:2013mfa} performed a more detailed analysis and found that the tension is actually much less significant. The authors of Ref.~\cite{Neveu:2013mfa} pointed out that Ref.~\cite{Appleby:2012ba} did not take into account the scaling degeneracy of the Galileon parameters; furthermore Ref.~\cite{Okada:2012mn} used data that is not corrected for the Alcock-Paczynski effect \cite{1979Natur.281..358A}, having assumed also that the shape of the linear matter power spectrum of Galileon models is the same as in $\Lambda$CDM, which is not guaranteed \cite{Barreira:2012kk, Barreira:2013jma}. Moreover, as acknowledged in Ref.~\cite{Neveu:2013mfa}, the constraints obtained by confronting linear theory predictions with growth data assume the validity of linear theory on the scales used to measure the growth rate. For example, the growth rate measurements of the WiggleZ Dark Energy Survey \cite{Blake:2011rj} are obtained by estimating redshift space distortions in the galaxy power spectrum measured down to scales of $k \sim 0.3 h{\rm Mpc}^{-1}$. On these scales, however, the impact of nonlinear effects, galaxy bias, and of the Vainshtein mechanism can be significant. N-body simulations of the Quartic Galileon model \cite{Li:2013tda, Barreira:2014zza} have shown that, for $k \sim 0.3 h{\rm Mpc}^{-1}$ at $z = 0$, the nonlinear matter power spectrum can differ from the linear theory prediction by $\sim 10 \mbox{--} 15 \%$. For the nonlinear velocity divergence power spectrum the difference is even larger reaching up to $40\%$ on these scales, which can have a big impact on RSDs.

The first observational constraints using the full shape of the CMB temperature anisotropy power spectrum were presented by us in Ref.~\cite{Barreira:2013jma}. We found that the amplitude of the low-$l$ region of the CMB power spectrum, which is mostly determined by the Integrated Sachs-Wolfe (ISW) effect, plays a decisive role in constraining the parameter space of the Galileon model. The use of the full shape of the CMB temperature data results in much tighter constraints than those obtained by using only the information encoded in the CMB distance priors \cite{Neveu:2014vua}. Reference~\cite{Barreira:2013jma} finds that the best-fitting Galileon models to the WMAP9 data \cite{Hinshaw:2012fq} have a lower ISW power relative to $\Lambda$CDM, which results in a better fit. However, some observational tensions become apparent when background data from SNIa and BAO is added to the analysis. We discuss these further in Sec.~\ref{sec:resultscubic}. In the constraints of Ref.~\cite{Barreira:2013jma}, the impact of massive neutrinos was not considered.

The best-fitting Galileon models to the WMAP9 data \cite{Barreira:2013jma} predict a relatively large amplitude for the matter fluctuations, $\sigma_8 \sim 1$. This has raised some concern about the ability of the model to match the observed amplitude of the galaxy power spectrum. This issue has been addressed by Ref.~\cite{Barreira:2014zza} by performing a Halo Occupation Distribution (HOD) \cite{Kravtsov:2003sg} analysis, using N-body simulations of the Cubic and Quartic Galileon models. It was found that, despite the enhanced clustering amplitude on large scales, the modifications to the halo and galaxy bias allow the models to match the observed clustering amplitude of LRG galaxies \cite{Reid:2009xm}, with realistic HODs for these galaxies. Therefore, the amplitude of the LRG power spectrum is not a problem in Galileon gravity.

In the Quartic and Quintic Galileon models, the direct couplings to $R$ and $G_{\mu\nu}$ in $\mathcal{L}_4$ and $\mathcal{L}_5$ in Eqs.~(\ref{L's}) give rise to modifications to gravity that may not be totally suppressed by the Vainshtein mechanism \cite{Babichev:2011iz, Kimura:2011dc}. In fact, Refs.~\cite{Li:2013tda, Barreira:2013xea} have shown that the Quartic Galileon models that best fit the WMAP9 data predict a non-negligible time variation of the effective gravitational strength near massive bodies like the Sun. The physical solutions of the Quintic model in the nonlinear regime are much more challenging to obtain, but it is likely that similar behavior arises. One then expects that Solar System tests \cite{Will:2014kxa} should be able to constrain the values of the $c_4$ and $c_5$ parameters to be very close to zero. We shall return to this discussion in subsection \ref{subsec:quartic-chains}.

\bigskip

Here, the main improvements over the above studies are: (i) the update of the CMB temperature data from WMAP to Planck; (ii) the use of the CMB lensing data to constrain Galileon models for the first time; and (iii) allowing $\Sigma m_\nu \neq 0$ to be a free parameter in the MCMC runs. The latter, as we shall see next, plays an important role in the observational viability of the Galileon model.

\section{Results: Cubic Galileon}\label{sec:resultscubic}

\begin{figure*}
	\centering
	\includegraphics[scale=0.390]{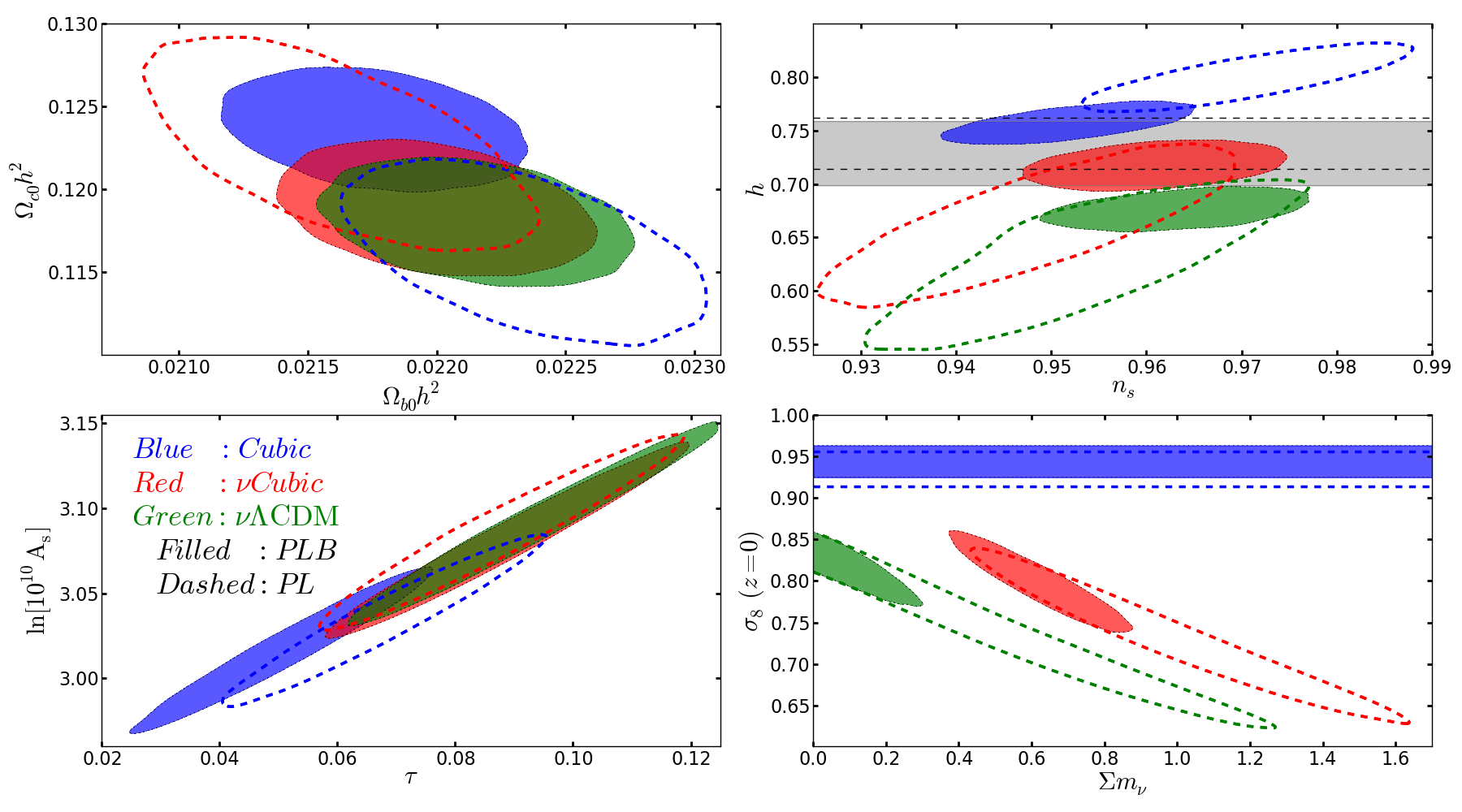}
	\caption{Marginalized two-dimensional $95\%$ confidence level contours obtained using the {\it PL} (open dashed) and {\it PLB} (filled) datasets for the base Cubic Galileon (blue), $\nucubic$ Galileon (red) and $\nulcdm$ (green) models. In the top right panel, the horizontal bands indicate the $68\%$ confidence limits of the direct measurements of $h$ presented in Ref.~\cite{Riess:2011yx} (open dashed) and Ref.~\cite{Humphreys:2013eja} (grey filled). In the lower right panel, the horizontal dashed bands indicate the $95\%$ confidence interval on $\sigma_8$ for the base Galileon model, for which $\Sigma m_\nu = 0$.}
\label{fig:cubic-contours}\end{figure*}

\begin{figure*}
	\centering
	\includegraphics[scale=0.390]{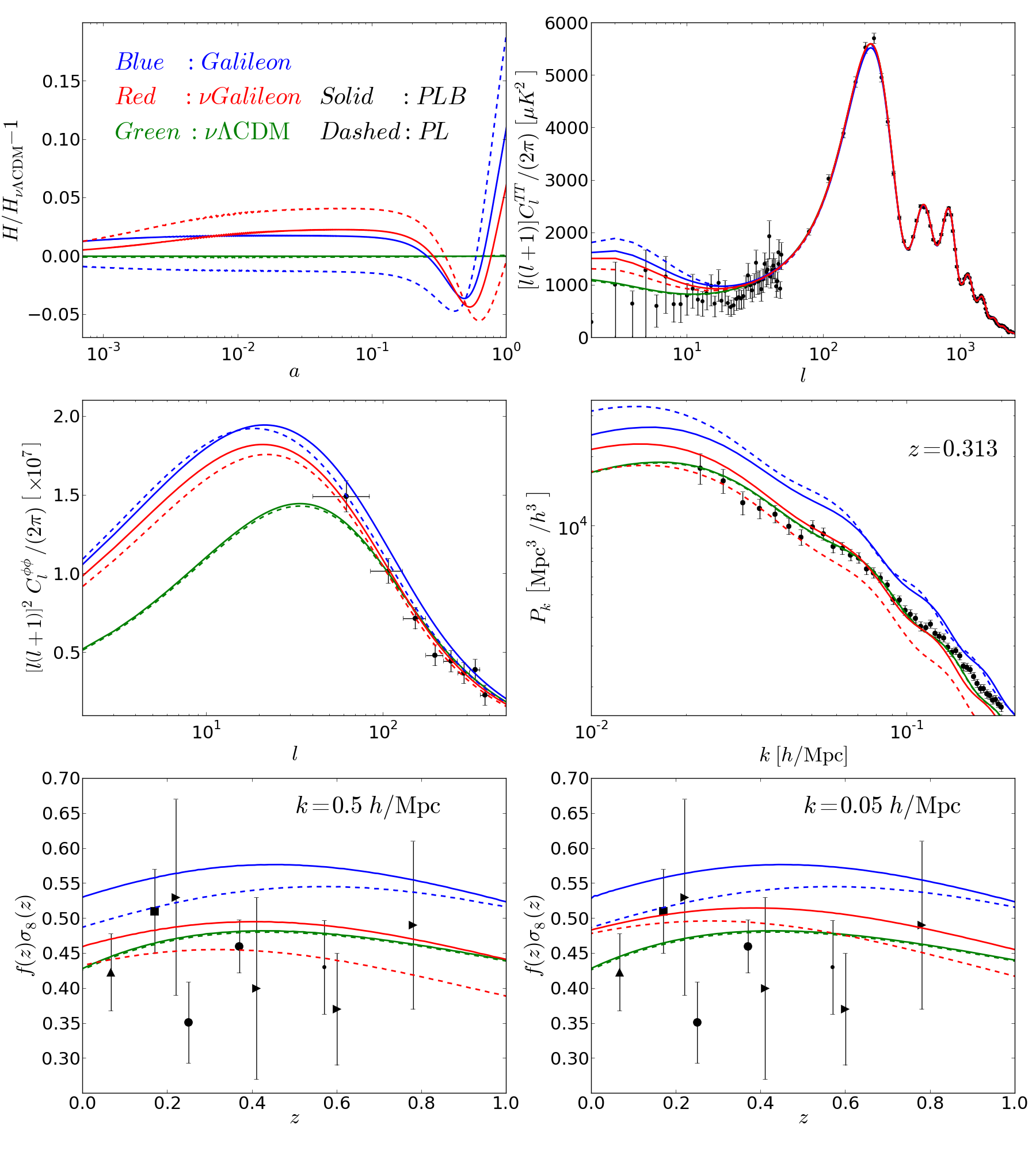}
	\caption{Time evolution of the Hubble expansion rate (top left), CMB temperature power spectrum (top right), CMB lensing potential power spectrum (middle left), linear matter power spectrum (middle right) and time evolution of $f\sigma_8$ for $k = 0.5 h/\rm{Mpc}$ (bottom left) and $k = 0.005 h/\rm{Mpc}$ (bottom right) for the best-fitting base Cubic (blue), $\nucubic$ (red) and $\nulcdm$ (green) models obtained using the {\it PL} (dashed) and {\it PLB} (solid) datasets. In the top left panel, the $\nulcdm$ model used in the denominator is the corresponding best-fitting model to the {\it PLB} dataset. In the top right and middle left panels, the data points show the power spectrum measured by the Planck satellite\cite{Ade:2013zuv, Ade:2013tyw}. In the middle right panel, the data points show the SDSS-DR7 Luminous Red Galaxy host halo power spectrum from Ref.~\cite{Reid:2009xm}, but scaled down by a constant factor to match approximately the amplitude of the best-fitting $\nucubic$ ({\it PLB}) model. In the lower panels, the data points show the measurements extracted by using the data from the 2dF \cite{1475-7516-2009-10-004} (square), 6dF \cite{Beutler:2012px} (triangle), SDSS DR7 (LRG) \cite{Samushia01032012} (circle), BOSS \cite{Reid:2012sw} (dot) and WiggleZ \cite{Blake:2011ep} (side triangles) galaxy surveys.}
\label{fig:cubic-bfs}\end{figure*}

\begin{figure*}
	\centering
	\includegraphics[scale=0.39]{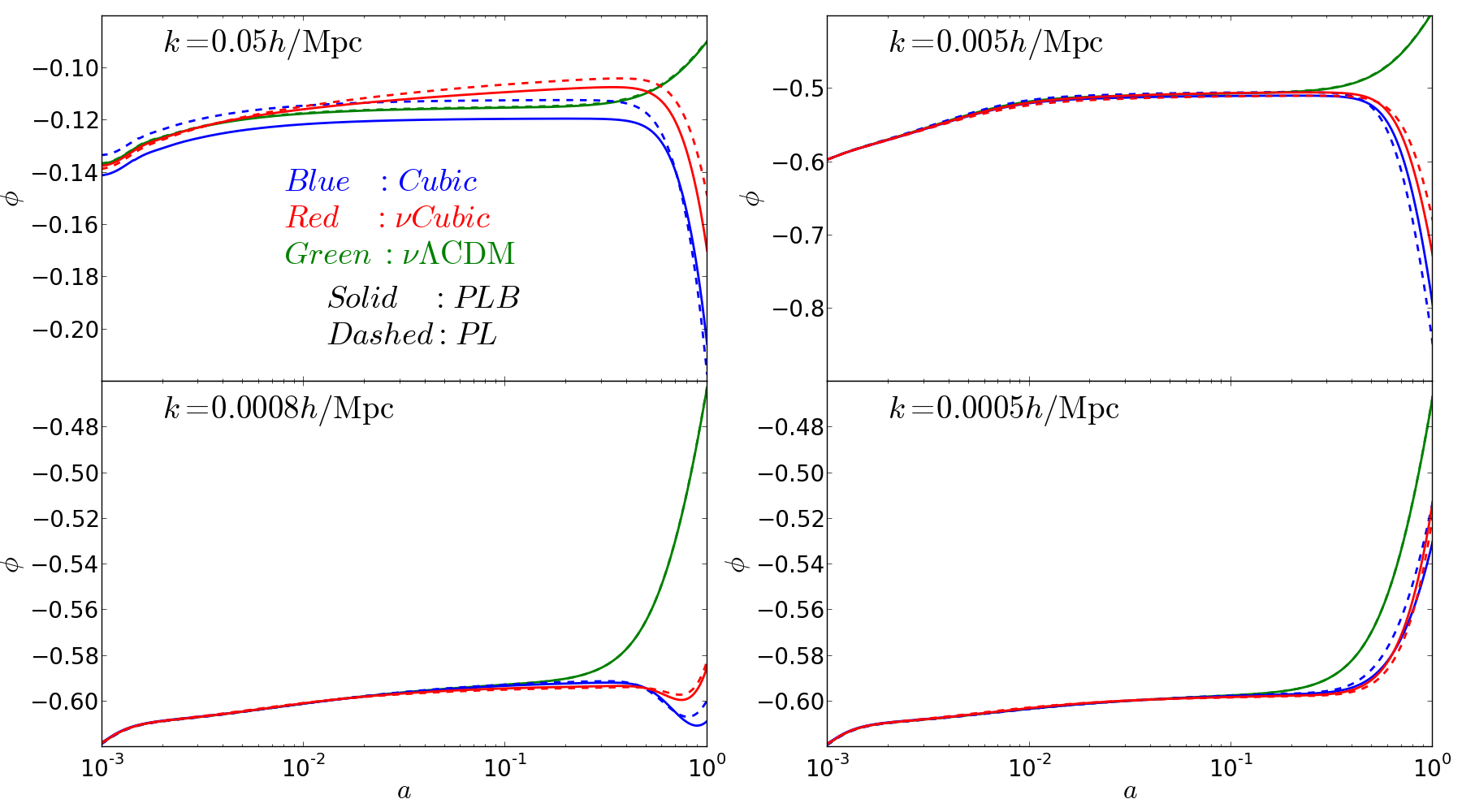}
	\caption{Time evolution of the lensing potential, $\phi$, for the best-fitting base Cubic (blue), $\nucubic$ (red) and $\nulcdm$ (green) for the {\it PL} (dashed) and {\it PLB} (solid) datasets for $k = 0.05 h/{\rm Mpc}$, $k = 0.005 h/{\rm Mpc}$, $k = 0.0008 h/{\rm Mpc}$ and $k = 0.0005 h/{\rm Mpc}$, as labelled in each panel.}
\label{fig:cubic-lenspot}\end{figure*}

\begin{figure}
	\centering
	\includegraphics[scale=0.43]{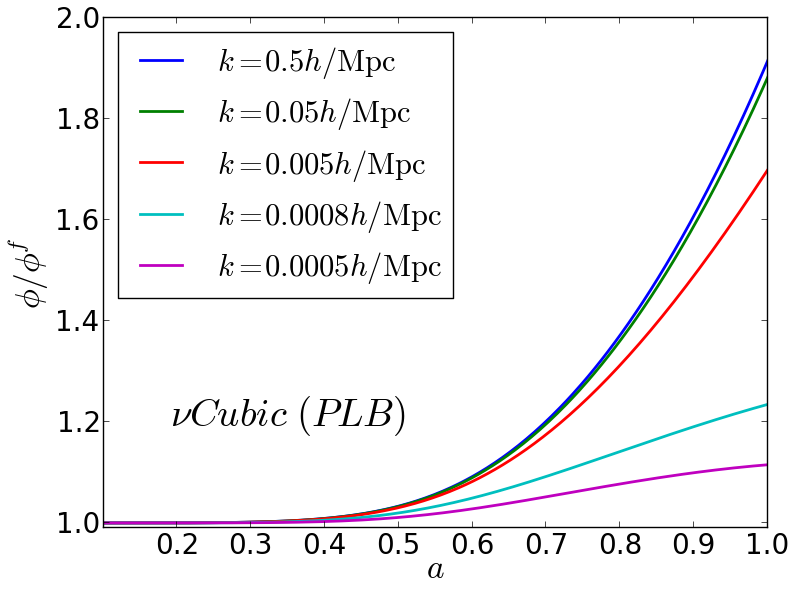}
	\caption{Time evolution of the ratio of the total lensing potential, $\phi$, and the lensing potential caused by the matter fluid only, $\phi^{f}$, for the $\nucubic$ ({\it PLB}) model and for a range of $k$ scales, as labelled. This quantity gives a measure of fifth force induced by the Galileon field.}
\label{fig:cubic-geff}\end{figure}

\begin{table*}
\caption{Summary of the one-dimensional marginalized likelihood distributions. The upper part of the table shows the best-fitting $\chi^2 = -2\rm{ln}\mathcal{L}$ values of the components of the {\it P}, {\it PL} and {\it PLB} datasets. The goodness of fit of the Galileon models can be inferred by comparing the respective $\chi^2$ values with those for $\nulcdm$, which has been shown to be a good fit to these data in \cite{Ade:2013zuv}. The CMB lensing and BAO datasets contain, respectively, eight and six degrees of freedom. It is not straightforward to quote the number of degrees of freedom for the CMB temperature data, due to the way in which the low-$l$ data are analyzed. The middle part of the table shows the corresponding best-fitting model parameters. In the lower part of the table, we show the $1\sigma$ limits on the cosmological parameters obtained for the {\it PL} and {\it PLB} datasets ($h$ and $\sigma_8$ are derived parameters). Recall that for the Cubic sector of Galileon gravity $c_4 = c_5 = 0$, $c_3$ and $\xi$ are given by Eqs.~(\ref{eq:cubic-fixings}) and $c_2 = -1$ to break the scaling degeneracy.}
\begin{tabular}{@{}lccccccccccc}
\hline
\hline
\rule{0pt}{3ex}  Parameter/Dataset  & \ \ ${\rm Base\ Cubic}$ & $\nucubic$ &  $\nulcdm$ &\ \ 
\\
\hline
\rule{0pt}{3ex} $(\chi^2_{P} ; --  ; --) $                                   &\ \  $(9829.8\ ; -- \ ; -- )$        &  \ \ $(9811.5\ ; -- \ ; -- )$      &  $(9805.5\ ; -- \ ; -- )$&\ \ 
\\
$(\chi^2_{P} ; \chi^2_{L} ; -- )$                 &\ \  $(9834.6\ ; 8.0\ ; --)$      &  \ \ $(9811.6\ ; 4.4\ ; --)$    &  $(9805.3\ ; 8.8\ ; --)$ &\ \ 
\\
$(\chi^2_{P} ; \chi^2_{L} ; \chi^2_{B})$                              &\ \  $(9836.7\ ; 19.5\ ; 8.4)$ &  \ \ $(9814.0\ ; 5.1\ ; 2.4)$ &  $(9805.7\ ;8.7\ ; 1.4)$ &\ \ 
\\
\hline
\rule{0pt}{3ex}$100\Omega_{b0} h^2$: {\ \ \ ({\it P, PL, PLB})}                   &\ \  $(2.216; 2.257; 2.173)$ &  \ \ $(2.166; 2.176; 2.202)$ & $(2.217; 2.208; 2.213)$ \ \ 
\\
$\Omega_{c0} h^2$: {\ \ \ \ \ \ \ \ \  ({\it P, PL, PLB})}          &\ \  $(0.121; 0.115; 0.124)$ &  \ \ $(0.123; 0.122; 0.120)$ & $(0.120; 0.118; 0.119)$ \ \ 
\\
$10^4\theta_{\rm MC}$: {\ \ \ \ \ \ ({\it P, PL, PLB})}                  &\ \  $(104.12; 104.18; 104.05)$ &  \ \ $(104.03; 104.06; 104.11)$ & $(104.15; 104.12; 104.10)$ \ \ 
\\
$\tau$: {\ \ \ \ \ \ \ \ \ \ \ \ \ \ \ \ \ ({\it P, PL, PLB})}             &\ \  $(0.070; 0.072; 0.052)$ &  \ \ $(0.099; 0.084; 0.088)$ & $(0.092; 0.086; 0.088)$ \ \ 
\\
$n_s$: {\ \ \ \ \ \ \ \ \ \ \ \ \ \ \ ({\it P, PL, PLB})}                   &\ \  $(0.963; 0.980; 0.955)$ &  \ \ $(0.949; 0.953; 0.958)$ & $(0.960; 0.964; 0.961)$ \ \ 
\\
${\rm ln}(10^{10}A_s)$: {({\it P, PL, PLB})}                            &\ \  $(3.054; 3.040; 3.021)$ &  \ \ $(3.111; 3.081; 3.084)$ & $(3.096; 3.079; 3.081)$ \ \ 
\\
$\Sigma m_\nu \ [\rm{eV}]$: {\ \ ({\it P, PL, PLB})}             &\ \  $(0\ {\rm fixed}; 0\ {\rm fixed}; 0\ {\rm fixed})$ &  \ \ $(1.043; 0.875; 0.538)$ & $(0.061; 0.043; 0.033)$ \ \ 
\\
\rule{0pt}{3ex}$h$: {\ \ \ \ \ \ \ \ \ \ \ \ \ \ \ \ \ ({\it P, PL, PLB})}                  &\ \  $(0.774; 0.810; 0.755)$ &  \ \ $(0.656; 0.677; 0.722)$ & $(0.674; 0.680; 0.679)$ \ \ 
\\
$\sigma_8(z = 0)$: {\ \ ({\it P, PL, PLB})}                              &\ \  $(0.959; 0.935; 0.949)$ &  \ \ $(0.729; 0.749; 0.822)$ & $(0.835; 0.827; 0.829)$ \ \ 
\\
\hline
\rule{0pt}{3ex}$100\Omega_{b0} h^2$: {\ \ \ ({\it PL, PLB})}                   &\ \  $(2.233 \pm 0.028\ ; 2.175 \pm 0.024)$ &  \ \ $(2.161 \pm 0.030\ ; 2.198 \pm 0.024)$ & $(2.182 \pm 0.035\ ; 2.215 \pm 0.025)$ \ \ 
\\
$\Omega_{c0} h^2$: {\ \ \ \ \ \ \ \ \  ({\it PL, PLB})}          &\ \  $(0.116 \pm 0.002\ ; 0.124 \pm 0.002)$ &  \ \ $(0.123 \pm 0.003\ ; 0.119 \pm 0.002)$ & $(0.121 \pm 0.003\ ; 0.118 \pm 0.002)$ \ \ 
\\
$10^4\theta_{\rm MC}$: {\ \ \ \ \ \ ({\it PL, PLB})}                  &\ \  $(104.17 \pm 0.061\ ; 104.05 \pm 0.055)$ &  \ \ $(104.04 \pm 0.066\ ; 104.10 \pm 0.056)$ & $(104.08 \pm 0.073\ ; 104.14 \pm 0.057)$ \ \ 
\\
$\tau$: {\ \ \ \ \ \ \ \ \ \ \ \ \ \ \ \ \ ({\it PL, PLB})}             &\ \  $(0.067 \pm 0.011\ ; 0.051 \pm 0.010)$ &  \ \ $(0.087 \pm 0.012\ ; 0.088 \pm 0.012)$ & $(0.091 \pm 0.013\ ; 0.092 \pm 0.013)$ \ \ 
\\
$n_s$: {\ \ \ \ \ \ \ \ \ \ \ \ \ \ \ ({\it PL, PLB})}                   &\ \  $(0.970 \pm 0.007\ ; 0.952 \pm 0.005)$ &  \ \ $(0.948 \pm 0.009\ ; 0.961 \pm 0.006)$ & $(0.954 \pm 0.009\ ; 0.963 \pm 0.006)$ \ \ 
\\
${\rm ln}(10^{10}A_s)$: {({\it PL, PLB})}                            &\ \  $(3.034 \pm 0.020\ ; 3.019 \pm 0.019)$ &  \ \ $(3.085 \pm 0.023\ ; 3.080 \pm 0.023)$ & $(3.093 \pm 0.024\ ; 3.090 \pm 0.024)$ \ \ 
\\
$\Sigma m_\nu \ [\rm{eV}]$: {\ \ ({\it PL, PLB})}             &\ \  $(0\ {\rm fixed}            \ ;  0\ {\rm fixed})$ &  \ \ $(0.980 \pm 0.237\ ; 0.618 \pm 0.101)$ & $( < 0.551\ ;  < 0.120)$ \ \ 
\\
\rule{0pt}{3ex}$h$: {\ \ \ \ \ \ \ \ \ \ \ \ \ \ \ \ \ ({\it PL, PLB})}                  &\ \  $(0.800 \pm 0.013\ ; 0.758 \pm 0.008)$ &  \ \ $(0.663 \pm 0.030\ ; 0.718 \pm 0.009)$ & $(0.634 \pm 0.036\ ; 0.678 \pm 0.008)$ \ \ 
\\
$\sigma_8(z = 0)$: {\ \ ({\it PL, PLB})}                              &\ \ $(0.935 \pm 0.010\ ; 0.944 \pm 0.010)$ &  \ \ $(0.733 \pm 0.042\ ; 0.798 \pm 0.024)$ & $(0.757 \pm 0.056\ ; 0.817 \pm 0.017)$ \ \ 
\\
\hline
\hline
\end{tabular}
\label{table:cubicdists}
\end{table*} 

The Cubic Galileon sector is defined by $c_4 = c_5 = 0$, which leaves $c_3$ and $\xi$ in Eq.~(\ref{eq:gali-params}) as potential free parameters (recall $c_2 = -1$ to break the scaling degeneracy). However, one can use Eqs.~(\ref{eq:constraint1}) and (\ref{eq:constraint2}) to fix two more Galileon parameters. For the case of the Cubic model we then get
\bq\label{eq:cubic-fixings}
c_3 = 1/\left(6\sqrt{6\Omega_{\varphi 0}}\right)\ ;\ \ \ \ \ \xi = \sqrt{6\Omega_{\varphi 0}}.
\eq
In this way, the only free parameters in the Cubic Galileon model are those in Eq.~(\ref{eq:cosmo-params}), just like in $\Lambda$CDM. This contrasts with popular modified gravity theories (with $f(R)$ gravity \cite{Sotiriou:2008rp, Motohashi:2012wc, He:2013qha, Baldi:2013iza} being perhaps the leading example), for which there are, in general, extra functions and parameters to tune, compared to $\Lambda$CDM.

\subsection{Cosmological constraints}

Figure \ref{fig:cubic-contours} shows two-dimensional $95\%$ confidence level marginalized contours obtained with the {\it PL} (dashed) {\it PLB} (filled) datasets for the base Cubic (blue), $\nucubic$ (red) and $\nulcdm$ (green) models. Table \ref{table:cubicdists} summarizes the one-dimensional marginalized likelihood ($\mathcal{L}$) statistics, showing also the best-fitting parameters and corresponding values of $\chi^2 = -2{\rm ln}\mathcal{L}$. Figure \ref{fig:cubic-bfs} shows $H(a)$, the CMB temperature and lensing power spectrum, the linear matter power spectrum and time evolution of the linear growth rate, $f = {\rm dln}D/{\rm dln}a$, expressed as $f\sigma_8$, for the best-fitting models.

\subsubsection{Observational tensions in the base Cubic model}

The $\chi^2$ values in the base Cubic model are significantly larger than those in the $\nucubic$ and $\nulcdm$ models, which indicates the markedly poorer fits of the base Cubic model. Moreover, the quality of the fit becomes worse as one combines the different datasets. In particular, when constrained with the {\it PLB} dataset, the base Cubic model fails to provide a reasonable fit to any of the likelihood components: $\chi^2_{Lensing} \sim 22$, for $8$ degrees of freedom (dof); $\chi^2_{BAO} \sim 8$, for $6$ dof \footnote{In Ref.~\cite{Barreira:2014ija}, without including the WiggleZ measurements in the BAO data, it was found that $\chi^2_{BAO} \sim 8$, for $3$ dof.}. This poorer fit to the data by the base Cubic model is primarily driven by the difficulty of the model in fitting, simultaneously, the BAO and the CMB peak positions.

The angular acoustic scale of the CMB fluctuations, $\theta^*$, is essentially what determines the CMB peak positions. It is given by $\theta^* = r_s^*/d_A^*$, where
\bq\label{eq:theta-cmb}
r_s^* &=& \int_{z_*}^\infty {\frac{c_s}{H(z)}{\rm d}z}, \\ 
d_A^* &=& \int_0^{z_*} {\frac{1}{H(z)}{\rm d}z},
\eq
are, respectively, the sound horizon and the comoving angular diameter distance to the redshift of recombination $z^*$; $c_s = 1/\sqrt{3\left(1 + 3\bar{\rho}_b/(4\bar{\rho}_\gamma)\right)}$ and $\bar{\rho}_b$ and $\bar{\rho}_\gamma$ are the background energy densities of baryons ($b$) and photons ($\gamma$). The constraints on $\theta^*$ tend to be fairly model independent, since they depend mostly on the peak positions, rather than the amplitude of the power spectrum (cf.~Table \ref{table:cubicdists}). From Eq.~(\ref{eq:tracker_H}), one can show that, at early times, $H(a)$ evolves in the same way in the Galileon and $\Lambda{\rm CDM}$ models. Hence, for fixed cosmological parameters, $r_s^*$ is also the same in these two models. However, at late times, $H(a)$ is smaller in the Galileon models compared to $\Lambda$CDM. This can be seen, again, by inspecting Eq.~(\ref{eq:tracker_H}) or by noting the late-time "dips" in $H/H_{\nulcdm} - 1$ in the top left panel of Fig.~\ref{fig:cubic-bfs} (although in this plot the cosmological parameters differ from model to model). The point here is that the smaller late-time expansion rate increases $d_A^*$, which in turn decreases $\theta^*$. In order to fit the CMB peak positions, the "intrinsically" smaller expansion rate at late times is compensated for by larger values of the expansion rate today, $h$, in such a way as to preserve the values of $d_A^*$, and hence $\theta^*$. The preference of the CMB data for high values of $h$ in the base Cubic model is illustrated in top right panel of Fig.~\ref{fig:cubic-bfs}. The lensing data lowers the matter density slightly to reduce the amplitude of the lensing power spectrum (the $C_l^{\phi\phi}$ for the base Cubic ({\it P}) model is not shown in Fig.~\ref{fig:cubic-bfs}, but is similar to that of the base Cubic ({\it PLB}) model). This increases both $r_s^*$ and $d_A^*$, but affects the latter more. As a result, and by the above reasoning, the addition of the CMB lensing to the CMB temperature data helps to push $h$ to even higher values (cf.~Table {\ref{table:cubicdists}}).

The inclusion of the BAO data counteracts the preference of the CMB data for higher values of $h$. The significance of this tension is illustrated by the offset between the contours obtained with the {\it PL} and {\it PLB} datasets for the base Cubic model. The addition of the BAO data also pushes the total matter density to higher values, which has an impact on the amplitude of both the CMB temperature and lensing spectra. This triggers a number of slight shifts in the remaining cosmological parameters in order to optimize the fit. Nevertheless, this optimization is not perfect, and the base Cubic model ultimately fails to fit the combined data well. In addition to the poor BAO fit, the base Cubic model predicts a high amplitude for $C_l^{TT}$ at low-$l$ (top right panel of Fig.~\ref{fig:cubic-bfs}), caused by a rapid late-time deepening of the lensing potential (cf.~top panels of Fig.~\ref{fig:cubic-lenspot}). The amplitude of the lensing power spectrum, $C_l^{\phi\phi}$ is also visibly larger than the data (middle left panel of Fig.~\ref{fig:cubic-bfs}).

\subsubsection{Alleviating the tensions with $\Sigma m_\nu$}

Although at sufficiently early times massive neutrinos act as an extra source of radiation, at late times (after becoming non-relativistic), they will effectively raise the total matter density, modifying the evolution of $H(a)$ accordingly. In particular, higher values of $\Sigma m_\nu$ increase $H(a)$ at late times, and therefore have the same impact as increasing $h$ on the value of $d_A^*$. This degeneracy between $\Sigma m_\nu$ and $h$ eliminates the preference of the CMB data for large values of $h$, as shown by the $\nucubic$ contours for the {\it PL} dataset in Fig.~\ref{fig:cubic-contours}. An important consequence of this is that, in the $\nucubic$ model, there is no longer a tension between the CMB and the BAO data, as illustrated by the overlap between the contours for the {\it PL} and {\it PLB} datasets and by the acceptable $\chi^2$ values listed in Table \ref{table:cubicdists}.

The presence of the massive neutrinos causes the lensing potential to deepen less rapidly with time, which reduces the amplitude of the CMB temperature power spectrum at large angular scales. On these scales, there is still an excess of power compared to $\nulcdm$, but the large errorbars do not allow tight constraints to be derived. Note also that in the $\nucubic$ model for $k \gtrsim 0.05 h/{\rm Mpc}$, the presence of the massive neutrinos causes the gravitational potentials to decay slightly during the matter era, rather than remaining constant. The massive neutrinos also lower substantially the amount of matter clustering (lower $\sigma_8$), which results in a better fit to the CMB lensing power spectrum. Compared to the $\nulcdm$ model, the $\nucubic$ model provides a slightly better fit, as it predicts more power at $l \sim 40-80$ and the amplitude decreases more rapidly at higher $l$. 

We note, for completeness, that relaxing the assumption that the universe is spatially flat may also help to alleviate the tension between the CMB and BAO peak positions. In particular, $\Omega_k < 0$ also lowers $d_A^*$, and as a result, may mimic to some extent the effect of $\Sigma m_\nu > 0$ on $H(a)$. We leave for future work the study of the impact of $\Omega_k$ on the ISW effect (see next subsection), and CMB temperature and lensing spectra.

\subsection{Sign of the ISW effect}\label{sub:isw}

The ISW effect is a secondary anisotropy on the CMB temperature maps induced by time-evolving gravitational potentials. Consider for instance a photon travelling through a supercluster whose potential is getting shallower with time. This photon will get blueshifted (increase of temperature) as it goes into the center of the potential well, but redshifted (decrease of temperature) as it comes out of it. Since the potential was deeper at the time the photon was entering it, overall the temperature of the photon will increase. This causes a so-called "hot spot" in the CMB maps. If the potential of the supercluster is getting deeper with time, then one would end up with a "cold spot" instead.

The amplitude of the ISW effect is proportional to the time derivative of the lensing potential, ${\rm d}\phi/{\rm d}t$, integrated along the line of sight. In Fourier space, $\phi$ is given by the equation \footnote{In terms of the $\Psi$ and $\Phi$ potentials of the linearly perturbed FRW line element in the Newtonian gauge ${\rm d}s^2 = \left(1 + 2\Psi\right){\rm d}t^2 - \left(1 - 2\Phi\right){\rm d}{x}^i{\rm d}{x}_i$, one has $\phi = \left(\Psi - \Phi\right)/2$.}

\bq\label{eq:phieq}
-k^3\phi = 4\pi Ga^2\left[k\left(\Pi + \chi\right) + 2aH(a)q\right],
\eq
where $\chi$, $q$ and $\Pi$ are, respectively, the Fourier modes of the total density perturbation, heat flux and anisotropic stress (see Ref.~\cite{Barreira:2012kk} for more details) \footnote{The $q$ term is subdominant on small length scales (large $k$)  and for matter $\Pi = 0$. In this case, one then recovers the standard Poisson equation $-k^2\phi = 4\pi G a^2\chi$.}. Figure \ref{fig:cubic-lenspot} shows the time evolution of $\phi$ for the best-fitting models for four different scales $k = 0.05 h/{\rm Mpc}$, $k = 0.005 h/{\rm Mpc}$, $k = 0.0008 h/{\rm Mpc}$ and $k = 0.0005 h/{\rm Mpc}$. In the standard $\Lambda{\rm CDM}$ picture, $\phi$ grows at the transition from the radiation to the matter dominated eras, stays approximately constant during the matter era ($\Omega_m \sim 1$), and starts decaying (note the negative sign on the $y$-axis) at the onset of the dark energy era. The physical picture in the Cubic Galileon models is more complex. During the matter era, $\phi$ also remains approximately constant, although on smaller length scales $k \gtrsim 0.05 h/{\rm Mpc}$, the presence of the massive neutrinos can cause $\phi$ to decay slightly. The modifications induced by the Galileon field become apparent at later times ($a \gtrsim 0.5$) and are scale-dependent. For $k \gtrsim 0.005 h/{\rm Mpc}$, $\phi$ deepens at late times, whereas for $k \lesssim 0.0005 h/{\rm Mpc}$ it decays. On intermediate scales ($k \sim 0.0008 h/{\rm Mpc}$) the potential can remain approximately constant, even at late times, undergoing only small amplitude oscillations. To help understand the scale-dependent behaviour of $\phi$ in the Cubic model, we plot the time evolution of $\phi/\phi^{f}$ for a range of scales $k$ in Fig.~\ref{fig:cubic-geff}. The quantity $\phi^{f}$ is given by Eq.~(\ref{eq:phieq}), but considering only the contribution from the matter fluid in $\chi$, $q$ and $\Pi$. This isolates the impact of the Galileon field, and as such $\phi/\phi^{f}$ provides a measure of the fifth force modifications to the lensing potential. Firstly, we note that the Galileon field contribution only becomes nonnegligible at late times, i.e., $\phi/\phi^{f} \approx 1$ for $a \lesssim 0.4$. At late times, on smaller length scales (larger $k$), the Galileon field contributes significantly to $\phi$, making it deeper. On the other hand, on larger length scales (smaller $k$), the Galileon terms become less important, which leads to a gradual recovery of the $\Lambda$CDM behaviour, i.e., $\phi$ decays at late times. 

The physical picture depicted in Fig.~\ref{fig:cubic-lenspot} suggests that the Cubic Galileon and $\Lambda$CDM models predict opposite signs for the ISW effect on sub-horizon scales, a fact that can potentially be used to distinguish between them. The CMB temperature power spectrum is sensitive to $\left({\rm d}\phi/{\rm d}t\right)^2$, and hence it cannot probe the sign of the ISW effect. There are however a number of different techniques that can be used to determine ${\rm d}\phi/{\rm d}t$. One of these consists of stacking CMB maps at the locations of known superclusters and supervoids. Given their size, these superstructures are not yet virialized, and hence constitute good probes of the ISW effect since their potentials are still evolving. A recent analysis of this type was performed by the Planck collaboration \cite{Ade:2013dsi} who claimed to have found a detection of a positive ISW effect using the superstructure catalogue of Refs.~\cite{Granett:2008ju, 2008arXiv0805.2974G}. The significance of this detection becomes, however, substantially weaker when the catalogues of Refs.~\cite{Sutter:2012wh} and \cite{Pan:2011hx} are used instead. Moreover, all these signals are typically higher than the standard $\Lambda$CDM expectation \cite{HernandezMonteagudo:2012ms, Cai:2013ik}. This fact, together with the differences between using different cluster and void catalogues, may raise concerns about the presence of unknown systematics in the analysis, such as selection effects. More recently, Refs.~\cite{Finelli:2014yha, Szapudi:2014zha} claimed the detection of a supervoid aligned with a prominent cold spot in the Planck CMB maps, as one would expect in models with positive ISW effect.

The cross-correlation of the CMB with tracers of large-scale structure (LSS) provides another way to probe the ISW effect. A positive amplitude for this cross correlation was first detected in Ref.~\cite{2004Natur.427...45B}, and later confirmed by Refs.~\cite{2008PhRvD..78d3519H, Giannantonio:2012aa, Ade:2013dsi}, although with different significances. The cross correlation functions obtained by using different galaxy catalogues typically show positive correlation at smaller angular scales, and become consistent with zero at large angular scales (see e.g.~Fig.~3 of Ref.~\cite{Giannantonio:2012aa}). This trend is consistent with the $\Lambda{\rm CDM}$ expectation, but Refs.~\cite{Francis:2009pt, Francis:2009ps, HernandezMonteagudo:2009fb, Sawangwit:2009gd, LopezCorredoira:2010rr} have raised some skepticism about the significance of these claims for a positive detection (some of this skepticism is addressed in Ref.~\cite{Giannantonio:2012aa}).

The potential $\phi$ is responsible for both the lensing of the CMB photons and the ISW effect. As a result, cross correlating CMB temperature maps with maps of the lensing potential (used to measure $C_l^{\phi\phi}$ in Fig.~\ref{fig:cubic-bfs}) can potentially be used to probe the sign and amplitude of the ISW effect. This has been made possible after the data release by the Planck collaboration \cite{Ade:2013dsi}, who found a signal that is consistent with the $\Lambda$CDM expectation that $\phi$ decays at late times.

Taken at face value, the above-mentioned measurements seem to be inconsistent with the predictions of the base Cubic and $\nucubic$ Galileon models. Note that this is not a question of matching the amplitude of the signal, but instead its sign, and as a result, it may be hard to reconcile the model predictions with the claims of a positive ISW effect. Note also that although $\phi$ can decay in the Cubic models, this happens only on horizon-like scales, which do not affect the observational measurements. However, there is still some ongoing discussion about the understanding of the systematics in these measurements of the ISW effect. This makes us reluctant to add these data to the constraints at present. Moreover, in the case of Galileon gravity there is also the potential impact of the Vainshtein screening mechanism, which is unaccounted for in linear perturbation theory studies. For instance, on smaller scales, where the ISW detections are more significant, the screening mechanism may suppress the modifications to gravity, making the potentials decay as in $\Lambda$CDM. For the time being, we limit ourselves to noting that the positiveness of the ISW effect may turn out to be a crucial observational tension of the $\nucubic$ model. In the future, one will be able to say more about it, as more data become available and the discussion about the role of systematic effects is settled, and also when fully nonlinear predictions are used to model the signal.

\subsection{Future constraints}

We now discuss briefly the impact that additional data can have in further constraining the $\nucubic$ model.

For $l \lesssim 40$, the $\nucubic$ and $\Lambda{\rm CDM}$ models make quite distinct predictions for $C_l^{\phi\phi}$. As a result, future measurements of the lensing potential on these angular scales have a strong potential to discriminate between these two models, provided the errorbars are small enough \cite{Ade:2013tyw}.

The horizontal bands in the top right panel of Fig.~\ref{fig:cubic-contours} show the $1\sigma$ limits of the direct determinations of the Hubble constant $h$ using Cepheid variables reported in Ref.~\cite{Riess:2011yx} (open dashed) and Ref.~\cite{Humphreys:2013eja} (grey filled). As one can see in the figure, these determinations are in tension with the CMB constraints for $\Lambda$CDM models. This fact has been the subject of discussions about the role that systematic effects can play in these direct measurements of $h$ (see e.g.~Ref.~\cite{Efstathiou:2013via}). This is why we did not include them in our constraints. Here, we simply note that $\nucubic$ models avoid the tensions apparent in $\Lambda{\rm CDM}$, and therefore, adding a prior for $h$ would favour the $\nucubic$ over $\Lambda{\rm CDM}$.

Another $\Lambda{\rm CDM}$ tension that has become apparent after the release of the Planck data is associated with the normalization of the matter density fluctuations. In short, the values of $\sigma_8$ inferred from probes such as galaxy shear \cite{Heymans:2013fya} and cluster number counts \cite{Ade:2013lmv} seem to be smaller than the values preferred by the CMB constraints. Massive neutrinos have been shown to alleviate some of these problems \cite{Wyman:2013lza, Battye:2013xqa}. However, some residual tensions between datasets seem to remain. In the case of the $\nucubic$ model, the presence of the massive neutrinos lowers substantially the value of $\sigma_8$. Compared to $\Lambda{\rm CDM}$, the constraints on $\sigma_8$ are rather similar, although they can extend to slightly lower values (cf.~Fig.~\ref{fig:cubic-contours}). This happens despite the enhanced gravitational strength driven by the Galileon field. It is therefore interesting to investigate whether or not the $\nucubic$ model can evade the above-mentioned $\Lambda$CDM tensions. This requires the modelling of nonlinear structure formation in the $\nucubic$ model, which is left for future work.

In the context of the $\nucubic$ model, the {\it PLB} dataset suggests that $\Sigma m_\nu > 0$ at more than $6 \sigma$ significance (cf.~Table \ref{table:cubicdists}). This contrasts with the constraints on $\nulcdm$, for which $\Sigma m_\nu < 0.3\ {\rm eV}$ (at $2\sigma$). This opens an interesting window for upcoming terrestrial determinations of the absolute neutrino mass scale (see e.g.~Ref.~\cite{Drexlin:2013lha} for a review) to distinguish between these two models. For instance, the high energy part of the Tritium $\beta$-decay spectrum provides a robust and model-independent way to measure the mass of the electron neutrino directly. The MAINZ \cite{Kraus:2004zw} and TROITSK \cite{Aseev:2011dq} experiments have set $\Sigma m_\nu \lesssim 6.6 {\rm eV}$ (at $2\sigma$), but near-future experiments such as KATRIN \cite{katrin} are expected to improve the mass sensitivity down to $\Sigma m_\nu \lesssim 0.6 {\rm eV}$. In the case that neutrinos are Majorana particles and provide the dominant contribution in the neutrinoless double $\beta$-decay of heavy nuclei \cite{Vergados:2012xy}, then one may achieve even higher sensitivity: in case of nondetection, these type of experiments are expected to constrain $\Sigma m_\nu \lesssim 0.3 {\rm eV}$. This would completely probe the quasi-degenerate neutrino hierarchy spectrum ($m_1 \sim m_2 \sim m_3 \sim m_\nu > 0.1\ {\rm eV}$). The forecast sensitivity of these experiments should be reached in a few years time and will say more about the viability of the $\nulcdm$ and $\nucubic$ models.

The lower panels of Fig.~\ref{fig:cubic-bfs} show the time evolution of $f\sigma_8$ in the best-fitting models (computed using linear theory), together with the measurements from the 2dF \cite{1475-7516-2009-10-004} (square), 6dF \cite{Beutler:2012px} (triangle), SDSS DR7 (LRG) \cite{Samushia01032012} (circle), BOSS \cite{Reid:2012sw} (dot) and WiggleZ \cite{Blake:2011ep} (side triangles). In principle, these data can be used to further constrain the $\nucubic$ model. However, such a comparison between theory and observation may not be straightforward for at least three reasons. The first one was already addressed in Sec.~\ref{sec:pastconstraints} and it relates to the validity of linear theory on the length scales probed by the surveys. As discussed in Sec.~\ref{sec:pastconstraints}, on these scales, nonlinearity can affect the statistics of both the density and velocity fields, and hence modify significantly the linear theory expectations (see e.g.~\cite{Samushia01032012} for some discussion). The growth measurements are extracted from the data by analyzing the redshift space distortions induced by galaxy peculiar motions. This is usually achieved by assuming a model for how these peculiar velocities modify the true (unobserved) real space statistics. These models are typically calibrated and tested against N-body simulations, most of which are performed assuming GR (see however \cite{2011MNRAS.410.2081J, Jennings:2012pt, Wyman:2013jaa}). Here lies the second nontrivial aspect: to avoid obtaining results biased towards standard gravity, it seems reasonable to demand first the development of a self-consistent RSD model for modified gravity to see how it can have an impact on the extraction of the $f\sigma_8$ values from the galaxy catalogues. The third complication has to do with the scale-dependent growth introduced by the massive neutrinos, even at the linear level. In the lower panels of Fig.~\ref{fig:cubic-bfs}, one notes that the predictions of the $\nucubic$ model are, indeed, scale dependent, due to the relatively large massive neutrino fraction, compared to $\nulcdm$. However, the measured values of $f\sigma_8$ obtained from the different surveys are derived from the clustering signal of galaxies measured over a range of different scales at once. Future constraints on the $\nucubic$ model using these data have therefore to take this scale dependence into account. We note that this third complication also applies to $\Lambda$CDM models with a large value of $\Sigma m_\nu$, such as those found in Refs.~\cite{Wyman:2013lza, Battye:2013xqa}; and to $f(R)$ gravity models (see e.g. Ref~\cite{Jennings:2012pt} for a study of RSD in $f(R)$ gravity).

\section{Results: Quartic and Quintic Galileon}\label{sec:results45}

\begin{figure*}
	\centering
	\includegraphics[scale=0.390]{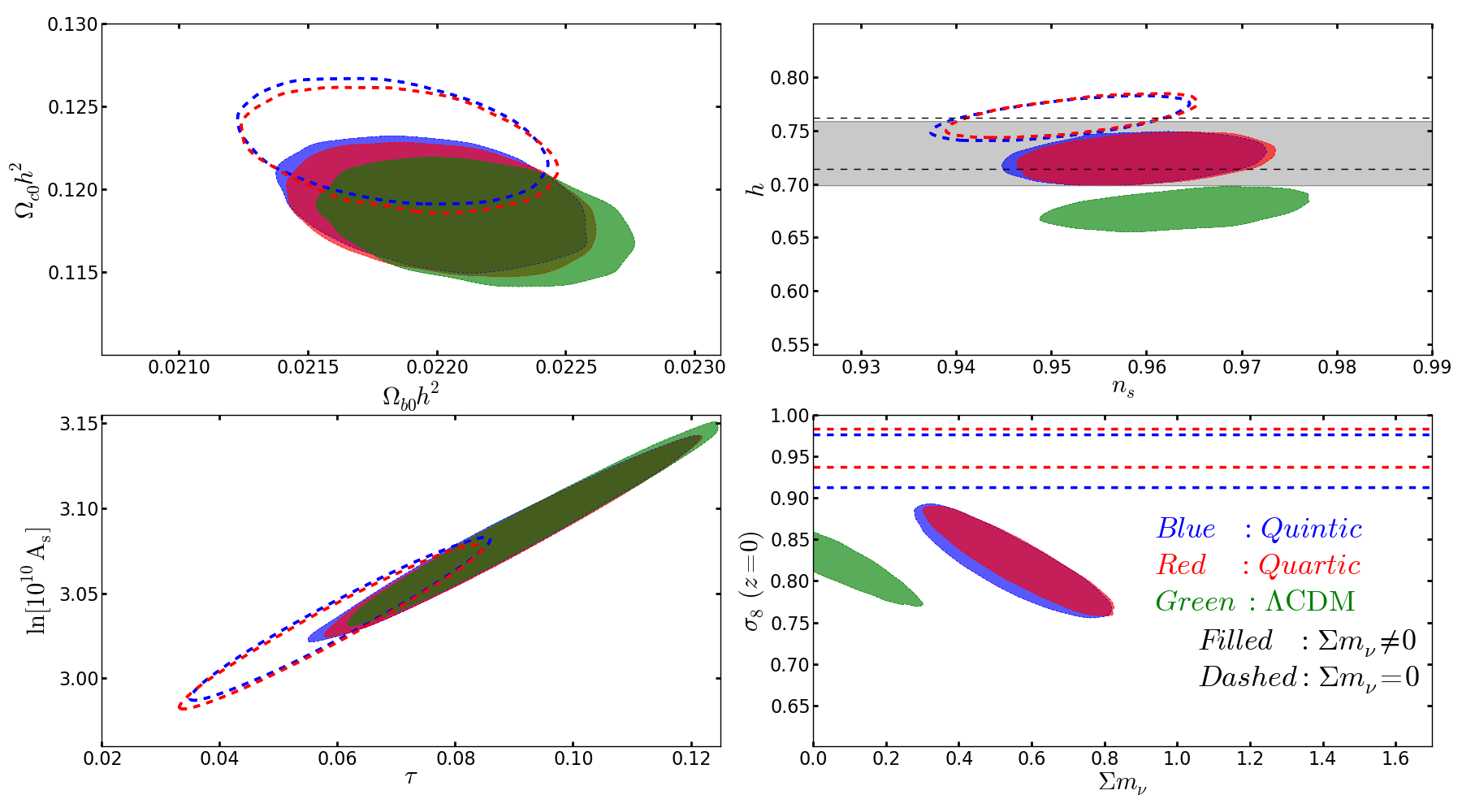}
	\caption{Same as Fig.~\ref{fig:cubic-contours} but for the base Quartic (red dashed), $\nuquartic$ (red filled), base Quintic (blue dashed) and $\nuquintic$ (blue filled) models, using the {\it PLB} dataset.}
\label{fig:45-contours}\end{figure*}

\begin{figure*}
	\centering
	\includegraphics[scale=0.390]{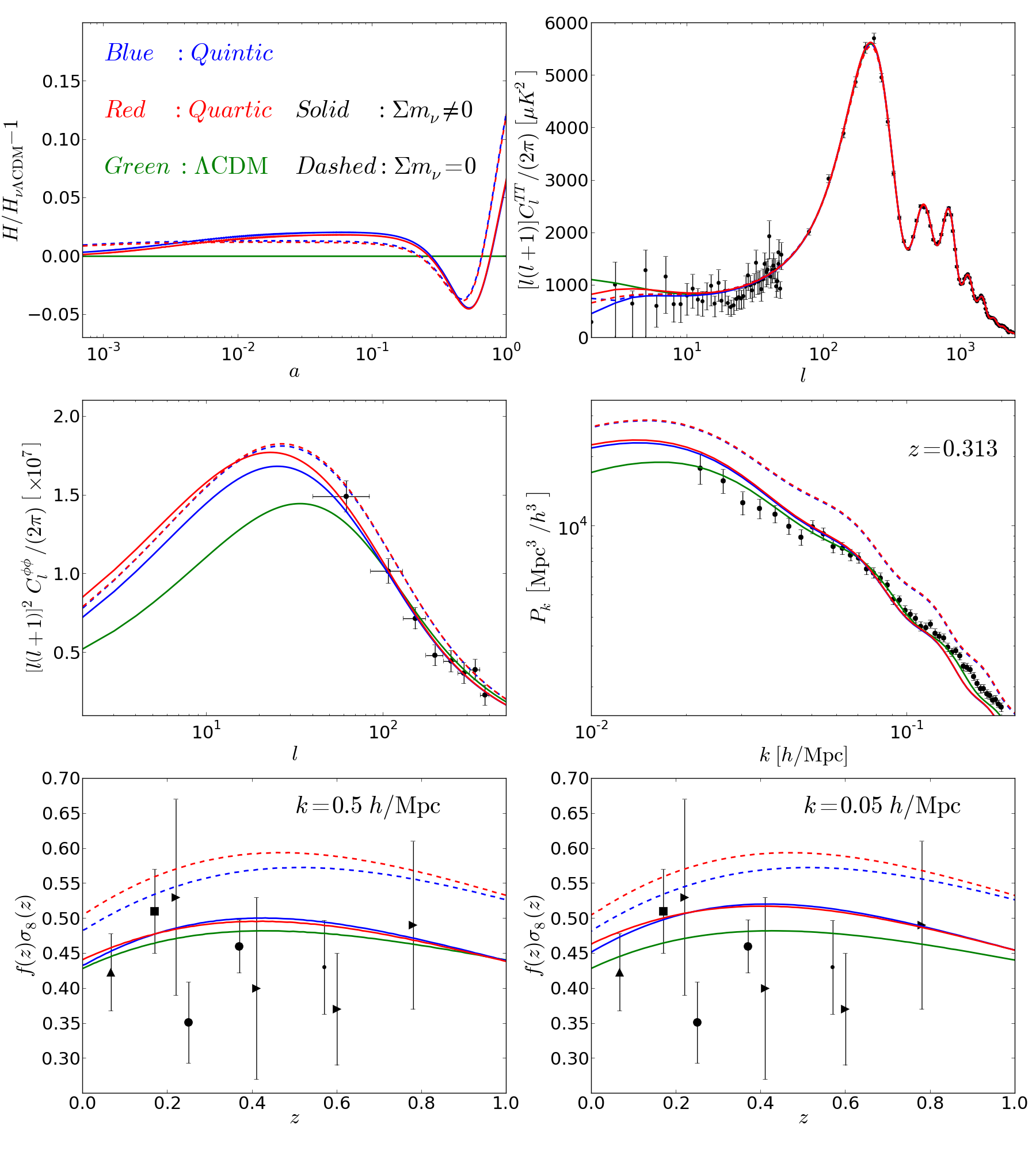}
	\caption{Same as Fig.~\ref{fig:cubic-bfs} but for the base Quartic (red dashed), $\nuquartic$ (red solid), base Quintic (blue dashed) and $\nuquintic$ (blue solid) models that best fit the {\it PLB} dataset.}
\label{fig:45-bfs}\end{figure*}

\begin{figure*}
	\centering
	\includegraphics[scale=0.39]{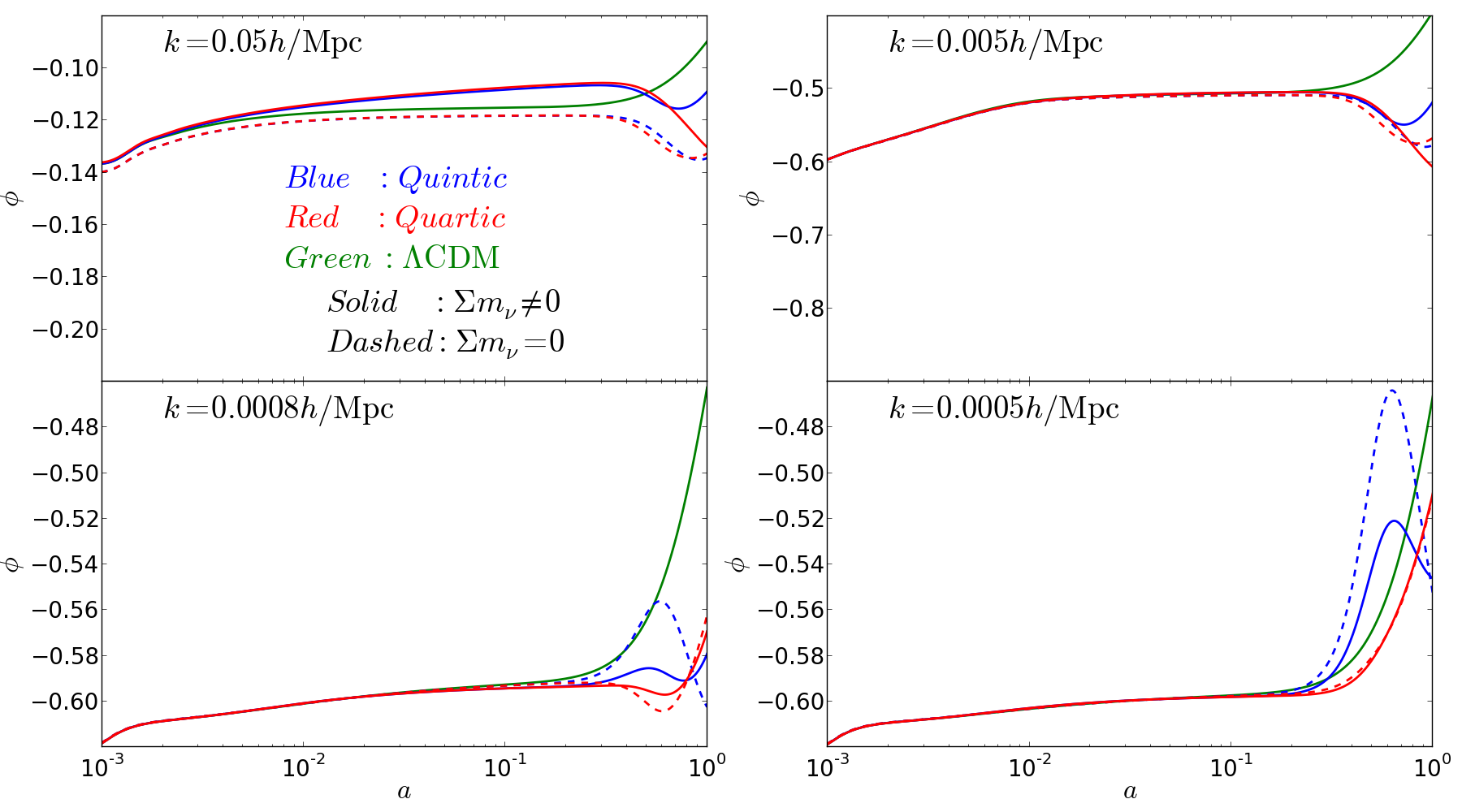}
	\caption{Same as Fig.~\ref{fig:45-lenspot} but for the base Quartic (red dashed), $\nuquartic$ (red solid), base Quintic (blue dashed) and $\nuquintic$ (blue solid) models that best fit the {\it PLB} dataset.}
\label{fig:45-lenspot}\end{figure*}

\begin{figure}
	\centering
	\includegraphics[scale=0.43]{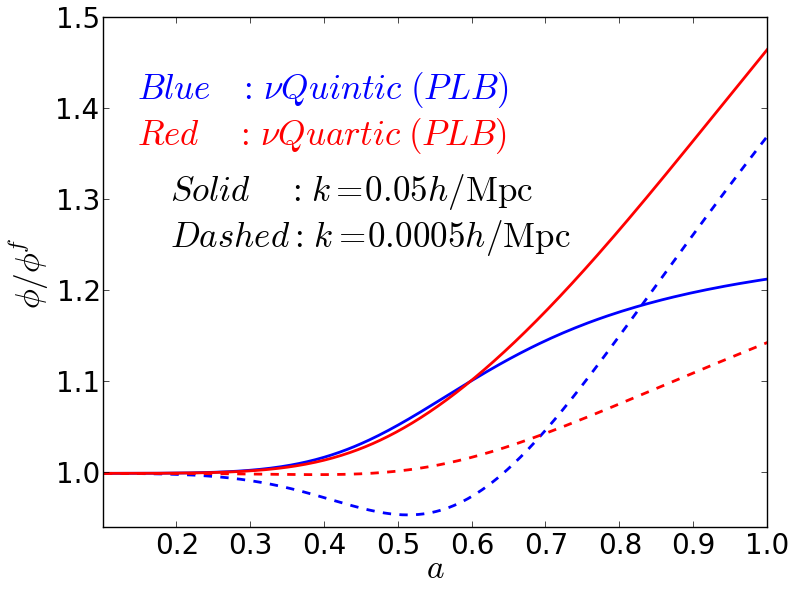}
	\caption{Same as Fig.~\ref{fig:cubic-geff} but for the $\nuquartic$ (red) and $\nuquintic$ (blue) Galileon models that best-fit the {\it PLB} dataset. The solid and dashed lines correspond to $k = 0.05 h/{\rm Mpc}$ and $k = 0.0005 h/{\rm Mpc}$, respectively.}
\label{fig:45-geff}\end{figure}

\begin{figure}
	\centering
	\includegraphics[scale=0.44]{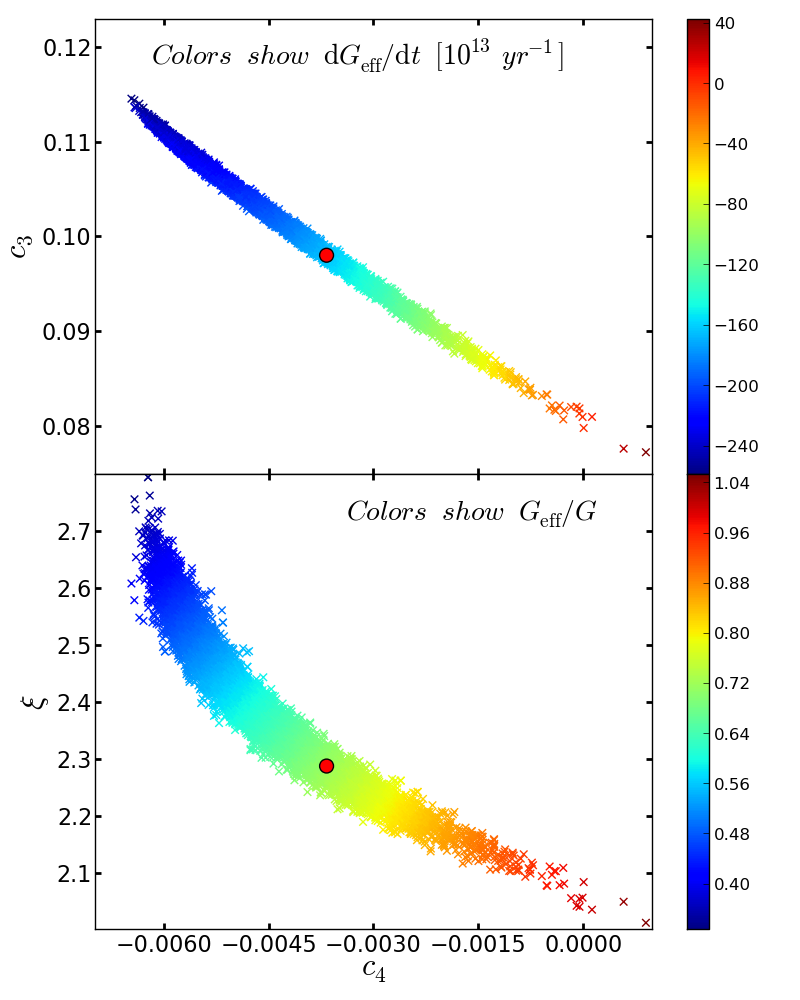}
	\caption{Sample of $10^4$ randomly selected points from the chains used to constrain the $\nuquartic$ model with the {\it PLB} dataset, projected onto the Galileon subspace of parameters. The points are coloured according to their respective values of $G_{\rm eff}/G(a, \delta)$ (lower panel) and $\dot{G}_{\rm eff}/G(a, \delta)$ (upper panel), at $a = 1$ and for top-hat profiles with $\delta = 10^7$ (see text). The big red dot indicates the position of the best-fitting point of (cf.~Table \ref{table:45dists}).}
\label{fig:quartic-chains}\end{figure}

\begin{figure}
	\centering
	\includegraphics[scale=0.44]{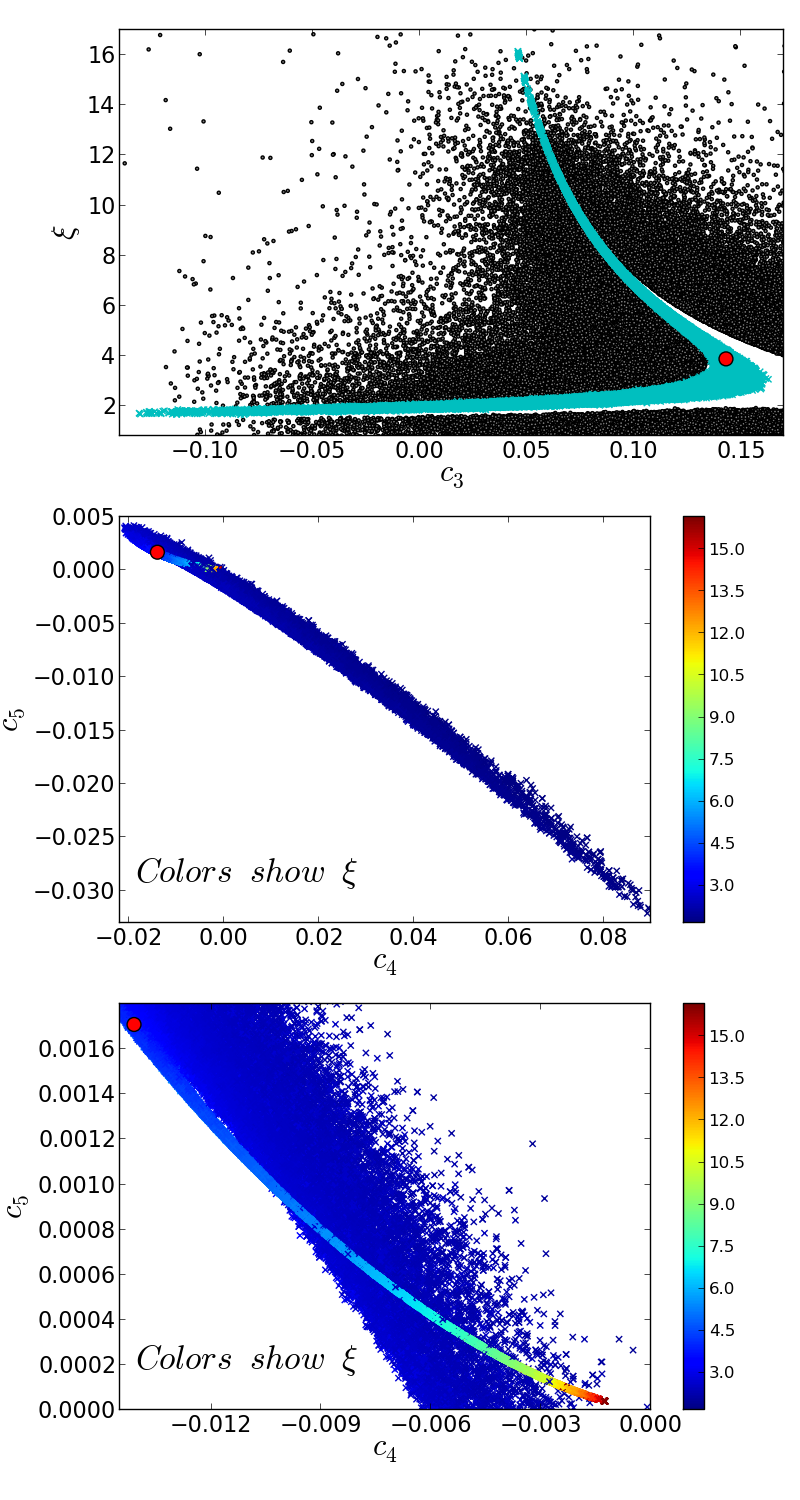}
	\caption{Accepted MCMC points (after the burn-in period) obtained in the constraints of the $\nuquintic$ model using the {\it PLB} dataset, projected onto the $\xi-c_3$ (top panels) and $c_5-c_4$ (middle and lower panels) planes. In the top panel, the black dots indicate those points that were tried during the sampling but failed to meet the conditions for the absence of ghost and Laplace instabilities of the scalar fluctuations. In the middle and lower panels, the points are color coded according to their value of $\xi$. The lower panel zooms into a region of the middle panel. The big red dot indicates the best-fitting point.}
\label{fig:quintic-chains}\end{figure}

\begin{table*}
\caption{Same as Table \ref{table:cubicdists}, but for the Quartic and Quintic models and using the {\it PLB} dataset. Recall that for the Quartic model constraints, $c_5 = 0$, and $c_3$, $c_4$ are derived parameters. For the Quintic model, the derived parameters are $c_4$ and $c_5$. For both models, $c_2 = -1$ to break the scaling degeneracy.}
\begin{tabular}{@{}lccccccccccc}
\hline
\hline
\rule{0pt}{3ex}  Parameter/Dataset  & \ \ ${\rm Base\ Quartic}$ & $\nuquartic$ &  ${\rm Base\ Quintic}$ & $\nuquintic$ &\ \ 
\\
\hline
\\
$(\chi^2_{P} ; \chi^2_{L} ; \chi^2_{B})$                              &\ \  $(9813.6\ ; 15.6\ ; 13.3)$ &  \ \ $(9805.9\ ; 4.7\ ; 2.2)$ &  $(9805.0\ ;20.5\ ; 13.4)$ & $(9800.4\ ;5.2\ ; 2.2)$ &\ \ 
\\
\hline
\rule{0pt}{3ex}$100\Omega_{b0} h^2$: {\ \ \ ({\it PLB})}                   &\ \  $(2.175)$ &  \ \ $(2.200)$ & $(2.219)$ & $(2.211)$ \ \ 
\\
$\Omega_{c0} h^2$: {\ \ \ \ \ \ \ \ \  ({\it PLB})}          &\ \  $(0.123)$ &  \ \ $(0.120)$ & $(0.123)$ & $(0.119)$\ \ 
\\
$10^4\theta_{\rm MC}$: {\ \ \ \ \ \ ({\it PLB})}                  &\ \  $(104.07)$ &  \ \ $(104.13)$ & $(104.07)$ & $(104.08)$\ \ 
\\
$\tau$: {\ \ \ \ \ \ \ \ \ \ \ \ \ \ \ \ \ ({\it PLB})}             &\ \  $(0.058)$ &  \ \ $(0.095)$ & $(0.064)$  & $(0.082)$\ \ 
\\
$n_s$: {\ \ \ \ \ \ \ \ \ \ \ \ \ \ \ ({\it PLB})}                   &\ \  $(0.951)$ &  \ \ $(0.957)$ & $(0.950)$ & $0.958)$\ \ 
\\
${\rm ln}(10^{10}A_s)$: {({\it PLB})}                            &\ \  $(3.027)$ &  \ \ $(3.095)$ & $(3.043)$ & $(3.071)$\ \ 
\\
$\Sigma m_\nu \ [\rm{eV}]$: {\ \ ({\it PLB})}             &\ \  $(0\ {\rm fixed})$ &  \ \ $(0.576)$ & $(0\ {\rm fixed})$ & $(0.556)$\ \ 
\\
\rule{0pt}{3ex}$\xi$: {\ \ \ \ \ \ \ \ \ \ \ \ \ \ \ \ \ ({\it PLB})}             &\ \  $(2.43)$ &  \ \ $(2.29)$ & $(4.50)$ & $(3.89)$\ \ 
\\
$c_3$: {\ \ \ \ \ \ \ \ \ \ \ \ \ \ \ ({\it PLB})}             &\ \  $(0.101)$ &  \ \ $(0.098)$ & $(0.134)$ & $(0.143)$\ \ 
\\
$c_4$: {\ \ \ \ \ \ \ \ \ \ \ \ \ \ \ ({\it PLB})}             &\ \  $(-0.0045)$ &  \ \ $(-0.0037)$ & $(-0.012)$ & $(0.014)$\ \ 
\\
$c_5$: {\  \ \ \ \ \ \ \ \ \ \ \ \ \ \ ({\it PLB})}             &\ \  $(0\ {\rm fixed})$ &  \ \ $(0\ {\rm fixed})$ & $(0.0013)$ & $(0.0017)$\ \ 
\\
\rule{0pt}{3ex}$h$: {\ \ \ \ \ \ \ \ \ \ \ \ \ \ \ \ \ ({\it PLB})}                  &\ \  $(0.763)$ &  \ \ $(0.725)$ & $(0.764)$ & $(0.723)$\ \ 
\\
$\sigma_8(z = 0)$: {\ \ ({\it PLB})}                              &\ \  $(0.956)$ &  \ \ $(0.816)$ & $(0.945)$ & $(0.814)$\ \ 
\\
\hline
\rule{0pt}{3ex}$100\Omega_{b0} h^2$: {\ \ \ ({\it PLB})}                   &\ \  $(2.185 \pm 0.024)$ &  \ \ $(2.201 \pm 0.024)$ & $(2.218 \pm 0.024)$ & $(2.220 \pm 0.024)$\ \ 
\\
$\Omega_{c0} h^2$: {\ \ \ \ \ \ \ \ \  ({\it PLB})}          &\ \  $(0.122 \pm 0.002)$ &  \ \ $( 0.119 \pm 0.002)$ & $(0.123 \pm 0.002)$ & $(0.119 \pm 0.002)$\ \ 
\\
$10^4\theta_{\rm MC}$: {\ \ \ \ \ \ ({\it PLB})}                  &\ \  $(104.07 \pm 0.057)$ &  \ \ $(104.11 \pm 0.057)$ & $(104.07 \pm 0.058)$ & $(104.10 \pm 0.056)$\ \ 
\\
$\tau$: {\ \ \ \ \ \ \ \ \ \ \ \ \ \ \ \ \ ({\it PLB})}             &\ \  $(0.060 \pm 0.010)$ &  \ \ $(0.088 \pm 0.013)$ & $(0.060 \pm 0.010)$ & $(0.087 \pm 0.013)$\ \ 
\\
$n_s$: {\ \ \ \ \ \ \ \ \ \ \ \ \ \ \ ({\it PLB})}                   &\ \  $(0.952 \pm 0.005)$ &  \ \ $(0.960 \pm 0.006)$ & $(0.951 \pm 0.006)$ & $(0.959 \pm 0.006)$\ \ 
\\
${\rm ln}(10^{10}A_s)$: {({\it PLB})}                            &\ \  $(3.032 \pm 0.019)$ &  \ \ $(3.082 \pm 0.024)$ & $(3.035 \pm 0.019)$ & $(3.081 \pm 0.024)$\ \ 
\\
$\Sigma m_\nu \ [\rm{eV}]$: {\ \ ({\it PLB})}             &\ \  $(0\ {\rm fixed})$ &  \ \ $(0.560 \pm 0.101)$ & $ (0\ {\rm fixed})$ & $ (0.540 \pm 0.108)$\ \ 
\\
\rule{0pt}{3ex}$\xi$: {\ \ \ \ \ \ \ \ \ \ \ \ \ \ \ \ \ ({\it PLB})}             &\ \  $(2.46^{+0.10}_{-0.12})$ &  \ \ $(2.40 \pm 0.13)$ & $ (4.3^{+0.52}_{-1.58})$ & $ (4.23^{+0.53}_{-2.14})$\ \ 
\\
$c_3$: {\ \ \ \ \ \ \ \ \ \ \ \ \ \ \ ({\it PLB})}             &\ \  $( 0.101^{+0.006}_{-0.003})$ &  \ \ $(0.103^{+0.008}_{-0.004})$ & $(0.132^{+0.019}_{-0.004})$ & $(0.123^{+0.033}_{-0.006})$\ \ 
\\
$c_4$: {\ \ \ \ \ \ \ \ \ \ \ \ \ \ \ ({\it PLB})}             &\ \  $( -0.0045^{+0.0005}_{-0.0010})$ &  \ \ $(-0.0046^{+0.0007}_{-0.0014})$ & $(-0.012^{+0.002}_{-0.004})$ & $(-0.010^{+0.002}_{-0.008})$\ \ 
\\
$c_5$: {\  \ \ \ \ \ \ \ \ \ \ \ \ \ \ ({\it PLB})}             &\ \  $(0\ {\rm fixed})$ &  \ \ $(0\ {\rm fixed})$ & $(0.0015^{+0.0009}_{-0.0005})$ & $(0.0009^{+0.0017}_{-0.0005})$\ \ 
\\
\rule{0pt}{3ex}$h$: {\ \ \ \ \ \ \ \ \ \ \ \ \ \ \ \ \ ({\it PLB})}                  &\ \  $(0.764 \pm 0.008)$ &  \ \ $(0.724 \pm 0.010)$ & $(0.762 \pm 0.008)$ & $(0.724 \pm 0.010)$\ \ 
\\
$\sigma_8(z = 0)$: {\ \ ({\it PLB})}                              &\ \ $(0.960 \pm 0.012)$ &  \ \ $(0.824 \pm 0.027)$ & $(0.945 \pm 0.016)$ & $(0.823 \pm 0.027)$\ \ 
\\
\hline
\hline
\end{tabular}
\label{table:45dists}
\end{table*} 

The parameter space of the Quartic Galileon model is the same as the Cubic model, but with $c_4 \neq 0$. In our constraints, we use Eqs.~(\ref{eq:constraint1}) and (\ref{eq:constraint2}) to derive $c_3$ and $c_4$ as

\bq
c_3 &=& \frac{1}{2}\xi^{-1} - 2\Omega_{\varphi 0}\xi^{-3}, \nonumber \\
c_4 &=& -\frac{1}{9}\xi^{-2}  + \frac{2}{3}\Omega_{\varphi 0}\xi^{-4},
\eq
with $\xi$ being the free parameter varied in the chains. In the case of the Quintic model ($c_5 \neq 0$), we vary $\xi$ and $c_3$ in the chains, and derive $c_4$ and $c_5$:

\bq\label{eq:c4c5deriv}
c_4 &=& \frac{1}{3}\xi^{-2} - \frac{8}{9}c_3\xi^{-1} - \frac{10}{9}\Omega_{\varphi0}\xi^{-4}, \nonumber \\
c_5 &=& -\frac{1}{3}\xi^{-3} + \frac{2}{3}c_3\xi^{-2} + \frac{4}{3}\Omega_{\varphi0}\xi^{-5}.
\eq

\subsection{Cosmological constraints}

Figures \ref{fig:45-contours}, \ref{fig:45-bfs} and \ref{fig:45-lenspot} show the same as Figs.~\ref{fig:cubic-contours}, \ref{fig:cubic-bfs} and \ref{fig:cubic-lenspot}, respectively, but for the base Quartic (red dashed), $\nuquartic$ (red filled/solid), base Quintic (blue dashed) and $\nuquintic$ (blue filled/solid) Galileon models and using the {\it PLB} dataset. Table \ref{table:45dists} summarizes the one-dimensional marginalized statistics.

Just as in the case of the Cubic model, the presence of massive neutrinos in the Quartic and Quintic models also alleviates substantially the observational tensions between the different datasets in {\it PLB} (cf. Table \ref{table:45dists}). The situation here is completely analogous to the case of the Cubic Galileon model discussed in the last section. Recall that the origin of these observational tensions lies in the specifics of the late-time evolution of $H(a)$, which does not depend on the values of the $c_i$. Consequently, the same degeneracy between $h$ and $\Sigma m_\nu$ exists in the Quartic and Quintic Galileon models, which leads to good fits to the BAO, CMB temperature and CMB lensing data (cf. Fig.~\ref{fig:45-bfs}). It is also noteworthy that the constraints on the cosmological parameters of Eq.~(\ref{eq:cosmo-params}) are roughly the same in the Cubic, Quartic and Quintic models. Since these models differ in the Galileon subspace of parameters, this indicates that, to a reasonable extent, the constraints on the cosmological parameters do not correlate with those on the Galileon parameters.

One noticeable difference w.r.t.~the Cubic Galileon case relates to the lower amplitude of the CMB temperature spectrum at low-$l$ in both the Quartic and Quintic models. This is explained by the milder late-time evolution of $\phi$, as shown in Fig.~\ref{fig:45-lenspot}. The extra Galileon terms in the Quartic and Quintic models help to reduce the magnitudes of the fifth force, and hence $\phi$ is less affected by the Galileon field. This is illustrated in Fig.~\ref{fig:45-geff}, which shows the same as Fig.~\ref{fig:cubic-geff} but for the $\nuquartic$ and $\nuquintic$ ({\it PLB}) models. For instance, for $a = 1$ and $k = 0.05h/{\rm Mpc}$, $\phi/\phi^{f} \sim 1.21$ in the $\nuquintic$ ({\it PLB}), whereas $\phi/\phi^{f} \sim 1.9$ in the $\nucubic$ ({\it PLB}). It is interesting to note the nontrivial time evolution of $\phi/\phi^{f}$ in the $\nuquintic$ ({\it PLB}) model for $k = 0.0005 h/{\rm Mpc}$, which indicates that the fifth force terms can be repulsive ($\phi/\phi^{f} < 1$) rather than attractive. This shows that in the more general Quintic models there is more freedom to tune the modifications to gravity, in such a way as to reduce substantially the ISW power in the low-$l$ part of the CMB spectrum (blue lines in the top right panel of Fig.~\ref{fig:45-bfs}). 

In Sec.~\ref{sub:isw}, we discussed the possible role that an observational determination of the sign of the ISW effect could play in determining the viability of the $\nucubic$ Galileon model. The physical picture depicted in Fig.~\ref{fig:45-lenspot} suggests that any observational tension that might fall upon the Cubic Galileon model (due to its negative ISW effect) should be less severe in the $\nuquartic$ and $\nuquintic$ models. In particular, for $k = 0.005 h/{\rm Mpc}$, the $\nuquintic$ ({\it PLB}) model predicts that the lensing potential should even decay at late times ($a \gtrsim 0.7$), after a period of deepening ($0.4 \lesssim a \lesssim 0.7$). Therefore, it might be of interest to investigate the signatures that such a nontrivial time evolution of the lensing potential can have on the ISW observations discussed in Sec.~\ref{sub:isw}. However, such an investigation is beyond the scope of this paper.

\subsection{Local time variation of $G_{\rm eff}$ in the Quartic model}\label{subsec:quartic-chains}

As pointed out by Refs.~\cite{Babichev:2011iz, Kimura:2011dc}, the implementation of the Vainshtein screening effect in models like the Quartic and Quintic Galileons may not be enough to suppress all local modifications to gravity. Without loss of generality, the modified Poisson equation in Galileon gravity can be written as (see e.g. Ref.~\cite{Barreira:2013xea} for more details):

\bq\label{eq:modpoisson}
\nabla^2 \Psi = \left[A(t) + B(t, \nabla^2\varphi)\right]\nabla^2\Psi^{\rm GR} + C(t, \nabla^2\varphi),
\eq
where $\nabla^2$ is the three-dimensional Laplace operator, $\Psi$ is the total modified gravitational potential and $\Psi^{\rm GR}$ is the GR potential that satisfies the standard Poisson equation, $\nabla^2\Psi^{\rm GR} = 4\pi G\delta\rho_m$, where $\delta\rho$ is the total matter perturbation. The shapes of the functions $A, B, C$ depend on whether one assumes the Cubic, Quartic or Quintic models. An important aspect of the functions $B$ and $C$ is that they can be neglected if the spatial variations of the Galileon field are small compared to the variations in the gravitational potential, i.e., if $\nabla^2\varphi/\nabla^2\Psi^{\rm GR} \rightarrow 0$, then $B, C \rightarrow 0$.

The Vainshtein mechanism is implemented through nonlinear terms in the Galileon field equation of motion, which effectively suppress $\nabla^2\varphi$ (compared to $\nabla^2\Psi^{\rm GR}$) near overdense objects like our Sun. As a result, in the Solar System, Eq.~(\ref{eq:modpoisson}) reduces to

\bq\label{eq:sspoisson}
\nabla^2 \Psi = A(t)\nabla^2\Psi^{\rm GR}.
\eq
In the case of the Cubic Galileon model, $A(t) \equiv 1$ \cite{Barreira:2013xea, Barreira:2014zza} and one recovers exactly the standard Poisson equation in GR. However, in the Quartic and Quintic models, $A(t)$ depends on the time evolution of $\bar{\varphi}$ (which cannot be screened), and hence, residual modifications remain, even after the implementation of the Vainshtein mechanism. Figure \ref{fig:quartic-chains} shows $10^4$ randomly selected points from the chains used to constrain the $\nuquartic$ model with the {\it PLB} dataset, projected onto the $c_3-c_4$ and $\xi-c_4$ planes. The points are coloured according to the value of $G_{\rm eff}/G$ (lower panel) and $\dot{G}_{\rm eff}/G$ (upper panel) today. These two quantities were evaluated by following the strategy presented in Refs.~\cite{Barreira:2013xea, Barreira:2014zza}. In short, assuming spherical symmetry, one evaluates 

\bq\label{eq:geffexp}
\frac{G_{\rm eff}}{G}(a, \delta) = \frac{\Psi,_r/r}{\Psi,_r^{\rm GR}/r},
\eq
where $_,r$ denotes a partial derivative w.r.t. the radial coordinate $r$ and $\delta = \delta\rho_m/\bar{\rho}_m$ is the density contrast of the (top-hat) matter fluctuation. In Fig.~\ref{fig:quartic-chains}, we have assumed that in our Solar System $\delta = 10^7$, although this is not critical for our mostly qualitative discussion\footnote{To first approximation, we assume also that all of the matter components (baryons, CDM and massive neutrinos) contribute equally to $\delta$.}. The value of $\dot{G}_{\rm eff}/G$ was evaluated by taking finite differences at two consecutive times close to the present day (we have ensured that the time step used is small enough to be accurate). The figure shows that if $c_4$ is not sufficiently close to zero, then $G_{\rm eff}/G \neq 1$ and  $\dot{G}_{\rm eff}/G \neq 0$, contrary to what one would expect in standard gravity. These modifications are caused by the function $A(t)$, whose origin can be traced back to the explicit coupling to the Ricci scalar $R$ in $\mathcal{L}_4$ (cf.~Eq.~(\ref{L's})), which is needed to ensure that the theory is free from Ostrogradski ghosts. As a result, the requirement that the theory remains free from pathologies ultimately leads to nonvanishing local modifications to gravity. For the reasons listed in Ref.~\cite{Barreira:2013xea}, the same calculations for the Quintic model are much more challenging to perform due to the extra level of nonlinearity in the equations. However, the direct coupling to $G_{\mu\nu}$ in $\mathcal{L}_5$ is likely to give rise to the same qualitative behavior. Finally, we note that in Minkowski space, the couplings to curvature tensors are not needed to keep the theory ghost-free, and hence $A(t) = 1$, just like in the Cubic model. This shows the importance of taking the time-varying cosmological background into account when studying modified gravity models locally.

The phenomenology of the Quartic and Quintic models near massive bodies like our Sun can be used to further constrain their parameter space. The best-fitting $\nuquartic$ ({\it PLB}) model predicts that the effective local gravitational strength is varying at a rate $\dot{G}_{\rm eff}/G \sim -150 \times 10^{-13} {\rm yr}^{-1}$. However, Lunar Laser Ranging experiments constrain $\dot{G}_{\rm eff}/G = (4 \pm 9)\times 10^{-13} {\rm yr}^{-1}$ \cite{Williams:2004qba}. From the figure we see that this is only allowed provided $c_4$ is very close to zero. From the above reasoning, these constraints are also likely to set $c_5 \approx 0$,  in which case one recovers the Cubic Galileon studied in the last section. We leave for future work a formal and more detailed use of local gravity experiments to constrain this and other modified gravity models.

\subsection{The Galileon subspace of parameters in Quintic model}\label{subsec:quintic-chains}

Figure \ref{fig:quintic-chains} shows the points accepted in the chains (after the burn-in period) used to constrain the $\nuquintic$ model with the {\it PLB} dataset, projected onto the $c_3-\xi$ and $c_4-c_5$ planes. The grey dots indicate the points that were tried during the sampling, but which failed to meet the conditions of no ghost and Laplace instabilities of the scalar fluctuations. It is noteworthy that these stability conditions can, on their own, rule out a significant portion of the parameter space. 

In the $\xi-c_3$ plane, one can identify two branches that develop along stable but increasingly narrow regions of the parameter space, and that intersect at the location of the best-fitting regions.  The narrowness of these branches may raise concerns about the fairness of the Monte Carlo sampling. Consider, for instance, a chain that is currently in the upper branch (which goes through $\xi \sim 10$ to guide the eye). Since there are only two possible directions that do not lead to instabilities, the majority of the MCMC trials will be rejected and the chain will remain at the same point for a large number of steps. The narrowness of the gap between the unstable points therefore makes it harder for the chains to explore the regions that lie along the direction of the gap. Consequently, the "end point" of the branches may be determined not only by its poorer fit to the data, but also by these limitations of the numerical sampling.

To address the above concerns, as a test, we have run chains with priors on $\xi$ to force the chains to sample only the lower ($\xi \lesssim 3$) and the upper branches ($\xi \gtrsim 3$). These runs have shown that the length of the branches may extend just slightly (compared to Fig.~\ref{fig:quintic-chains}). This is expected since the chains spend more time in each branch, and hence, have a better chance of probing the limits of the branches. To learn more about the likelihood surface along the direction of the branches, we have further forced the chains to sample only the branch regions that are sufficiently far away from the intersection (to explore the far end of the lower branch we have imposed $\xi \lesssim 3$ and $c_3 \lesssim 0.0$; and for the upper branch we have imposed $\xi \gtrsim 12$). Also, in this second test, we have fixed all of the remaining cosmological parameters to their best-fitting values from Table \ref{table:45dists}. Again, as expected, these chains extended a bit more compared to Fig.~\ref{fig:quintic-chains}. In all these tests, however, the value of $\chi^2$ increases along these branches, indicating that the far end of the branches are indeed worse fits to the data. We have also looked at the CMB power spectrum for points located deep in the branches to confirm that the CMB spectra becomes visibly worse, compared to the best-fitting point. We therefore conclude that, despite some sampling difficulties that may arise due to the narrow stable regions, the "end points" of the branches are mostly determined by their poorer fit to the data. We stress that these complications in sampling the branches of the top panel of Fig.~\ref{fig:quintic-chains} are only important in determining the exact limit of confidence contours. For the purpose of identifying the best-fitting parameters, and subsequent analysis of its cosmology, these issues are not important as the best-fitting regions lie sufficiently far away from the end of the branches.

In the constraints of Ref.~\cite{Barreira:2013jma}, the allowed values of $c_4/c_3^{4/3}$ and $c_5/c_3^{5/3}$ ($c_3 = 10$) could not cross zero due to the constraints imposed by the ghost and Laplace stability conditions (cf.~Fig.~2 of Ref.~\cite{Barreira:2013jma}). This is different from what it is shown in Fig.~\ref{fig:quintic-chains}. In Ref.~\cite{Barreira:2013jma}, the background evolution was obtained numerically for general cases, which follow the tracker typically only at late times. Consequently, at early times and for fixed cosmological parameters, the exact time evolution of $\bar{\varphi}$ in this work and in Ref.~\cite{Barreira:2013jma} is not the same (see the discussion in Appendix \ref{ap:tracker}). Although these early time differences do not affect the CMB spectrum nor $H(a)$ (as the Galileon is subdominant), they can be important in the stability conditions. In particular, in the chains of Ref.~\cite{Barreira:2013jma}, the points with $c_5 < 0$ and $c_4>0$ were rejected due to violations of the ghost and Laplace stability conditions at early times, when $\bar{\varphi}(a)$ was not yet following the tracker. If the tracker solution is assumed at all cosmological epochs, then these points no longer develop instabilities and can in fact be accepted by the chains, as shown in Fig.~\ref{fig:quintic-chains}. Note also that the constraints of Ref.~\cite{Barreira:2013jma} have not missed the best-fitting points, since the latter are characterized by $c_5 > 0$ and $c_4 < 0$. In Ref.~\cite{Barreira:2013jma}, if all of the Galileon parameters were allowed to vary to explore the scaling degeneracy, then the stability conditions prevent $c_3$ from changing its sign, which is also different from the case in this paper (cf. subsection \ref{subsub:scaling}). This is why $c_3$ could be used as the pivot parameter to break the scaling degeneracy in Ref.~\cite{Barreira:2013jma}, but it is not a good option when one assumes the tracker evolution.

The lower panel of Fig.~\ref{fig:quintic-chains} zooms into the best-fitting regions of the $c_5-c_4$ plane. The points are color coded according to their values of $\xi$, which helps to identify the branches in the top panel. The projection along the $\xi$ direction gives rise to overlap of the points for which $\xi \gtrsim 3$ and for which $\xi \lesssim 3$. We also note that the high-$\xi$ points lie on a much narrower region of the $c_5-c_4$ plane, compared to those with lower $\xi$. This can be understood by recalling that $c_4$ and $c_5$ are derived parameters that depend on $c_3$, $\xi$ and $\Omega_{\varphi0}$ (cf.~Eqs.(\ref{eq:c4c5deriv})). When $\xi$ is sufficiently large, the terms $\propto \Omega_{\varphi0}$ in Eqs.~(\ref{eq:c4c5deriv}) can be neglected. This way, the narrow constraints imposed by the stability conditions on the $c_3$ and $\xi$ parameters (upper panel of Fig.~\ref{fig:quintic-chains}) lead directly to narrow constraints on $c_4$ and $c_5$, as well. On the other hand, when $\xi$ is smaller, the terms $\propto \Omega_{\varphi0}$ are no longer negligible. Consequently, the different sampled values of $\Omega_{\varphi0}$ (which are not as tightly constrained as $c_3$ and $\xi$ by the stability conditions) introduce extra scatter, which broadens the shape of the region of accepted points. A closer inspection shows also that an empty (unsampled) region forms at $(c_4, c_5) \sim (-0.012, 0.0013)$ (barely visible at the resolution of the figure).  The same empty region was also found in Ref.~\cite{Barreira:2013jma}. This serves to show the rather nontrivial shape of the parameter space in Quintic Galileon model. For instance, the {\tt CosmoMC} routines that evaluate the confidence contours from chain samples cannot resolve all these details clearly.

\section{Summary}\label{sec:summary}

We have studied and constrained the parameter space of the covariant Galileon gravity model using the latest observational CMB (temperature and lensing) data from the Planck satellite and BAO measurements from the 2dF, 6dF, SDSS-DR7, BOSS and WiggleZ galaxy redshift surveys. The parameter space in the Galileon model can be divided into a set of cosmological parameters, Eq.~(\ref{eq:cosmo-params}), and a set of five Galileon parameters, Eq.~(\ref{eq:gali-params}). However, we have shown that the dimensionality of the Galileon subspace of parameters can be reduced from five to two (i) by taking advantage of a set of scaling relations between the Galileon parameters, Eq.~(\ref{eq:scaling}); (ii) by assuming a spatially flat Universe, Eq.~(\ref{eq:constraint1}) and (iii) by assuming that the background evolution of the Galileon field follows the so-called tracker solution from sufficiently early times, Eq.~(\ref{eq:constraint2}). The latter assumption is justified, as otherwise, the model fails to provide a good fit to the CMB data. Moreover, the tracker solution admits analytical expressions for $H(a)$ and $\bar{\varphi}(a)$. The exploration of the parameter space was performed using MCMC methods with the aid of suitably modified versions of the publicly available {\tt CAMB} and {\tt CosmoMC} codes.

The action of the Galileon model we considered is made up of four Lagrangian densities, $\mathcal{L}_{2-5}$, which contain nonlinear derivative self-couplings of the Galileon scalar field. These Lagrangian densities are named after the power with which the field appears in them. Since the different levels of complexity can lead to different phenomenologies, we have analyzed separately the three main branches of the Galileon model parameter space. These are the so-called Cubic, $\left\{\mathcal{L}_2, \mathcal{L}_3\right\}$; Quartic $\left\{\mathcal{L}_2, \mathcal{L}_3, \mathcal{L}_4\right\}$ and Quintic $\left\{\mathcal{L}_2, \mathcal{L}_3, \mathcal{L}_4, \mathcal{L}_5\right\}$ Galileon models. A major goal of this paper was to investigate the impact that massive neutrinos have on the observational viability of Galileon gravity. We have therefore constrained "base Galileon models", for which $\Sigma m_\nu = 0$; and ${\nu}{\rm Galileon}$ models, for which $\Sigma m_\nu$ is a free parameter to be constrained by the data. Our main findings can be summarized as follows:

\begin{itemize}

\item When $\Sigma m_\nu = 0$, all sectors of Galileon gravity have difficulties in fitting the BAO and the CMB peak positions simultaneously. This tension is related to the specific late-time evolution of $H(a)$ (cf. Eq.~(\ref{eq:tracker_H})), which leads to different constraints on the value of $h$ by the CMB (higher $h$) and BAO (lower $h$) data (cf. Fig.~\ref{fig:cubic-contours} and \ref{fig:45-contours}). This tension applies to all sectors of the Galileon model, since $H(a)$ does not depend on the values of the $c_i$ parameters. In addition to this observational tension, these best-fitting models also predict too much power for the CMB lensing potential spectrum (left middle panels of Figs.~\ref{fig:cubic-bfs} and \ref{fig:45-bfs}). In the case of the Cubic Galileon model, there is also an excess of ISW power in the low-$l$ region of the CMB temperature power spectrum (top right panel of Fig.~\ref{fig:cubic-bfs}).

\item If neutrinos are sufficiently massive, then they modify the late-time expansion history in such a way that the CMB data no longer prefers high values for $h$. This completely eliminates the tension with the BAO data if $\Sigma m_\nu \gtrsim 0.4\ {\rm eV}$ ($2\sigma$) in the case of the Cubic Galileon (cf. Fig.~\ref{fig:cubic-bfs}), and $\Sigma m_\nu \gtrsim 0.3\ {\rm eV}$ ($2\sigma$) in the case of the Quartic and Quintic models (cf. Fig.~\ref{fig:45-bfs}). These best-fitting ${\nu}{\rm Galileon}$ models also reproduce much better the CMB lensing power spectrum. This fit can be even slightly better than in $\Lambda$CDM models, mainly due to a better fit to the data at $l \sim 60$ (left middle panels of Figs.~\ref{fig:cubic-bfs} and \ref{fig:45-bfs}). For the Cubic Galileon model, massive neutrinos also help to lower the excess of ISW power in the CMB. The neutrino mass constraints in the $\nu{\rm Galileon}$ models leave room for upcoming terrestrial neutrino experiments to help distinguish between these models and $\Lambda{\rm CDM}$.

\item In Galileon gravity, the time evolution of the lensing potential $\phi$ differs from the $\Lambda{\rm CDM}$ result at late times, and its qualitative behaviour is also scale-dependent (cf.~Figs.~\ref{fig:cubic-lenspot} and \ref{fig:45-lenspot}). In the case of the Cubic models, $\phi$ deepens considerably at late times on scales $k \gtrsim 0.005 h/{\rm Mpc}$, but decays (as in $\Lambda$CDM) on scales $k \lesssim 0.0005 h/{\rm Mpc}$. This behaviour follows from the scale dependence of the magnitude of the modifications to gravity induced by the Galileon field, which becomes weaker on horizon-like scales (cf.~Fig.~\ref{fig:cubic-geff}). The extra Galileon terms in the Quartic and Quintic sectors of the model allow for milder and smoother time evolution of $\phi$ for $k \gtrsim 0.005 h/{\rm Mpc}$, but the potential can still decay for $k \lesssim 0.0005 h/{\rm Mpc}$.

\item The fact that $\phi$ deepens at late times for $k \gtrsim 0.005 h/{\rm Mpc}$ in the Cubic model implies a negative ISW sign. This is opposite to what has been found recently by a number of observational studies that claimed the detection of a positive sign for the ISW effect (cf. Sec.~\ref{sub:isw}). There is still ongoing discussion about the impact that systematics (such as selection effects) might play in the significance of these observations. In Galileon gravity there are also additional complications associated with how the ISW might be affected by the nonlinearities of the screening mechanism. If in the future, these potential issues with the observations turn out to be unimportant, then the sign of the ISW effect can play a crucial role in determining the viability of Cubic Galileon models, potentially ruling them out. The same may hold for our best-fitting Quartic models. The situation is a bit more unclear in the Quintic sector, since $\phi$ can decay and deepen at different epochs (cf.~Fig.~\ref{fig:45-lenspot}).

\item The modified expansion rate in Galileon gravity compared to $\Lambda$CDM (cf.~Eq.~(\ref{eq:tracker_H})) leads to CMB constraints that are compatible with local determinations of $h$ (cf.~Figs.~\ref{fig:cubic-contours} and \ref{fig:45-contours}). $\nu{\rm Galileon}$ models therefore avoid this observational tension that is currently plaguing $\Lambda{\rm CDM}$.

\item The large-scale modifications to gravity, together with the small-scale screening effects and the impact of the large neutrino density, may lead to clear imprints in observables that are sensitive to the growth rate of structure. The latter include the shape and amplitude of the galaxy power spectrum (cf.~middle right panels of Figs.~\ref{fig:cubic-bfs} and \ref{fig:45-bfs}), cluster abundance, galaxy weak-lensing, measurements of $f\sigma_8$ (cf.~lower panels of Figs.~\ref{fig:cubic-bfs} and \ref{fig:45-bfs}), etc. A proper constraint study using these observables requires the use of N-body simulations, which is left for future work.

\item The explicit couplings between the Galileon field derivatives and curvature tensors in the $\mathcal{L}_4$ and $\mathcal{L}_5$ Lagrangian densities (cf.~Eq.~(\ref{L's})) give rise to modifications to gravity that cannot be totally suppressed by the Vainshtein mechanism. In particular, the effective local gravitational strength is time varying in the Quartic and Quintic models, which puts these models into significant tension with Solar System constraints. We have shown how, by imposing a prior for the time variation of $G_{\rm eff}$ obtained from Lunar Laser experiments, one can essentially constrain the Quartic (and very likely the Quintic) model to look almost like the Cubic model (cf.~Fig.~\ref{fig:quartic-chains}). A more quantitative use of Solar System tests to constrain Galileon gravity is left for future work.

\end{itemize}

\begin{acknowledgments}

We thank Luca Amendola and Yan-Chuan Cai for useful comments and discussions. We are also grateful to Antony Lewis for help with the {\tt CosmoMC} code, and to Lydia Heck for valuable numerical support. This work was supported by the Science and Technology Facilities Council [grant number ST/F001166/1]. This work used the DiRAC Data Centric system at Durham University, operated by the Institute for Computational Cosmology on behalf of the STFC DiRAC HPC Facility (www.dirac.ac.uk). This equipment was funded by BIS National E-infrastructure capital grant ST/K00042X/1, STFC capital grant ST/H008519/1, and STFC DiRAC Operations grant ST/K003267/1 and Durham University. DiRAC is part of the National E-Infrastructure. AB is supported by FCT-Portugal through grant SFRH/BD/75791/2011. BL is supported by the Royal Astronomical Society and Durham University. The research leading to these results has received funding from the European Research Council under the European Union's Seventh Framework Programme (FP/2007-2013) / ERC Grant NuMass Agreement n. [617143]. This work has been partially supported by the European Union FP7  ITN INVISIBLES (Marie Curie Actions, PITN- GA-2011- 289442) and STFC. 

\end{acknowledgments}

\appendix

\begin{figure}
	\centering
	\includegraphics[scale=0.44]{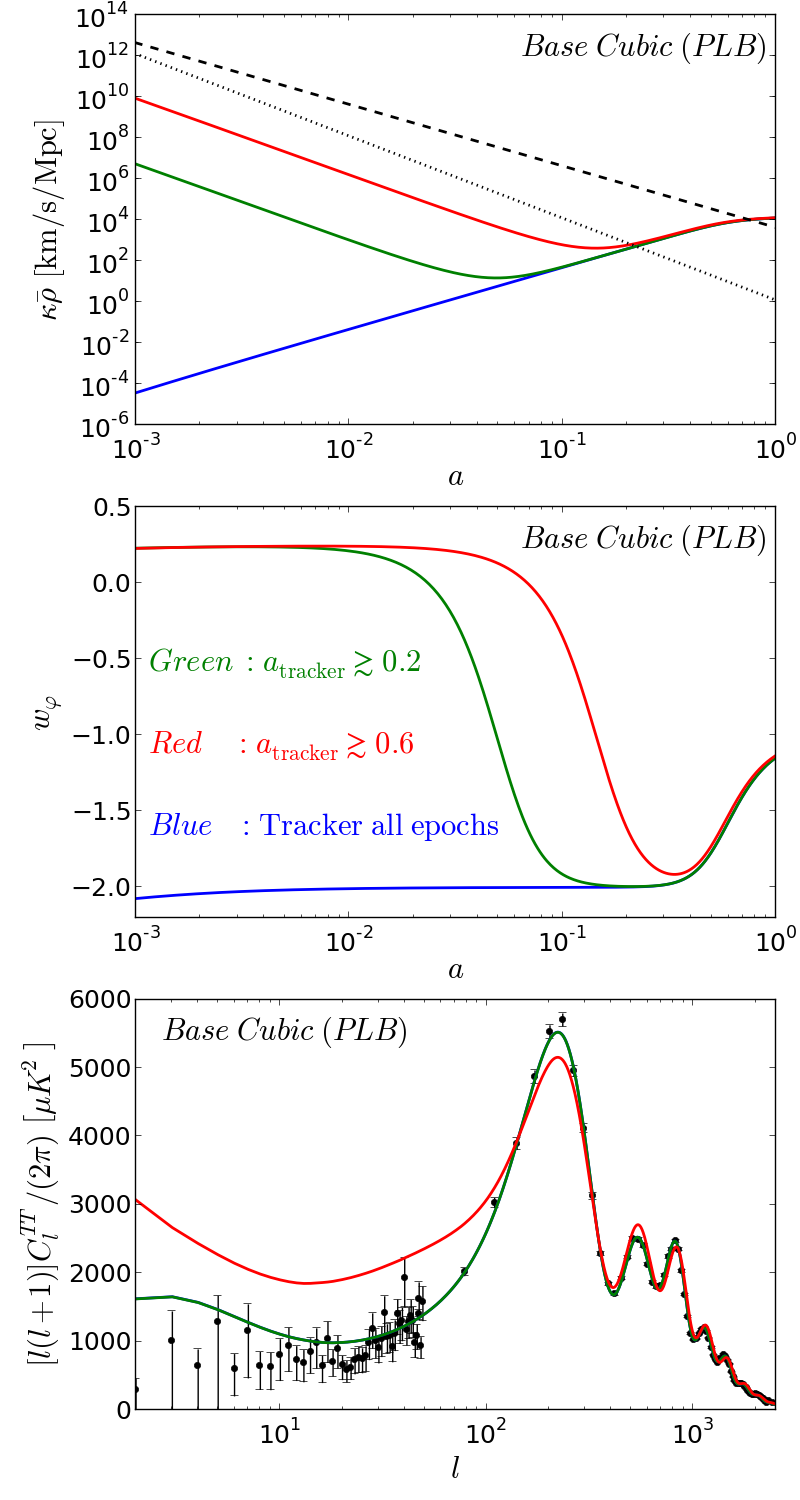}
	\caption{From top to bottom, the first two panels show the time evolution of the background energy density and of the Galileon field equation-of-state parameter $w_{\varphi}$ for the base Cubic (PLB) model. The time evolution is shown for three cases that differ in the time when the background evolution follows the tracker solution: all epochs (blue), $a \gtrsim 0.2$ (green) and $a \gtrsim 0.6$ (red). The bottom panels shows the corresponding CMB power spectrum (here, the blue and green curves are overlapping). In the top panel, the dashed and dotted curves correspond, respectively, to the total matter (baryons and CDM for this model) and radiation (photons and massless neutrinos for this model).}
\label{fig:cubic-trackertime}\end{figure}

\begin{figure}
	\centering
	\includegraphics[scale=0.44]{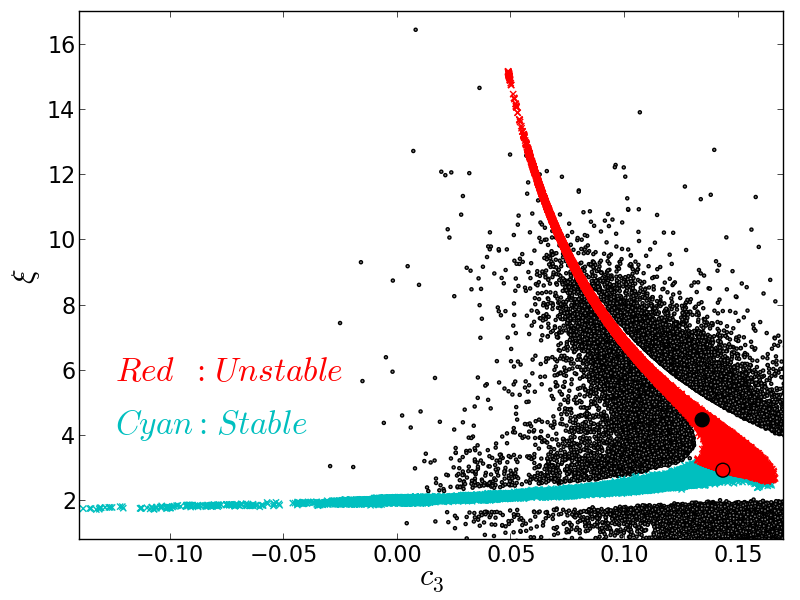}
	\caption{Same as the upper panel of Fig.~\ref{fig:quintic-chains}, but for the base Quintic model and with the accepted points colored according to their stability of the tensor perturbations. The red crosses indicate points which are associated with tensor Laplace instabilities, as labelled. The big black dot indicates the best-fitting point found in the chains. The big red dot indicates the best-fitting point considering only tensor-stable points (the fact that this point looks like it lies in the tensor-unstable region is purely due to the resolution of the figure).}
\label{fig:quintictensorchains}\end{figure}

\begin{figure*}
	\centering
	\includegraphics[scale=0.39]{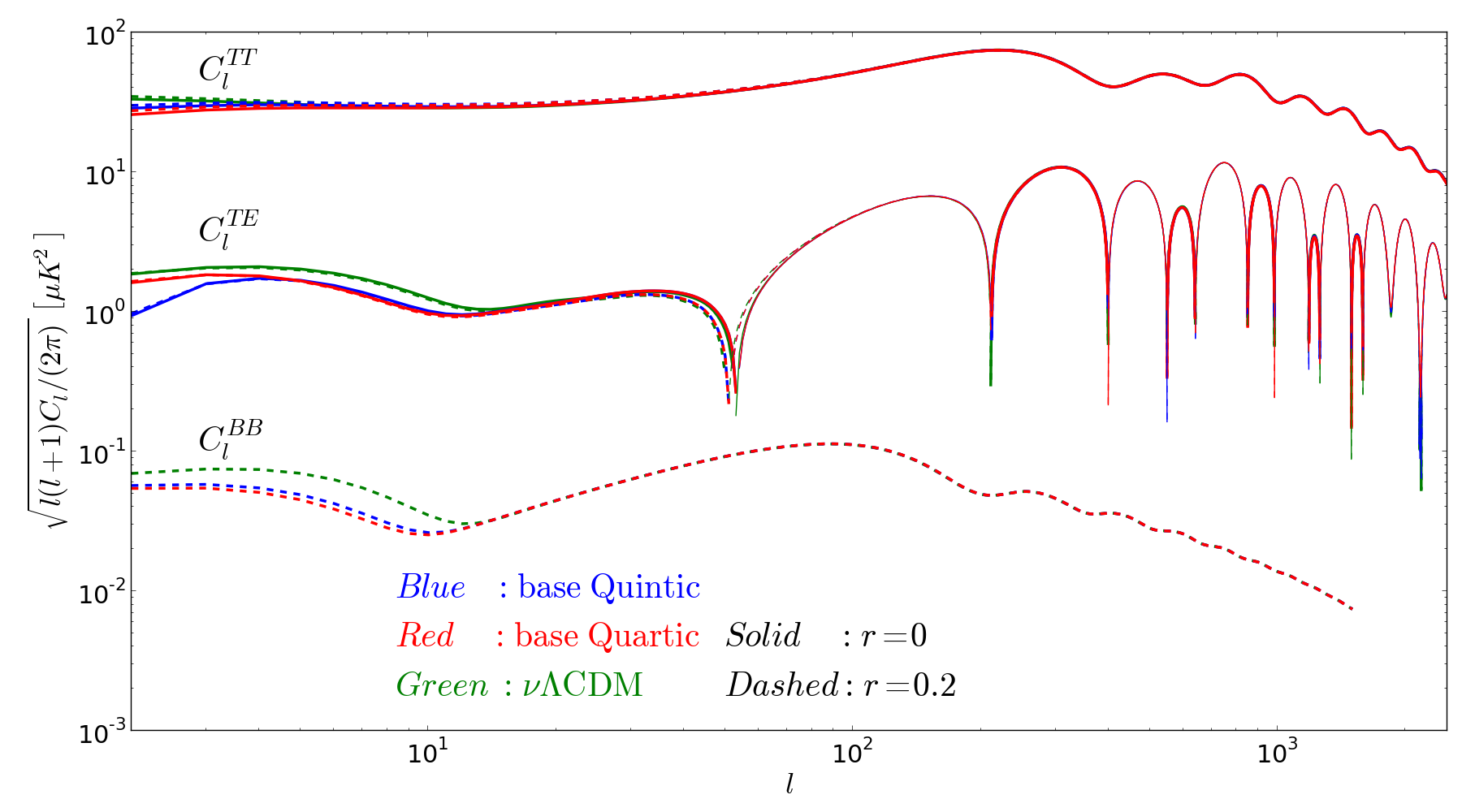}
	\caption{From top to bottom, the lines show, respectively, the CMB temperature power spectrum,  the cross-correlation of the CMB temperature with the E-mode polarization and the B-mode polarization power spectrum for the $\nulcdm$ (green), base Quartic (red) and base Quintic (blue) models, for $r_{0.05} = 0$ (solid) and $r_{0.05} = 0.2$ (dashed). For the $C_l^{TE}$ curves, the lines are thiner at the values of $l$ for which the cross-correlation becomes negative. The curves for the Quintic model correspond to the best-fitting model that is tensor-stable (c.f.~Fig.~\ref{fig:quintictensorchains}).}
\label{fig:tensorspectra}\end{figure*}

\section{The time of tracker solution}\label{ap:tracker}

Figure \ref{fig:cubic-trackertime} shows the time evolution of the Galileon field background energy density $\bar{\rho}_{\varphi}$ (upper) and equation-of-state parameter $w_{\varphi}$ (middle), for the best-fitting base Cubic Galileon model ({\it PLB}) for three different epochs where the background evolution follows the tracker solution: all cosmic epochs (blue), $a \gtrsim 0.2$ (green) and $a \gtrsim 0.6$ (red). The lower panel shows the respective CMB angular power spectrum. One notes that, indeed, models with different time evolution at early times eventually start following the tracker solution. The latter is characterized by a phantom evolution $w_{\varphi} < -1$, i.e., the dark energy density grows with the expansion of the Universe. Also, the tracker is reached sooner if the Galileon density is smaller at earlier times. The initial conditions are set up by different values of $\bar{\varphi}_i$ at some initial time, when solving Eqs.~(\ref{background1} - \ref{background EoM}) numerically.

The figure shows that, if the background evolution is not on the tracker sufficiently before the start of the dark energy era ($a \sim 0.5$), then this leads to a poor fit to the CMB temperature data. Also, the physical predictions do not depend on the time the tracker is reached, provided it does so at $a \lesssim 0.5$. This shows that by assuming the tracker solution at all epochs we are not at risk of missing any best-fitting regions of the parameter space of the model. This justifies the strategy adopted in this paper.

\section{Tensor perturbations}\label{ap:tensor}

In our constraints, we are mainly interested in the role played by scalar fluctuations, and as a result, we have set the amplitude of tensor fluctuations to zero (vector perturbations play a negligible role as they decay very quickly after their generation). In this appendix, we briefly investigate the impact that the evolution of tensor fluctuations can have in Galileon gravity models.

The relevant equations for the evolution of the tensor modes can be written as \cite{Challinor:1998xk, Challinor:1999xz, Challinor:2000as}:

\bq
\label{p2}
0 &=& \dot{\sigma}_{\mu\nu} + \frac{2}{3}\theta\sigma_{\mu\nu} + \mathcal{E}_{\mu\nu} + \frac{\kappa}{2}\pi_{\mu\nu}, \\
\label{p4}
0 &=& \frac{\kappa}{2}\left[ \dot{\pi}_{\mu\nu} + \frac{1}{3}\theta\pi_{\mu\nu}\right] - \frac{\kappa}{2}(\rho + p)\sigma_{\mu\nu} \nonumber \\
&& - \left[ \dot{\mathcal{E}}_{\mu\nu} + \theta\mathcal{E}_{\mu\nu} - \hat{\nabla}^\alpha\mathcal{B}_{\beta(\mu}\epsilon_{\nu)\gamma\alpha}^{\ \ \ \ \ \beta}u^\gamma\right],
\eq
where, $\mathcal{E}_{\mu\nu}$ and $\mathcal{B}_{\mu\nu}$ are, respectively, the electric and magnetic parts of the Weyl tensor, $\mathcal{W}_{\mu\nu\alpha\beta}$, defined by $\mathcal{E}_{\mu\nu} = u^\alpha u^\beta\mathcal{W}_{\mu\alpha\nu\beta}$ and $\mathcal{B}_{\mu\nu} = -\frac{1}{2}u^\alpha u^\beta \epsilon_{\mu\alpha}^{\ \ \gamma\delta}\mathcal{W}_{\gamma\delta\nu\beta}$. $\epsilon_{\mu\nu\alpha\beta}$ is the covariant permutation tensor. The Galileon field contributes to the tensor modes evolution via its modifications to the background dynamics, but also via its anisotropic stress (Eq.~(\ref{perturbed4})), both of which are only important at late times. Explicitly, the relevant terms from Eq.~(\ref{perturbed4}) that enter Eqs.~(\ref{p2}) and (\ref{p4}) are

\begin{widetext}
\bq
&&\pi_{\mu\nu}^{\varphi,\ {\rm tensor}} \doteq \frac{c_4}{M^6}\left[  -\dot{\varphi}^4 \left( \dot{\sigma}_{\mu\nu} - \mathcal{E}_{\mu\nu}\right) - \left(6\ddot{\varphi}\dot{\varphi}^3 + \frac{4}{3}\dot{\varphi}^4\theta\right)\sigma_{\mu\nu}\right] \nonumber \\
&&\ \ \ \ \ \ \ \ \ + \frac{c_5}{M^9}\left[ -\left(\dot{\varphi}^5\dot{\theta} +\dot{\varphi}^5\theta^2 + 6\ddot{\varphi}\dot{\varphi}^4\theta\right)\sigma_{\mu\nu} -  \left(\dot{\varphi}^5\theta + 3\ddot{\varphi}\dot{\varphi}^4\right)\dot{\sigma}_{\mu\nu} - 6\ddot{\varphi}\dot{\varphi}^4\mathcal{E}_{\mu\nu} \right],
\eq
\end{widetext}
where the superscript $^{\rm tensor}$ indicates we are only considering the terms that contribute to the tensor fluctuations. Recall that for the Cubic Galileon model, $c_4 = c_5 = 0$, and as a result, there is no explicit contribution from the Galileon field to the tensor perturbations.

As in the case of scalar fluctuations, when studying the evolution of the tensor perturbations of the Galileon field, one must also ensure that they do not develop ghost nor Laplace instabilities. The conditions for the avoidance of these pathologies were derived in Ref.~\cite{DeFelice:2010nf}. With our notation, the no-ghost condition is given by

\bq
Q_T/M_{\rm Pl}^2 = \frac{1}{2} - \frac{3}{4}\frac{c_4}{M^6}\dot{\varphi}^4 - \frac{3}{2}\frac{c_5}{M^9}\dot{\varphi}^5H > 0,
\eq
whereas the no-Laplace instability condition is given by

\bq
c_T^2 = \frac{1}{Q_T}\left[\frac{M_{\rm Pl}^2}{2} + \frac{1}{4}\frac{c_4}{M^6}\dot{\varphi}^4 - \frac{3}{2}\frac{c_5}{M^9}\dot{\varphi}^4\ddot{\varphi} \right] > 0.
\eq

During the sampling of the parameter space, we have only checked for the stability of the scalar fluctuations. Consequently, it is possible that some of the accepted points are associated with tensor instabilities. For the case of the Quartic model, we have checked that all of the accepted points in the chains are tensor-stable. The same however is not true for the Quintic model. This is illustrated in Fig.~\ref{fig:quintictensorchains}, which shows the points accepted in the chains used to constrain the base Quintic model with the {\it PLB} dataset. The red crosses, which contain the best-fitting point of Table \ref{table:45dists} (black circle), indicate the regions of the parameter space which develop Laplace instabilities of the tensor perturbations. It is remarkable that taking the tensor stability conditions into account rules out more than half of the parameter space space allowed by the {\it PLB} dataset and scalar stability conditions. The red circle indicates the best-fitting point that is tensor-stable, for which $\Delta\chi^2 = 1.7$ compared to the best-fitting point of Table \ref{table:45dists}. Hence, although the stability conditions rule out a significant portion of the parameter space, they still leave behind regions which can provide a similar fit to the data, compared to the case where only scalar stability conditions are considered.

Figure \ref{fig:tensorspectra} shows the CMB temperature power spectrum, the cross-correlation of the temperature and E-mode polarization of the CMB and the B-mode polarization power spectrum for the tensor-stable $\nulcdm$ (green), base Quartic (red) and base Quintic (blue) models that best-fit the {\it PLB} dataset. The dashed curves show the spectra obtained by setting $r_{0.05} = 0.2$, where $r_{0.05}$ is the tensor-to-scalar ratio of primordial power (at a pivot scale $k = 0.05\rm{Mpc}^{-1}$). Our choice of $r_{0.05}$ is merely illustrative. The solid curves show the spectra for $r_{0.05} = 0$. We also assume a zero tensor spectral index with no running. One notes that the modifications driven by setting $r_{0.05} = 0.2$ are roughly of the same size for the three models. This shows that the tensor perturbations from the Galileon field are not affecting the overall spectra in a nontrivial and sizeable way. This justifies the approach in our model constraints, where we have neglected the role of the tensor modes. The differences between the Quartic, Quintic and $\nulcdm$ models are only visible at low-$l$. In the particular case of the B-mode power spectrum, it is interesting to note that for $l \gtrsim 10$, the Galileon models predict essentially the same amplitude as standard $\Lambda$CDM. As a result, any detections of the B-mode signal at $l \sim 80$ such as those reported by BICEP-2 \cite{Ade:2014xna} are unlikely to be directly related to the Galileon field {\it per se}.

\bibliography{planck-galileon.bib}

\begin{thebibliography}{100}%
\makeatletter
\providecommand \@ifxundefined [1]{%
 \ifx #1\undefined \expandafter \@firstoftwo
 \else \expandafter \@secondoftwo
\fi
}%
\providecommand \@ifnum [1]{%
 \ifnum #1\expandafter \@firstoftwo
 \else \expandafter \@secondoftwo
\fi
}%
\providecommand \enquote [1]{``#1''}%
\providecommand \bibnamefont  [1]{#1}%
\providecommand \bibfnamefont [1]{#1}%
\providecommand \citenamefont [1]{#1}%
\providecommand\href[0]{\@sanitize\@href}%
\providecommand\@href[1]{\endgroup\@@startlink{#1}\endgroup\@@href}%
\providecommand\@@href[1]{#1\@@endlink}%
\providecommand \@sanitize [0]{\begingroup\catcode`\&12\catcode`\#12\relax}%
\@ifxundefined \pdfoutput {\@firstoftwo}{%
 \@ifnum{\z@=\pdfoutput}{\@firstoftwo}{\@secondoftwo}%
}{%
 \providecommand\@@startlink[1]{\leavevmode\special{html:<a href="#1">}}%
 \providecommand\@@endlink[0]{\special{html:</a>}}%
}{%
 \providecommand\@@startlink[1]{%
  \leavevmode
  \pdfstartlink
   attr{/Border[0 0 1 ]/H/I/C[0 1 1]}%
   user{/Subtype/Link/A<</Type/Action/S/URI/URI(#1)>>}%
  \relax
 }%
 \providecommand\@@endlink[0]{\pdfendlink}%
}%
\providecommand \url  [0]{\begingroup\@sanitize \@url }%
\providecommand \@url [1]{\endgroup\@href {#1}{\urlprefix}}%
\providecommand \urlprefix [0]{URL }%
\providecommand \Eprint[0]{\href }%
\@ifxundefined \urlstyle {%
  \providecommand \doi [1]{doi:\discretionary{}{}{}#1}%
}{%
  \providecommand \doi [0]{doi:\discretionary{}{}{}\begingroup
  \urlstyle{rm}\Url }%
}%
\providecommand \doibase [0]{http://dx.doi.org/}%
\providecommand \Doi[1]{\href{\doibase#1}}%
\providecommand \bibAnnote [3]{%
  \BibitemShut{#1}%
  \begin{quotation}\noindent
    \textsc{Key:}\ #2\\\textsc{Annotation:}\ #3%
  \end{quotation}%
}%
\providecommand \bibAnnoteFile [2]{%
  \IfFileExists{#2}{\bibAnnote {#1} {#2} {\input{#2}}}{}%
}%
\providecommand \typeout [0]{\immediate \write \m@ne }%
\providecommand \selectlanguage [0]{\@gobble}%
\providecommand \bibinfo [0]{\@secondoftwo}%
\providecommand \bibfield [0]{\@secondoftwo}%
\providecommand \translation [1]{[#1]}%
\providecommand \BibitemOpen[0]{}%
\providecommand \bibitemStop [0]{}%
\providecommand \bibitemNoStop [0]{.\EOS\space}%
\providecommand \EOS [0]{\spacefactor3000\relax}%
\providecommand \BibitemShut [1]{\csname bibitem#1\endcsname}%
\bibitem{Clifton:2011jh}%
  \BibitemOpen
  \bibfield{author}{%
  \bibinfo {author} {\bibfnamefont{T.}~\bibnamefont{Clifton}}, \bibinfo
  {author} {\bibfnamefont{P.~G.}\ \bibnamefont{Ferreira}}, \bibinfo {author}
  {\bibfnamefont{A.}~\bibnamefont{Padilla}},\ and\ \bibinfo {author}
  {\bibfnamefont{C.}~\bibnamefont{Skordis}},\ }%
  \bibfield{journal}{%
  \Doi{10.1016/j.physrep.2012.01.001}{\bibinfo {journal} {Phys.Rept.}}\ }%
  \textbf{\bibinfo {volume} {513}},\ \bibinfo {pages} {1} (\bibinfo {year}
  {2012}),\ \Eprint{http://arxiv.org/abs/1106.2476}{arXiv:1106.2476
  [astro-ph.CO]}%
  \bibAnnoteFile{NoStop}{Clifton:2011jh}%
\bibitem{Will:2014kxa}%
  \BibitemOpen
  \bibfield{author}{%
  \bibinfo {author} {\bibfnamefont{C.~M.}\ \bibnamefont{Will}}}%
   (\bibinfo {year} {2014}),\
  \Eprint{http://arxiv.org/abs/1403.7377}{arXiv:1403.7377 [gr-qc]}%
  \bibAnnoteFile{NoStop}{Will:2014kxa}%
\bibitem{PhysRevD.79.064036}%
  \BibitemOpen
  \bibfield{author}{%
  \bibinfo {author} {\bibfnamefont{A.}~\bibnamefont{Nicolis}}, \bibinfo
  {author} {\bibfnamefont{R.}~\bibnamefont{Rattazzi}},\ and\ \bibinfo {author}
  {\bibfnamefont{E.}~\bibnamefont{Trincherini}},\ }%
  \bibfield{journal}{%
  \Doi{10.1103/PhysRevD.79.064036}{\bibinfo {journal} {Phys. Rev. D}}\ }%
  \textbf{\bibinfo {volume} {79}},\ \bibinfo {pages} {064036} (\bibinfo {year}
  {2009})%
  \bibAnnoteFile{NoStop}{PhysRevD.79.064036}%
\bibitem{Horndeski:1974wa}%
  \BibitemOpen
  \bibfield{author}{%
  \bibinfo {author} {\bibfnamefont{G.~W.}\ \bibnamefont{Horndeski}},\ }%
  \bibfield{journal}{%
  \Doi{10.1007/BF01807638}{\bibinfo {journal} {Int.J.Theor.Phys.}}\ }%
  \textbf{\bibinfo {volume} {10}},\ \bibinfo {pages} {363} (\bibinfo {year}
  {1974})%
  \bibAnnoteFile{NoStop}{Horndeski:1974wa}%
\bibitem{Woodard:2006nt}%
  \BibitemOpen
  \bibfield{author}{%
  \bibinfo {author} {\bibfnamefont{R.~P.}\ \bibnamefont{Woodard}},\ }%
  \bibfield{journal}{%
  \Doi{10.1007/978-3-540-71013-4_14}{\bibinfo {journal} {Lect.Notes Phys.}}\ }%
  \textbf{\bibinfo {volume} {720}},\ \bibinfo {pages} {403} (\bibinfo {year}
  {2007}),\
  \Eprint{http://arxiv.org/abs/astro-ph/0601672}{arXiv:astro-ph/0601672
  [astro-ph]}%
  \bibAnnoteFile{NoStop}{Woodard:2006nt}%
\bibitem{PhysRevD.79.084003}%
  \BibitemOpen
  \bibfield{author}{%
  \bibinfo {author} {\bibfnamefont{C.}~\bibnamefont{Deffayet}}, \bibinfo
  {author} {\bibfnamefont{G.}~\bibnamefont{Esposito-Far\`ese}},\ and\ \bibinfo
  {author} {\bibfnamefont{A.}~\bibnamefont{Vikman}},\ }%
  \bibfield{journal}{%
  \Doi{10.1103/PhysRevD.79.084003}{\bibinfo {journal} {Phys. Rev. D}}\ }%
  \textbf{\bibinfo {volume} {79}},\ \bibinfo {pages} {084003} (\bibinfo {year}
  {2009})%
  \bibAnnoteFile{NoStop}{PhysRevD.79.084003}%
\bibitem{Deffayet:2009mn}%
  \BibitemOpen
  \bibfield{author}{%
  \bibinfo {author} {\bibfnamefont{C.}~\bibnamefont{Deffayet}}, \bibinfo
  {author} {\bibfnamefont{S.}~\bibnamefont{Deser}},\ and\ \bibinfo {author}
  {\bibfnamefont{G.}~\bibnamefont{Esposito-Farese}},\ }%
  \bibfield{journal}{%
  \Doi{10.1103/PhysRevD.80.064015}{\bibinfo {journal} {Phys.Rev.}}\ }%
  \textbf{\bibinfo {volume} {D80}},\ \bibinfo {pages} {064015} (\bibinfo {year}
  {2009}),\ \Eprint{http://arxiv.org/abs/0906.1967}{arXiv:0906.1967 [gr-qc]}%
  \bibAnnoteFile{NoStop}{Deffayet:2009mn}%
\bibitem{Deffayet:2010qz}%
  \BibitemOpen
  \bibfield{author}{%
  \bibinfo {author} {\bibfnamefont{C.}~\bibnamefont{Deffayet}}, \bibinfo
  {author} {\bibfnamefont{O.}~\bibnamefont{Pujolas}}, \bibinfo {author}
  {\bibfnamefont{I.}~\bibnamefont{Sawicki}},\ and\ \bibinfo {author}
  {\bibfnamefont{A.}~\bibnamefont{Vikman}},\ }%
  \bibfield{journal}{%
  \Doi{10.1088/1475-7516/2010/10/026}{\bibinfo {journal} {JCAP}}\ }%
  \textbf{\bibinfo {volume} {1010}},\ \bibinfo {pages} {026} (\bibinfo {year}
  {2010}),\ \Eprint{http://arxiv.org/abs/1008.0048}{arXiv:1008.0048 [hep-th]}%
  \bibAnnoteFile{NoStop}{Deffayet:2010qz}%
\bibitem{Babichev:2012re}%
  \BibitemOpen
  \bibfield{author}{%
  \bibinfo {author} {\bibfnamefont{E.}~\bibnamefont{Babichev}}\ and\ \bibinfo
  {author} {\bibfnamefont{G.}~\bibnamefont{Esposito-Farèse}},\ }%
  \bibfield{journal}{%
  \Doi{10.1103/PhysRevD.87.044032}{\bibinfo {journal} {Phys.Rev.}}\ }%
  \textbf{\bibinfo {volume} {D87}},\ \bibinfo {pages} {044032} (\bibinfo {year}
  {2013}),\ \Eprint{http://arxiv.org/abs/1212.1394}{arXiv:1212.1394 [gr-qc]}%
  \bibAnnoteFile{NoStop}{Babichev:2012re}%
\bibitem{Vainshtein1972393}%
  \BibitemOpen
  \bibfield{author}{%
  \bibinfo {author} {\bibfnamefont{A.}~\bibnamefont{Vainshtein}},\ }%
  \bibfield{journal}{%
  \Doi{10.1016/0370-2693(72)90147-5}{\bibinfo {journal} {Phys.~Lett.~B}}\ }%
  \textbf{\bibinfo {volume} {39}},\ \bibinfo {pages} {393 } (\bibinfo {year}
  {1972}),\ ISSN \bibinfo {issn} {0370-2693}%
  \bibAnnoteFile{NoStop}{Vainshtein1972393}%
\bibitem{Babichev:2013usa}%
  \BibitemOpen
  \bibfield{author}{%
  \bibinfo {author} {\bibfnamefont{E.}~\bibnamefont{Babichev}}\ and\ \bibinfo
  {author} {\bibfnamefont{C.}~\bibnamefont{Deffayet}}}%
   (\bibinfo {year} {2013}),\
  \Eprint{http://arxiv.org/abs/1304.7240}{arXiv:1304.7240 [gr-qc]}%
  \bibAnnoteFile{NoStop}{Babichev:2013usa}%
\bibitem{Koyama:2013paa}%
  \BibitemOpen
  \bibfield{author}{%
  \bibinfo {author} {\bibfnamefont{K.}~\bibnamefont{Koyama}}, \bibinfo {author}
  {\bibfnamefont{G.}~\bibnamefont{Niz}},\ and\ \bibinfo {author}
  {\bibfnamefont{G.}~\bibnamefont{Tasinato}}}%
   (\bibinfo {year} {2013}),\
  \Eprint{http://arxiv.org/abs/1305.0279}{arXiv:1305.0279 [hep-th]}%
  \bibAnnoteFile{NoStop}{Koyama:2013paa}%
\bibitem{Gannouji:2010au}%
  \BibitemOpen
  \bibfield{author}{%
  \bibinfo {author} {\bibfnamefont{R.}~\bibnamefont{Gannouji}}\ and\ \bibinfo
  {author} {\bibfnamefont{M.}~\bibnamefont{Sami}},\ }%
  \bibfield{journal}{%
  \Doi{10.1103/PhysRevD.82.024011}{\bibinfo {journal} {Phys.Rev.}}\ }%
  \textbf{\bibinfo {volume} {D82}},\ \bibinfo {pages} {024011} (\bibinfo {year}
  {2010}),\ \Eprint{http://arxiv.org/abs/1004.2808}{arXiv:1004.2808 [gr-qc]}%
  \bibAnnoteFile{NoStop}{Gannouji:2010au}%
\bibitem{Chow:2009fm}%
  \BibitemOpen
  \bibfield{author}{%
  \bibinfo {author} {\bibfnamefont{N.}~\bibnamefont{Chow}}\ and\ \bibinfo
  {author} {\bibfnamefont{J.}~\bibnamefont{Khoury}},\ }%
  \bibfield{journal}{%
  \Doi{10.1103/PhysRevD.80.024037}{\bibinfo {journal} {Phys.Rev.}}\ }%
  \textbf{\bibinfo {volume} {D80}},\ \bibinfo {pages} {024037} (\bibinfo {year}
  {2009}),\ \Eprint{http://arxiv.org/abs/0905.1325}{arXiv:0905.1325 [hep-th]}%
  \bibAnnoteFile{NoStop}{Chow:2009fm}%
\bibitem{PhysRevD.82.103015}%
  \BibitemOpen
  \bibfield{author}{%
  \bibinfo {author} {\bibfnamefont{A.}~\bibnamefont{Ali}}, \bibinfo {author}
  {\bibfnamefont{R.}~\bibnamefont{Gannouji}},\ and\ \bibinfo {author}
  {\bibfnamefont{M.}~\bibnamefont{Sami}},\ }%
  \bibfield{journal}{%
  \Doi{10.1103/PhysRevD.82.103015}{\bibinfo {journal} {Phys. Rev. D}}\ }%
  \textbf{\bibinfo {volume} {82}},\ \bibinfo {pages} {103015} (\bibinfo {year}
  {2010})%
  \bibAnnoteFile{NoStop}{PhysRevD.82.103015}%
\bibitem{Neveu:2014vua}%
  \BibitemOpen
  \bibfield{author}{%
  \bibinfo {author} {\bibfnamefont{J.}~\bibnamefont{Neveu}}, \bibinfo {author}
  {\bibfnamefont{V.}~\bibnamefont{Ruhlmann-Kleider}}, \bibinfo {author}
  {\bibfnamefont{P.}~\bibnamefont{Astier}}, \bibinfo {author}
  {\bibfnamefont{M.}~\bibnamefont{Besançon}}, \bibinfo {author}
  {\bibfnamefont{A.}~\bibnamefont{Conley}}, \emph{et~al.}}%
   (\bibinfo {year} {2014}),\
  \Eprint{http://arxiv.org/abs/1403.0854}{arXiv:1403.0854 [gr-qc]}%
  \bibAnnoteFile{NoStop}{Neveu:2014vua}%
\bibitem{DeFelice:2010nf}%
  \BibitemOpen
  \bibfield{author}{%
  \bibinfo {author} {\bibfnamefont{A.}~\bibnamefont{De~Felice}}\ and\ \bibinfo
  {author} {\bibfnamefont{S.}~\bibnamefont{Tsujikawa}},\ }%
  \bibfield{journal}{%
  \Doi{10.1103/PhysRevD.84.124029}{\bibinfo {journal} {Phys.Rev.}}\ }%
  \textbf{\bibinfo {volume} {D84}},\ \bibinfo {pages} {124029} (\bibinfo {year}
  {2011}),\ \Eprint{http://arxiv.org/abs/1008.4236}{arXiv:1008.4236 [hep-th]}%
  \bibAnnoteFile{NoStop}{DeFelice:2010nf}%
\bibitem{kimura1}%
  \BibitemOpen
  \bibfield{author}{%
  \bibinfo {author} {\bibfnamefont{R.}~\bibnamefont{{Kimura}}}\ and\ \bibinfo
  {author} {\bibfnamefont{K.}~\bibnamefont{{Yamamoto}}},\ }%
  \bibfield{journal}{%
  \Doi{10.1088/1475-7516/2011/04/025}{\bibinfo {journal} {JCAP}}\ }%
  \textbf{\bibinfo {volume} {4}},\ \bibinfo {eid} {025} (\bibinfo {month}
  {Apr.}\ \bibinfo {year} {2011}),\
  \Eprint{http://arxiv.org/abs/1011.2006}{arXiv:1011.2006 [astro-ph.CO]}%
  \bibAnnoteFile{NoStop}{kimura1}%
\bibitem{kimura2}%
  \BibitemOpen
  \bibfield{author}{%
  \bibinfo {author} {\bibfnamefont{R.}~\bibnamefont{{Kimura}}}, \bibinfo
  {author} {\bibfnamefont{T.}~\bibnamefont{{Kobayashi}}},\ and\ \bibinfo
  {author} {\bibfnamefont{K.}~\bibnamefont{{Yamamoto}}},\ }%
  \bibfield{journal}{%
  \Doi{10.1103/PhysRevD.85.123503}{\bibinfo {journal} {PRD}}\ }%
  \textbf{\bibinfo {volume} {85}},\ \bibinfo {eid} {123503} (\bibinfo {month}
  {Jun.}\ \bibinfo {year} {2012}),\
  \Eprint{http://arxiv.org/abs/1110.3598}{arXiv:1110.3598 [astro-ph.CO]}%
  \bibAnnoteFile{NoStop}{kimura2}%
\bibitem{DeFelice:2010pv}%
  \BibitemOpen
  \bibfield{author}{%
  \bibinfo {author} {\bibfnamefont{A.}~\bibnamefont{De~Felice}}\ and\ \bibinfo
  {author} {\bibfnamefont{S.}~\bibnamefont{Tsujikawa}},\ }%
  \bibfield{journal}{%
  \Doi{10.1103/PhysRevLett.105.111301}{\bibinfo {journal} {Phys.Rev.Lett.}}\ }%
  \textbf{\bibinfo {volume} {105}},\ \bibinfo {pages} {111301} (\bibinfo {year}
  {2010}),\ \Eprint{http://arxiv.org/abs/1007.2700}{arXiv:1007.2700
  [astro-ph.CO]}%
  \bibAnnoteFile{NoStop}{DeFelice:2010pv}%
\bibitem{Nesseris:2010pc}%
  \BibitemOpen
  \bibfield{author}{%
  \bibinfo {author} {\bibfnamefont{S.}~\bibnamefont{Nesseris}}, \bibinfo
  {author} {\bibfnamefont{A.}~\bibnamefont{De~Felice}},\ and\ \bibinfo {author}
  {\bibfnamefont{S.}~\bibnamefont{Tsujikawa}},\ }%
  \bibfield{journal}{%
  \Doi{10.1103/PhysRevD.82.124054}{\bibinfo {journal} {Phys.Rev.}}\ }%
  \textbf{\bibinfo {volume} {D82}},\ \bibinfo {pages} {124054} (\bibinfo {year}
  {2010}),\ \Eprint{http://arxiv.org/abs/1010.0407}{arXiv:1010.0407
  [astro-ph.CO]}%
  \bibAnnoteFile{NoStop}{Nesseris:2010pc}%
\bibitem{DeFelice:2010as}%
  \BibitemOpen
  \bibfield{author}{%
  \bibinfo {author} {\bibfnamefont{A.}~\bibnamefont{De~Felice}}, \bibinfo
  {author} {\bibfnamefont{R.}~\bibnamefont{Kase}},\ and\ \bibinfo {author}
  {\bibfnamefont{S.}~\bibnamefont{Tsujikawa}},\ }%
  \bibfield{journal}{%
  \Doi{10.1103/PhysRevD.83.043515}{\bibinfo {journal} {Phys.Rev.}}\ }%
  \textbf{\bibinfo {volume} {D83}},\ \bibinfo {pages} {043515} (\bibinfo {year}
  {2011}),\ \Eprint{http://arxiv.org/abs/1011.6132}{arXiv:1011.6132
  [astro-ph.CO]}%
  \bibAnnoteFile{NoStop}{DeFelice:2010as}%
\bibitem{Appleby:2011aa}%
  \BibitemOpen
  \bibfield{author}{%
  \bibinfo {author} {\bibfnamefont{S.}~\bibnamefont{Appleby}}\ and\ \bibinfo
  {author} {\bibfnamefont{E.~V.}\ \bibnamefont{Linder}},\ }%
  \bibfield{journal}{%
  \Doi{10.1088/1475-7516/2012/03/043}{\bibinfo {journal} {JCAP}}\ }%
  \textbf{\bibinfo {volume} {1203}},\ \bibinfo {pages} {043} (\bibinfo {year}
  {2012}),\ \Eprint{http://arxiv.org/abs/1112.1981}{arXiv:1112.1981
  [astro-ph.CO]}%
  \bibAnnoteFile{NoStop}{Appleby:2011aa}%
\bibitem{Appleby:2012ba}%
  \BibitemOpen
  \bibfield{author}{%
  \bibinfo {author} {\bibfnamefont{S.~A.}\ \bibnamefont{Appleby}}\ and\
  \bibinfo {author} {\bibfnamefont{E.~V.}\ \bibnamefont{Linder}},\ }%
  \bibfield{journal}{%
  \Doi{10.1088/1475-7516/2012/08/026}{\bibinfo {journal} {JCAP}}\ }%
  \textbf{\bibinfo {volume} {1208}},\ \bibinfo {pages} {026} (\bibinfo {year}
  {2012}),\ \Eprint{http://arxiv.org/abs/1204.4314}{arXiv:1204.4314
  [astro-ph.CO]}%
  \bibAnnoteFile{NoStop}{Appleby:2012ba}%
\bibitem{Okada:2012mn}%
  \BibitemOpen
  \bibfield{author}{%
  \bibinfo {author} {\bibfnamefont{H.}~\bibnamefont{Okada}}, \bibinfo {author}
  {\bibfnamefont{T.}~\bibnamefont{Totani}},\ and\ \bibinfo {author}
  {\bibfnamefont{S.}~\bibnamefont{Tsujikawa}},\ }%
  \bibfield{journal}{%
  \Doi{10.1103/PhysRevD.87.103002}{\bibinfo {journal} {Phys.Rev.}}\ }%
  \textbf{\bibinfo {volume} {D87}},\ \bibinfo {pages} {103002} (\bibinfo {year}
  {2013}),\ \Eprint{http://arxiv.org/abs/1208.4681}{arXiv:1208.4681
  [astro-ph.CO]}%
  \bibAnnoteFile{NoStop}{Okada:2012mn}%
\bibitem{Neveu:2013mfa}%
  \BibitemOpen
  \bibfield{author}{%
  \bibinfo {author} {\bibfnamefont{J.}~\bibnamefont{Neveu}}, \bibinfo {author}
  {\bibfnamefont{V.}~\bibnamefont{Ruhlmann-Kleider}}, \bibinfo {author}
  {\bibfnamefont{A.}~\bibnamefont{Conley}}, \bibinfo {author}
  {\bibfnamefont{N.}~\bibnamefont{Palanque-Delabrouille}}, \bibinfo {author}
  {\bibfnamefont{P.}~\bibnamefont{Astier}}, \emph{et~al.},\ }%
  \bibfield{journal}{%
  \Doi{10.1051/0004-6361/201321256}{\bibinfo {journal} {Astron.Astrophys.}}\ }%
  \textbf{\bibinfo {volume} {555}},\ \bibinfo {pages} {A53} (\bibinfo {year}
  {2013}),\ \Eprint{http://arxiv.org/abs/1302.2786}{arXiv:1302.2786 [gr-qc]}%
  \bibAnnoteFile{NoStop}{Neveu:2013mfa}%
\bibitem{Barreira:2012kk}%
  \BibitemOpen
  \bibfield{author}{%
  \bibinfo {author} {\bibfnamefont{A.}~\bibnamefont{Barreira}}, \bibinfo
  {author} {\bibfnamefont{B.}~\bibnamefont{Li}}, \bibinfo {author}
  {\bibfnamefont{C.~M.}\ \bibnamefont{Baugh}},\ and\ \bibinfo {author}
  {\bibfnamefont{S.}~\bibnamefont{Pascoli}},\ }%
  \bibfield{journal}{%
  \Doi{10.1103/PhysRevD.86.124016}{\bibinfo {journal} {Phys.Rev.}}\ }%
  \textbf{\bibinfo {volume} {D86}},\ \bibinfo {pages} {124016} (\bibinfo {year}
  {2012}),\ \Eprint{http://arxiv.org/abs/1208.0600}{arXiv:1208.0600
  [astro-ph.CO]}%
  \bibAnnoteFile{NoStop}{Barreira:2012kk}%
\bibitem{Barreira:2013jma}%
  \BibitemOpen
  \bibfield{author}{%
  \bibinfo {author} {\bibfnamefont{A.}~\bibnamefont{Barreira}}, \bibinfo
  {author} {\bibfnamefont{B.}~\bibnamefont{Li}}, \bibinfo {author}
  {\bibfnamefont{A.}~\bibnamefont{Sanchez}}, \bibinfo {author}
  {\bibfnamefont{C.~M.}\ \bibnamefont{Baugh}},\ and\ \bibinfo {author}
  {\bibfnamefont{S.}~\bibnamefont{Pascoli}},\ }%
  \bibfield{journal}{%
  \Doi{10.1103/PhysRevD.87.103511}{\bibinfo {journal} {Phys.Rev.}}\ }%
  \textbf{\bibinfo {volume} {D87}},\ \bibinfo {pages} {103511} (\bibinfo {year}
  {2013}),\ \Eprint{http://arxiv.org/abs/1302.6241}{arXiv:1302.6241
  [astro-ph.CO]}%
  \bibAnnoteFile{NoStop}{Barreira:2013jma}%
\bibitem{Bartolo:2013ws}%
  \BibitemOpen
  \bibfield{author}{%
  \bibinfo {author} {\bibfnamefont{N.}~\bibnamefont{Bartolo}}, \bibinfo
  {author} {\bibfnamefont{E.}~\bibnamefont{Bellini}}, \bibinfo {author}
  {\bibfnamefont{D.}~\bibnamefont{Bertacca}},\ and\ \bibinfo {author}
  {\bibfnamefont{S.}~\bibnamefont{Matarrese}},\ }%
  \bibfield{journal}{%
  \Doi{10.1088/1475-7516/2013/03/034}{\bibinfo {journal} {JCAP}}\ }%
  \textbf{\bibinfo {volume} {1303}},\ \bibinfo {pages} {034} (\bibinfo {year}
  {2013}),\ \Eprint{http://arxiv.org/abs/1301.4831}{arXiv:1301.4831
  [astro-ph.CO]}%
  \bibAnnoteFile{NoStop}{Bartolo:2013ws}%
\bibitem{Barreira:2013eea}%
  \BibitemOpen
  \bibfield{author}{%
  \bibinfo {author} {\bibfnamefont{A.}~\bibnamefont{Barreira}}, \bibinfo
  {author} {\bibfnamefont{B.}~\bibnamefont{Li}}, \bibinfo {author}
  {\bibfnamefont{W.~A.}\ \bibnamefont{Hellwing}}, \bibinfo {author}
  {\bibfnamefont{C.~M.}\ \bibnamefont{Baugh}},\ and\ \bibinfo {author}
  {\bibfnamefont{S.}~\bibnamefont{Pascoli}},\ }%
  \bibfield{journal}{%
  \bibinfo {journal} {JCAP}\ }%
  \textbf{\bibinfo {volume} {2013}},\ \bibinfo {pages} {027} (\bibinfo {year}
  {2013}),\ \Eprint{http://arxiv.org/abs/1306.3219}{arXiv:1306.3219
  [astro-ph.CO]}%
  \bibAnnoteFile{NoStop}{Barreira:2013eea}%
\bibitem{Li:2013tda}%
  \BibitemOpen
  \bibfield{author}{%
  \bibinfo {author} {\bibfnamefont{B.}~\bibnamefont{Li}}, \bibinfo {author}
  {\bibfnamefont{A.}~\bibnamefont{Barreira}}, \bibinfo {author}
  {\bibfnamefont{C.~M.}\ \bibnamefont{Baugh}}, \bibinfo {author}
  {\bibfnamefont{W.~A.}\ \bibnamefont{Hellwing}}, \bibinfo {author}
  {\bibfnamefont{K.}~\bibnamefont{Koyama}}, \emph{et~al.},\ }%
  \bibfield{journal}{%
  \Doi{10.1088/1475-7516/2013/11/012}{\bibinfo {journal} {JCAP}}\ }%
  \textbf{\bibinfo {volume} {1311}},\ \bibinfo {pages} {012} (\bibinfo {year}
  {2013}),\ \Eprint{http://arxiv.org/abs/1308.3491}{arXiv:1308.3491
  [astro-ph.CO]}%
  \bibAnnoteFile{NoStop}{Li:2013tda}%
\bibitem{Barreira:2013xea}%
  \BibitemOpen
  \bibfield{author}{%
  \bibinfo {author} {\bibfnamefont{A.}~\bibnamefont{Barreira}}, \bibinfo
  {author} {\bibfnamefont{B.}~\bibnamefont{Li}}, \bibinfo {author}
  {\bibfnamefont{C.~M.}\ \bibnamefont{Baugh}},\ and\ \bibinfo {author}
  {\bibfnamefont{S.}~\bibnamefont{Pascoli}},\ }%
  \bibfield{journal}{%
  \Doi{10.1088/1475-7516/2013/11/056}{\bibinfo {journal} {JCAP}}\ }%
  \textbf{\bibinfo {volume} {1311}},\ \bibinfo {pages} {056} (\bibinfo {year}
  {2013}),\ \Eprint{http://arxiv.org/abs/1308.3699}{arXiv:1308.3699
  [astro-ph.CO]}%
  \bibAnnoteFile{NoStop}{Barreira:2013xea}%
\bibitem{Barreira:2014zza}%
  \BibitemOpen
  \bibfield{author}{%
  \bibinfo {author} {\bibfnamefont{A.}~\bibnamefont{Barreira}}, \bibinfo
  {author} {\bibfnamefont{B.}~\bibnamefont{Li}}, \bibinfo {author}
  {\bibfnamefont{W.~A.}\ \bibnamefont{Hellwing}}, \bibinfo {author}
  {\bibfnamefont{L.}~\bibnamefont{Lombriser}}, \bibinfo {author}
  {\bibfnamefont{C.~M.}\ \bibnamefont{Baugh}}, \emph{et~al.}}%
   (\bibinfo {year} {2014}),\ \doi{\bibinfo {doi}
  {10.1088/1475-7516/2014/04/029}},\
  \Eprint{http://arxiv.org/abs/1401.1497}{arXiv:1401.1497 [astro-ph.CO]}%
  \bibAnnoteFile{NoStop}{Barreira:2014zza}%
\bibitem{Hellwing:2014nma}%
  \BibitemOpen
  \bibfield{author}{%
  \bibinfo {author} {\bibfnamefont{W.~A.}\ \bibnamefont{Hellwing}}, \bibinfo
  {author} {\bibfnamefont{A.}~\bibnamefont{Barreira}}, \bibinfo {author}
  {\bibfnamefont{C.~S.}\ \bibnamefont{Frenk}}, \bibinfo {author}
  {\bibfnamefont{B.}~\bibnamefont{Li}},\ and\ \bibinfo {author}
  {\bibfnamefont{S.}~\bibnamefont{Cole}}}%
   (\bibinfo {year} {2014}),\
  \Eprint{http://arxiv.org/abs/1401.0706}{arXiv:1401.0706 [astro-ph.CO]}%
  \bibAnnoteFile{NoStop}{Hellwing:2014nma}%
\bibitem{GonzalezGarcia:2012sz}%
  \BibitemOpen
  \bibfield{author}{%
  \bibinfo {author} {\bibfnamefont{M.}~\bibnamefont{Gonzalez-Garcia}}, \bibinfo
  {author} {\bibfnamefont{M.}~\bibnamefont{Maltoni}}, \bibinfo {author}
  {\bibfnamefont{J.}~\bibnamefont{Salvado}},\ and\ \bibinfo {author}
  {\bibfnamefont{T.}~\bibnamefont{Schwetz}},\ }%
  \bibfield{journal}{%
  \Doi{10.1007/JHEP12(2012)123}{\bibinfo {journal} {JHEP}}\ }%
  \textbf{\bibinfo {volume} {1212}},\ \bibinfo {pages} {123} (\bibinfo {year}
  {2012}),\ \Eprint{http://arxiv.org/abs/1209.3023}{arXiv:1209.3023 [hep-ph
  (Latest results on Neutrino masses can be found at http://www.nu-fit.org/)]}%
  \bibAnnoteFile{NoStop}{GonzalezGarcia:2012sz}%
\bibitem{Kraus:2004zw}%
  \BibitemOpen
  \bibfield{author}{%
  \bibinfo {author} {\bibfnamefont{C.}~\bibnamefont{Kraus}}, \bibinfo {author}
  {\bibfnamefont{B.}~\bibnamefont{Bornschein}}, \bibinfo {author}
  {\bibfnamefont{L.}~\bibnamefont{Bornschein}}, \bibinfo {author}
  {\bibfnamefont{J.}~\bibnamefont{Bonn}}, \bibinfo {author}
  {\bibfnamefont{B.}~\bibnamefont{Flatt}}, \emph{et~al.},\ }%
  \bibfield{journal}{%
  \Doi{10.1140/epjc/s2005-02139-7}{\bibinfo {journal} {Eur.Phys.J.}}\ }%
  \textbf{\bibinfo {volume} {C40}},\ \bibinfo {pages} {447} (\bibinfo {year}
  {2005}),\ \Eprint{http://arxiv.org/abs/hep-ex/0412056}{arXiv:hep-ex/0412056
  [hep-ex]}%
  \bibAnnoteFile{NoStop}{Kraus:2004zw}%
\bibitem{Aseev:2011dq}%
  \BibitemOpen
  \bibfield{author}{%
  \bibinfo {author} {\bibfnamefont{V.}~\bibnamefont{Aseev}} \emph{et~al.}
  (\bibinfo {collaboration} {Troitsk Collaboration}),\ }%
  \bibfield{journal}{%
  \Doi{10.1103/PhysRevD.84.112003}{\bibinfo {journal} {Phys.Rev.}}\ }%
  \textbf{\bibinfo {volume} {D84}},\ \bibinfo {pages} {112003} (\bibinfo {year}
  {2011}),\ \Eprint{http://arxiv.org/abs/1108.5034}{arXiv:1108.5034 [hep-ex]}%
  \bibAnnoteFile{NoStop}{Aseev:2011dq}%
\bibitem{He:2013qha}%
  \BibitemOpen
  \bibfield{author}{%
  \bibinfo {author} {\bibfnamefont{J.-h.}\ \bibnamefont{He}},\ }%
  \bibfield{journal}{%
  \Doi{10.1103/PhysRevD.88.103523}{\bibinfo {journal} {Phys. Rev. D 88,}}\ }%
  \textbf{\bibinfo {volume} {103523}},\ \bibinfo {pages} {103523} (\bibinfo
  {year} {2013}),\ \Eprint{http://arxiv.org/abs/1307.4876}{arXiv:1307.4876
  [astro-ph.CO]}%
  \bibAnnoteFile{NoStop}{He:2013qha}%
\bibitem{Motohashi:2012wc}%
  \BibitemOpen
  \bibfield{author}{%
  \bibinfo {author} {\bibfnamefont{H.}~\bibnamefont{Motohashi}}, \bibinfo
  {author} {\bibfnamefont{A.~A.}\ \bibnamefont{Starobinsky}},\ and\ \bibinfo
  {author} {\bibfnamefont{J.}~\bibnamefont{Yokoyama}},\ }%
  \bibfield{journal}{%
  \Doi{10.1103/PhysRevLett.110.121302}{\bibinfo {journal} {Phys.Rev.Lett.}}\ }%
  \textbf{\bibinfo {volume} {110}},\ \bibinfo {pages} {121302} (\bibinfo {year}
  {2013}),\ \Eprint{http://arxiv.org/abs/1203.6828}{arXiv:1203.6828
  [astro-ph.CO]}%
  \bibAnnoteFile{NoStop}{Motohashi:2012wc}%
\bibitem{Baldi:2013iza}%
  \BibitemOpen
  \bibfield{author}{%
  \bibinfo {author} {\bibfnamefont{M.}~\bibnamefont{Baldi}}, \bibinfo {author}
  {\bibfnamefont{F.}~\bibnamefont{Villaescusa-Navarro}}, \bibinfo {author}
  {\bibfnamefont{M.}~\bibnamefont{Viel}}, \bibinfo {author}
  {\bibfnamefont{E.}~\bibnamefont{Puchwein}}, \bibinfo {author}
  {\bibfnamefont{V.}~\bibnamefont{Springel}}, \emph{et~al.}}%
   (\bibinfo {year} {2013}),\
  \Eprint{http://arxiv.org/abs/1311.2588}{arXiv:1311.2588 [astro-ph.CO]}%
  \bibAnnoteFile{NoStop}{Baldi:2013iza}%
\bibitem{Barreira:2014ija}%
  \BibitemOpen
  \bibfield{author}{%
  \bibinfo {author} {\bibfnamefont{A.}~\bibnamefont{Barreira}}, \bibinfo
  {author} {\bibfnamefont{B.}~\bibnamefont{Li}}, \bibinfo {author}
  {\bibfnamefont{C.}~\bibnamefont{Baugh}},\ and\ \bibinfo {author}
  {\bibfnamefont{S.}~\bibnamefont{Pascoli}}}%
   (\bibinfo {year} {2014}),\
  \Eprint{http://arxiv.org/abs/1404.1365}{arXiv:1404.1365 [astro-ph.CO]}%
  \bibAnnoteFile{NoStop}{Barreira:2014ija}%
\bibitem{Ade:2013zuv}%
  \BibitemOpen
  \bibfield{author}{%
  \bibinfo {author} {\bibfnamefont{P.}~\bibnamefont{Ade}} \emph{et~al.}
  (\bibinfo {collaboration} {Planck Collaboration})}%
   (\bibinfo {year} {2013}),\
  \Eprint{http://arxiv.org/abs/1303.5076}{arXiv:1303.5076 [astro-ph.CO]}%
  \bibAnnoteFile{NoStop}{Ade:2013zuv}%
\bibitem{Ade:2013tyw}%
  \BibitemOpen
  \bibfield{author}{%
  \bibinfo {author} {\bibfnamefont{P.}~\bibnamefont{Ade}} \emph{et~al.}
  (\bibinfo {collaboration} {Planck Collaboration})}%
   (\bibinfo {year} {2013}),\
  \Eprint{http://arxiv.org/abs/1303.5077}{arXiv:1303.5077 [astro-ph.CO]}%
  \bibAnnoteFile{NoStop}{Ade:2013tyw}%
\bibitem{PhysRevD.40.1804}%
  \BibitemOpen
  \bibfield{author}{%
  \bibinfo {author} {\bibfnamefont{G.~F.~R.}\ \bibnamefont{Ellis}}\ and\
  \bibinfo {author} {\bibfnamefont{M.}~\bibnamefont{Bruni}},\ }%
  \bibfield{journal}{%
  \Doi{10.1103/PhysRevD.40.1804}{\bibinfo {journal} {Phys. Rev. D}}\ }%
  \textbf{\bibinfo {volume} {40}},\ \bibinfo {pages} {1804} (\bibinfo {month}
  {Sep}\ \bibinfo {year} {1989}),\
  \url{http://link.aps.org/doi/10.1103/PhysRevD.40.1804}%
  \bibAnnoteFile{NoStop}{PhysRevD.40.1804}%
\bibitem{Challinor:1998xk}%
  \BibitemOpen
  \bibfield{author}{%
  \bibinfo {author} {\bibfnamefont{A.}~\bibnamefont{Challinor}}\ and\ \bibinfo
  {author} {\bibfnamefont{A.}~\bibnamefont{Lasenby}},\ }%
  \bibfield{journal}{%
  \Doi{10.1086/306841}{\bibinfo {journal} {Astrophys.J.}}\ }%
  \textbf{\bibinfo {volume} {513}},\ \bibinfo {pages} {1} (\bibinfo {year}
  {1999}),\
  \Eprint{http://arxiv.org/abs/astro-ph/9804301}{arXiv:astro-ph/9804301
  [astro-ph]}%
  \bibAnnoteFile{NoStop}{Challinor:1998xk}%
\bibitem{Challinor:1999xz}%
  \BibitemOpen
  \bibfield{author}{%
  \bibinfo {author} {\bibfnamefont{A.}~\bibnamefont{Challinor}},\ }%
  \bibfield{journal}{%
  \Doi{10.1088/0264-9381/17/4/309}{\bibinfo {journal} {Class.Quant.Grav.}}\ }%
  \textbf{\bibinfo {volume} {17}},\ \bibinfo {pages} {871} (\bibinfo {year}
  {2000}),\
  \Eprint{http://arxiv.org/abs/astro-ph/9906474}{arXiv:astro-ph/9906474
  [astro-ph]}%
  \bibAnnoteFile{NoStop}{Challinor:1999xz}%
\bibitem{Challinor:2000as}%
  \BibitemOpen
  \bibfield{author}{%
  \bibinfo {author} {\bibfnamefont{A.}~\bibnamefont{Challinor}},\ }%
  \bibfield{journal}{%
  \Doi{10.1103/PhysRevD.62.043004}{\bibinfo {journal} {Phys.Rev.}}\ }%
  \textbf{\bibinfo {volume} {D62}},\ \bibinfo {pages} {043004} (\bibinfo {year}
  {2000}),\
  \Eprint{http://arxiv.org/abs/astro-ph/9911481}{arXiv:astro-ph/9911481
  [astro-ph]}%
  \bibAnnoteFile{NoStop}{Challinor:2000as}%
\bibitem{camb_notes}%
  \BibitemOpen
  \bibfield{author}{%
  \bibinfo {author} {\bibfnamefont{A.}~\bibnamefont{Lewis}},\ }%
  \bibinfo {note} {\url{http://camb.info/}}%
  \bibAnnoteFile{NoStop}{camb_notes}%
\bibitem{Lewis:2002ah}%
  \BibitemOpen
  \bibfield{author}{%
  \bibinfo {author} {\bibfnamefont{A.}~\bibnamefont{Lewis}}\ and\ \bibinfo
  {author} {\bibfnamefont{S.}~\bibnamefont{Bridle}},\ }%
  \bibfield{journal}{%
  \Doi{10.1103/PhysRevD.66.103511}{\bibinfo {journal} {Phys.Rev.}}\ }%
  \textbf{\bibinfo {volume} {D66}},\ \bibinfo {pages} {103511} (\bibinfo {year}
  {2002}),\
  \Eprint{http://arxiv.org/abs/astro-ph/0205436}{arXiv:astro-ph/0205436
  [astro-ph]}%
  \bibAnnoteFile{NoStop}{Lewis:2002ah}%
\bibitem{Ade:2013kta}%
  \BibitemOpen
  \bibfield{author}{%
  \bibinfo {author} {\bibfnamefont{P.}~\bibnamefont{Ade}} \emph{et~al.}
  (\bibinfo {collaboration} {Planck Collaboration})}%
   (\bibinfo {year} {2013}),\
  \Eprint{http://arxiv.org/abs/1303.5075}{arXiv:1303.5075 [astro-ph.CO]}%
  \bibAnnoteFile{NoStop}{Ade:2013kta}%
\bibitem{Hinshaw:2012fq}%
  \BibitemOpen
  \bibfield{author}{%
  \bibinfo {author} {\bibfnamefont{G.}~\bibnamefont{Hinshaw}}, \bibinfo
  {author} {\bibfnamefont{D.}~\bibnamefont{Larson}}, \bibinfo {author}
  {\bibfnamefont{E.}~\bibnamefont{Komatsu}}, \bibinfo {author}
  {\bibfnamefont{D.}~\bibnamefont{Spergel}}, \bibinfo {author}
  {\bibfnamefont{C.}~\bibnamefont{Bennett}}, \emph{et~al.}}%
   (\bibinfo {year} {2012}),\
  \Eprint{http://arxiv.org/abs/1212.5226}{arXiv:1212.5226 [astro-ph.CO]}%
  \bibAnnoteFile{NoStop}{Hinshaw:2012fq}%
\bibitem{Beutler:2011hx}%
  \BibitemOpen
  \bibfield{author}{%
  \bibinfo {author} {\bibfnamefont{F.}~\bibnamefont{Beutler}}, \bibinfo
  {author} {\bibfnamefont{C.}~\bibnamefont{Blake}}, \bibinfo {author}
  {\bibfnamefont{M.}~\bibnamefont{Colless}}, \bibinfo {author}
  {\bibfnamefont{D.~H.}\ \bibnamefont{Jones}}, \bibinfo {author}
  {\bibfnamefont{L.}~\bibnamefont{Staveley-Smith}}, \emph{et~al.},\ }%
  \bibfield{journal}{%
  \Doi{10.1111/j.1365-2966.2011.19250.x}{\bibinfo {journal}
  {Mon.Not.Roy.Astron.Soc.}}\ }%
  \textbf{\bibinfo {volume} {416}},\ \bibinfo {pages} {3017} (\bibinfo {year}
  {2011}),\ \Eprint{http://arxiv.org/abs/1106.3366}{arXiv:1106.3366
  [astro-ph.CO]}%
  \bibAnnoteFile{NoStop}{Beutler:2011hx}%
\bibitem{Padmanabhan:2012hf}%
  \BibitemOpen
  \bibfield{author}{%
  \bibinfo {author} {\bibfnamefont{N.}~\bibnamefont{Padmanabhan}}, \bibinfo
  {author} {\bibfnamefont{X.}~\bibnamefont{Xu}}, \bibinfo {author}
  {\bibfnamefont{D.~J.}\ \bibnamefont{Eisenstein}}, \bibinfo {author}
  {\bibfnamefont{R.}~\bibnamefont{Scalzo}}, \bibinfo {author}
  {\bibfnamefont{A.~J.}\ \bibnamefont{Cuesta}}, \emph{et~al.},\ }%
  \bibfield{journal}{%
  \Doi{10.1111/j.1365-2966.2012.21888.x}{\bibinfo {journal}
  {Mon.Not.Roy.Astron.Soc.}}\ }%
  \textbf{\bibinfo {volume} {427}},\ \bibinfo {pages} {2132} (\bibinfo {year}
  {2012}),\ \Eprint{http://arxiv.org/abs/1202.0090}{arXiv:1202.0090
  [astro-ph.CO]}%
  \bibAnnoteFile{NoStop}{Padmanabhan:2012hf}%
\bibitem{Anderson:2012sa}%
  \BibitemOpen
  \bibfield{author}{%
  \bibinfo {author} {\bibfnamefont{L.}~\bibnamefont{Anderson}}, \bibinfo
  {author} {\bibfnamefont{E.}~\bibnamefont{Aubourg}}, \bibinfo {author}
  {\bibfnamefont{S.}~\bibnamefont{Bailey}}, \bibinfo {author}
  {\bibfnamefont{D.}~\bibnamefont{Bizyaev}}, \bibinfo {author}
  {\bibfnamefont{M.}~\bibnamefont{Blanton}}, \emph{et~al.},\ }%
  \bibfield{journal}{%
  \Doi{10.1111/j.1365-2966.2012.22066.x}{\bibinfo {journal}
  {Mon.Not.Roy.Astron.Soc.}}\ }%
  \textbf{\bibinfo {volume} {427}},\ \bibinfo {pages} {3435} (\bibinfo {year}
  {2013}),\ \Eprint{http://arxiv.org/abs/1203.6594}{arXiv:1203.6594
  [astro-ph.CO]}%
  \bibAnnoteFile{NoStop}{Anderson:2012sa}%
\bibitem{Blake:2011en}%
  \BibitemOpen
  \bibfield{author}{%
  \bibinfo {author} {\bibfnamefont{C.}~\bibnamefont{Blake}}, \bibinfo {author}
  {\bibfnamefont{E.}~\bibnamefont{Kazin}}, \bibinfo {author}
  {\bibfnamefont{F.}~\bibnamefont{Beutler}}, \bibinfo {author}
  {\bibfnamefont{T.}~\bibnamefont{Davis}}, \bibinfo {author}
  {\bibfnamefont{D.}~\bibnamefont{Parkinson}}, \emph{et~al.},\ }%
  \bibfield{journal}{%
  \Doi{10.1111/j.1365-2966.2011.19592.x}{\bibinfo {journal}
  {Mon.Not.Roy.Astron.Soc.}}\ }%
  \textbf{\bibinfo {volume} {418}},\ \bibinfo {pages} {1707} (\bibinfo {year}
  {2011}),\ \Eprint{http://arxiv.org/abs/1108.2635}{arXiv:1108.2635
  [astro-ph.CO]}%
  \bibAnnoteFile{NoStop}{Blake:2011en}%
\bibitem{Babichev:2007dw}%
  \BibitemOpen
  \bibfield{author}{%
  \bibinfo {author} {\bibfnamefont{E.}~\bibnamefont{Babichev}}, \bibinfo
  {author} {\bibfnamefont{V.}~\bibnamefont{Mukhanov}},\ and\ \bibinfo {author}
  {\bibfnamefont{A.}~\bibnamefont{Vikman}},\ }%
  \bibfield{journal}{%
  \Doi{10.1088/1126-6708/2008/02/101}{\bibinfo {journal} {JHEP}}\ }%
  \textbf{\bibinfo {volume} {0802}},\ \bibinfo {pages} {101} (\bibinfo {year}
  {2008}),\ \Eprint{http://arxiv.org/abs/0708.0561}{arXiv:0708.0561 [hep-th]}%
  \bibAnnoteFile{NoStop}{Babichev:2007dw}%
\bibitem{1979Natur.281..358A}%
  \BibitemOpen
  \bibfield{author}{%
  \bibinfo {author} {\bibfnamefont{C.}~\bibnamefont{{Alcock}}}\ and\ \bibinfo
  {author} {\bibfnamefont{B.}~\bibnamefont{{Paczynski}}},\ }%
  \bibfield{journal}{%
  \Doi{10.1038/281358a0}{\bibinfo {journal} {\nat}}\ }%
  \textbf{\bibinfo {volume} {281}},\ \bibinfo {pages} {358} (\bibinfo {month}
  {Oct.}\ \bibinfo {year} {1979})%
  \bibAnnoteFile{NoStop}{1979Natur.281..358A}%
\bibitem{Blake:2011rj}%
  \BibitemOpen
  \bibfield{author}{%
  \bibinfo {author} {\bibfnamefont{C.}~\bibnamefont{Blake}}, \bibinfo {author}
  {\bibfnamefont{S.}~\bibnamefont{Brough}}, \bibinfo {author}
  {\bibfnamefont{M.}~\bibnamefont{Colless}}, \bibinfo {author}
  {\bibfnamefont{C.}~\bibnamefont{Contreras}}, \bibinfo {author}
  {\bibfnamefont{W.}~\bibnamefont{Couch}}, \emph{et~al.},\ }%
  \bibfield{journal}{%
  \Doi{10.1111/j.1365-2966.2011.18903.x}{\bibinfo {journal}
  {Mon.Not.Roy.Astron.Soc.}}\ }%
  \textbf{\bibinfo {volume} {415}},\ \bibinfo {pages} {2876} (\bibinfo {year}
  {2011}),\ \Eprint{http://arxiv.org/abs/1104.2948}{arXiv:1104.2948
  [astro-ph.CO]}%
  \bibAnnoteFile{NoStop}{Blake:2011rj}%
\bibitem{Kravtsov:2003sg}%
  \BibitemOpen
  \bibfield{author}{%
  \bibinfo {author} {\bibfnamefont{A.~V.}\ \bibnamefont{Kravtsov}}, \bibinfo
  {author} {\bibfnamefont{A.~A.}\ \bibnamefont{Berlind}}, \bibinfo {author}
  {\bibfnamefont{R.~H.}\ \bibnamefont{Wechsler}}, \bibinfo {author}
  {\bibfnamefont{A.~A.}\ \bibnamefont{Klypin}}, \bibinfo {author}
  {\bibfnamefont{S.}~\bibnamefont{Gottloeber}}, \emph{et~al.},\ }%
  \bibfield{journal}{%
  \Doi{10.1086/420959}{\bibinfo {journal} {Astrophys.J.}}\ }%
  \textbf{\bibinfo {volume} {609}},\ \bibinfo {pages} {35} (\bibinfo {year}
  {2004}),\
  \Eprint{http://arxiv.org/abs/astro-ph/0308519}{arXiv:astro-ph/0308519
  [astro-ph]}%
  \bibAnnoteFile{NoStop}{Kravtsov:2003sg}%
\bibitem{Reid:2009xm}%
  \BibitemOpen
  \bibfield{author}{%
  \bibinfo {author} {\bibfnamefont{B.~A.}\ \bibnamefont{Reid}}, \bibinfo
  {author} {\bibfnamefont{W.~J.}\ \bibnamefont{Percival}}, \bibinfo {author}
  {\bibfnamefont{D.~J.}\ \bibnamefont{Eisenstein}}, \bibinfo {author}
  {\bibfnamefont{L.}~\bibnamefont{Verde}}, \bibinfo {author}
  {\bibfnamefont{D.~N.}\ \bibnamefont{Spergel}}, \emph{et~al.},\ }%
  \bibfield{journal}{%
  \Doi{10.1111/j.1365-2966.2010.16276.x}{\bibinfo {journal}
  {Mon.Not.Roy.Astron.Soc.}}\ }%
  \textbf{\bibinfo {volume} {404}},\ \bibinfo {pages} {60} (\bibinfo {year}
  {2010}),\ \Eprint{http://arxiv.org/abs/0907.1659}{arXiv:0907.1659
  [astro-ph.CO]}%
  \bibAnnoteFile{NoStop}{Reid:2009xm}%
\bibitem{Babichev:2011iz}%
  \BibitemOpen
  \bibfield{author}{%
  \bibinfo {author} {\bibfnamefont{E.}~\bibnamefont{Babichev}}, \bibinfo
  {author} {\bibfnamefont{C.}~\bibnamefont{Deffayet}},\ and\ \bibinfo {author}
  {\bibfnamefont{G.}~\bibnamefont{Esposito-Farese}},\ }%
  \bibfield{journal}{%
  \Doi{10.1103/PhysRevLett.107.251102}{\bibinfo {journal} {Phys.Rev.Lett.}}\ }%
  \textbf{\bibinfo {volume} {107}},\ \bibinfo {pages} {251102} (\bibinfo {year}
  {2011}),\ \Eprint{http://arxiv.org/abs/1107.1569}{arXiv:1107.1569 [gr-qc]}%
  \bibAnnoteFile{NoStop}{Babichev:2011iz}%
\bibitem{Kimura:2011dc}%
  \BibitemOpen
  \bibfield{author}{%
  \bibinfo {author} {\bibfnamefont{R.}~\bibnamefont{Kimura}}, \bibinfo {author}
  {\bibfnamefont{T.}~\bibnamefont{Kobayashi}},\ and\ \bibinfo {author}
  {\bibfnamefont{K.}~\bibnamefont{Yamamoto}},\ }%
  \bibfield{journal}{%
  \Doi{10.1103/PhysRevD.85.024023}{\bibinfo {journal} {Phys.Rev.}}\ }%
  \textbf{\bibinfo {volume} {D85}},\ \bibinfo {pages} {024023} (\bibinfo {year}
  {2012}),\ \Eprint{http://arxiv.org/abs/1111.6749}{arXiv:1111.6749
  [astro-ph.CO]}%
  \bibAnnoteFile{NoStop}{Kimura:2011dc}%
\bibitem{Riess:2011yx}%
  \BibitemOpen
  \bibfield{author}{%
  \bibinfo {author} {\bibfnamefont{A.~G.}\ \bibnamefont{Riess}}, \bibinfo
  {author} {\bibfnamefont{L.}~\bibnamefont{Macri}}, \bibinfo {author}
  {\bibfnamefont{S.}~\bibnamefont{Casertano}}, \bibinfo {author}
  {\bibfnamefont{H.}~\bibnamefont{Lampeitl}}, \bibinfo {author}
  {\bibfnamefont{H.~C.}\ \bibnamefont{Ferguson}}, \emph{et~al.},\ }%
  \bibfield{journal}{%
  \Doi{10.1088/0004-637X/732/2/129, 10.1088/0004-637X/730/2/119}{\bibinfo
  {journal} {Astrophys.J.}}\ }%
  \textbf{\bibinfo {volume} {730}},\ \bibinfo {pages} {119} (\bibinfo {year}
  {2011}),\ \Eprint{http://arxiv.org/abs/1103.2976}{arXiv:1103.2976
  [astro-ph.CO]}%
  \bibAnnoteFile{NoStop}{Riess:2011yx}%
\bibitem{Humphreys:2013eja}%
  \BibitemOpen
  \bibfield{author}{%
  \bibinfo {author} {\bibfnamefont{E.}~\bibnamefont{Humphreys}}, \bibinfo
  {author} {\bibfnamefont{M.~J.}\ \bibnamefont{Reid}}, \bibinfo {author}
  {\bibfnamefont{J.~M.}\ \bibnamefont{Moran}}, \bibinfo {author}
  {\bibfnamefont{L.~J.}\ \bibnamefont{Greenhill}},\ and\ \bibinfo {author}
  {\bibfnamefont{A.~L.}\ \bibnamefont{Argon}},\ }%
  \bibfield{journal}{%
  \Doi{10.1088/0004-637X/775/1/13}{\bibinfo {journal} {Astrophys.J.}}\ }%
  \textbf{\bibinfo {volume} {775}},\ \bibinfo {pages} {13} (\bibinfo {year}
  {2013}),\ \Eprint{http://arxiv.org/abs/1307.6031}{arXiv:1307.6031
  [astro-ph.CO]}%
  \bibAnnoteFile{NoStop}{Humphreys:2013eja}%
\bibitem{1475-7516-2009-10-004}%
  \BibitemOpen
  \bibfield{author}{%
  \bibinfo {author} {\bibfnamefont{Y.-S.}\ \bibnamefont{Song}}\ and\ \bibinfo
  {author} {\bibfnamefont{W.~J.}\ \bibnamefont{Percival}},\ }%
  \bibfield{journal}{%
  \bibinfo {journal} {Journal of Cosmology and Astroparticle Physics}\ }%
  \textbf{\bibinfo {volume} {2009}},\ \bibinfo {pages} {004} (\bibinfo {year}
  {2009}),\ \Eprint{http://arxiv.org/abs/0807.0810}{arXiv:0807.0810
  [astro-ph.CO]}%
  \bibAnnoteFile{NoStop}{1475-7516-2009-10-004}%
\bibitem{Beutler:2012px}%
  \BibitemOpen
  \bibfield{author}{%
  \bibinfo {author} {\bibfnamefont{F.}~\bibnamefont{Beutler}}, \bibinfo
  {author} {\bibfnamefont{C.}~\bibnamefont{Blake}}, \bibinfo {author}
  {\bibfnamefont{M.}~\bibnamefont{Colless}}, \bibinfo {author}
  {\bibfnamefont{D.~H.}\ \bibnamefont{Jones}}, \bibinfo {author}
  {\bibfnamefont{L.}~\bibnamefont{Staveley-Smith}}, \emph{et~al.},\ }%
  \bibfield{journal}{%
  \Doi{10.1111/j.1365-2966.2012.21136.x}{\bibinfo {journal}
  {Mon.Not.Roy.Astron.Soc.}}\ }%
  \textbf{\bibinfo {volume} {423}},\ \bibinfo {pages} {3430} (\bibinfo {year}
  {2012}),\ \Eprint{http://arxiv.org/abs/1204.4725}{arXiv:1204.4725
  [astro-ph.CO]}%
  \bibAnnoteFile{NoStop}{Beutler:2012px}%
\bibitem{Samushia01032012}%
  \BibitemOpen
  \bibfield{author}{%
  \bibinfo {author} {\bibfnamefont{L.}~\bibnamefont{Samushia}}, \bibinfo
  {author} {\bibfnamefont{W.~J.}\ \bibnamefont{Percival}},\ and\ \bibinfo
  {author} {\bibfnamefont{A.}~\bibnamefont{Raccanelli}},\ }%
  \bibfield{journal}{%
  \bibinfo {journal} {Monthly Notices of the Royal Astronomical Society}\ }%
  \textbf{\bibinfo {volume} {420}},\ \bibinfo {pages} {2102} (\bibinfo {year}
  {2012}),\ \Eprint{http://arxiv.org/abs/1102.1014}{arXiv:1102.1014
  [astro-ph.CO]}%
  \bibAnnoteFile{NoStop}{Samushia01032012}%
\bibitem{Reid:2012sw}%
  \BibitemOpen
  \bibfield{author}{%
  \bibinfo {author} {\bibfnamefont{B.~A.}\ \bibnamefont{Reid}}, \bibinfo
  {author} {\bibfnamefont{L.}~\bibnamefont{Samushia}}, \bibinfo {author}
  {\bibfnamefont{M.}~\bibnamefont{White}}, \bibinfo {author}
  {\bibfnamefont{W.~J.}\ \bibnamefont{Percival}}, \bibinfo {author}
  {\bibfnamefont{M.}~\bibnamefont{Manera}}, \emph{et~al.}}%
   (\bibinfo {year} {2012}),\
  \Eprint{http://arxiv.org/abs/1203.6641}{arXiv:1203.6641 [astro-ph.CO]}%
  \bibAnnoteFile{NoStop}{Reid:2012sw}%
\bibitem{Blake:2011ep}%
  \BibitemOpen
  \bibfield{author}{%
  \bibinfo {author} {\bibfnamefont{C.}~\bibnamefont{Blake}}, \bibinfo {author}
  {\bibfnamefont{K.}~\bibnamefont{Glazebrook}}, \bibinfo {author}
  {\bibfnamefont{T.}~\bibnamefont{Davis}}, \bibinfo {author}
  {\bibfnamefont{S.}~\bibnamefont{Brough}}, \bibinfo {author}
  {\bibfnamefont{M.}~\bibnamefont{Colless}}, \emph{et~al.},\ }%
  \bibfield{journal}{%
  \Doi{10.1111/j.1365-2966.2011.19606.x}{\bibinfo {journal}
  {Mon.Not.Roy.Astron.Soc.}}\ }%
  \textbf{\bibinfo {volume} {418}},\ \bibinfo {pages} {1725} (\bibinfo {year}
  {2011}),\ \Eprint{http://arxiv.org/abs/1108.2637}{arXiv:1108.2637
  [astro-ph.CO]}%
  \bibAnnoteFile{NoStop}{Blake:2011ep}%
\bibitem{Sotiriou:2008rp}%
  \BibitemOpen
  \bibfield{author}{%
  \bibinfo {author} {\bibfnamefont{T.~P.}\ \bibnamefont{Sotiriou}}\ and\
  \bibinfo {author} {\bibfnamefont{V.}~\bibnamefont{Faraoni}},\ }%
  \bibfield{journal}{%
  \Doi{10.1103/RevModPhys.82.451}{\bibinfo {journal} {Rev.Mod.Phys.}}\ }%
  \textbf{\bibinfo {volume} {82}},\ \bibinfo {pages} {451} (\bibinfo {year}
  {2010}),\ \Eprint{http://arxiv.org/abs/0805.1726}{arXiv:0805.1726 [gr-qc]}%
  \bibAnnoteFile{NoStop}{Sotiriou:2008rp}%
\bibitem{Ade:2013dsi}%
  \BibitemOpen
  \bibfield{author}{%
  \bibinfo {author} {\bibfnamefont{P.}~\bibnamefont{Ade}} \emph{et~al.}
  (\bibinfo {collaboration} {Planck Collaboration})}%
   (\bibinfo {year} {2013}),\
  \Eprint{http://arxiv.org/abs/1303.5079}{arXiv:1303.5079 [astro-ph.CO]}%
  \bibAnnoteFile{NoStop}{Ade:2013dsi}%
\bibitem{Granett:2008ju}%
  \BibitemOpen
  \bibfield{author}{%
  \bibinfo {author} {\bibfnamefont{B.~R.}\ \bibnamefont{Granett}}, \bibinfo
  {author} {\bibfnamefont{M.~C.}\ \bibnamefont{Neyrinck}},\ and\ \bibinfo
  {author} {\bibfnamefont{I.}~\bibnamefont{Szapudi}},\ }%
  \bibfield{journal}{%
  \Doi{10.1086/591670}{\bibinfo {journal} {Astrophys.J.}}\ }%
  \textbf{\bibinfo {volume} {683}},\ \bibinfo {pages} {L99} (\bibinfo {year}
  {2008}),\ \Eprint{http://arxiv.org/abs/0805.3695}{arXiv:0805.3695
  [astro-ph]}%
  \bibAnnoteFile{NoStop}{Granett:2008ju}%
\bibitem{2008arXiv0805.2974G}%
  \BibitemOpen
  \bibfield{author}{%
  \bibinfo {author} {\bibfnamefont{B.~R.}\ \bibnamefont{{Granett}}}, \bibinfo
  {author} {\bibfnamefont{M.~C.}\ \bibnamefont{{Neyrinck}}},\ and\ \bibinfo
  {author} {\bibfnamefont{I.}~\bibnamefont{{Szapudi}}},\ }%
  \bibfield{journal}{%
  \bibinfo {journal} {ArXiv e-prints}}%
   (\bibinfo {month} {May}\ \bibinfo {year} {2008}),\
  \Eprint{http://arxiv.org/abs/0805.2974}{arXiv:0805.2974}%
  \bibAnnoteFile{NoStop}{2008arXiv0805.2974G}%
\bibitem{Sutter:2012wh}%
  \BibitemOpen
  \bibfield{author}{%
  \bibinfo {author} {\bibfnamefont{P.}~\bibnamefont{Sutter}}, \bibinfo {author}
  {\bibfnamefont{G.}~\bibnamefont{Lavaux}}, \bibinfo {author}
  {\bibfnamefont{B.~D.}\ \bibnamefont{Wandelt}},\ and\ \bibinfo {author}
  {\bibfnamefont{D.~H.}\ \bibnamefont{Weinberg}},\ }%
  \bibfield{journal}{%
  \Doi{10.1088/0004-637X/761/1/44}{\bibinfo {journal} {Astrophys.J.}}\ }%
  \textbf{\bibinfo {volume} {761}},\ \bibinfo {pages} {44} (\bibinfo {year}
  {2012}),\ \Eprint{http://arxiv.org/abs/1207.2524}{arXiv:1207.2524
  [astro-ph.CO]}%
  \bibAnnoteFile{NoStop}{Sutter:2012wh}%
\bibitem{Pan:2011hx}%
  \BibitemOpen
  \bibfield{author}{%
  \bibinfo {author} {\bibfnamefont{D.~C.}\ \bibnamefont{Pan}}, \bibinfo
  {author} {\bibfnamefont{M.~S.}\ \bibnamefont{Vogeley}}, \bibinfo {author}
  {\bibfnamefont{F.}~\bibnamefont{Hoyle}}, \bibinfo {author}
  {\bibfnamefont{Y.-Y.}\ \bibnamefont{Choi}},\ and\ \bibinfo {author}
  {\bibfnamefont{C.}~\bibnamefont{Park}},\ }%
  \bibfield{journal}{%
  \Doi{10.1111/j.1365-2966.2011.20197.x}{\bibinfo {journal}
  {Mon.Not.Roy.Astron.Soc.}}\ }%
  \textbf{\bibinfo {volume} {421}},\ \bibinfo {pages} {926} (\bibinfo {year}
  {2012}),\ \Eprint{http://arxiv.org/abs/1103.4156}{arXiv:1103.4156
  [astro-ph.CO]}%
  \bibAnnoteFile{NoStop}{Pan:2011hx}%
\bibitem{HernandezMonteagudo:2012ms}%
  \BibitemOpen
  \bibfield{author}{%
  \bibinfo {author} {\bibfnamefont{C.}~\bibnamefont{Hernandez-Monteagudo}}\
  and\ \bibinfo {author} {\bibfnamefont{R.~E.}\ \bibnamefont{Smith}}}%
   (\bibinfo {year} {2012}),\
  \Eprint{http://arxiv.org/abs/1212.1174}{arXiv:1212.1174 [astro-ph.CO]}%
  \bibAnnoteFile{NoStop}{HernandezMonteagudo:2012ms}%
\bibitem{Cai:2013ik}%
  \BibitemOpen
  \bibfield{author}{%
  \bibinfo {author} {\bibfnamefont{Y.-C.}\ \bibnamefont{Cai}}, \bibinfo
  {author} {\bibfnamefont{M.~C.}\ \bibnamefont{Neyrinck}}, \bibinfo {author}
  {\bibfnamefont{I.}~\bibnamefont{Szapudi}}, \bibinfo {author}
  {\bibfnamefont{S.}~\bibnamefont{Cole}},\ and\ \bibinfo {author}
  {\bibfnamefont{C.~S.}\ \bibnamefont{Frenk}},\ }%
  \bibfield{journal}{%
  \Doi{10.1088/0004-637X/786/2/110}{\bibinfo {journal} {Astrophys.J.}}\ }%
  \textbf{\bibinfo {volume} {786}},\ \bibinfo {pages} {110} (\bibinfo {year}
  {2014}),\ \Eprint{http://arxiv.org/abs/1301.6136}{arXiv:1301.6136
  [astro-ph.CO]}%
  \bibAnnoteFile{NoStop}{Cai:2013ik}%
\bibitem{Finelli:2014yha}%
  \BibitemOpen
  \bibfield{author}{%
  \bibinfo {author} {\bibfnamefont{F.}~\bibnamefont{Finelli}}, \bibinfo
  {author} {\bibfnamefont{J.}~\bibnamefont{Garcia-Bellido}}, \bibinfo {author}
  {\bibfnamefont{A.}~\bibnamefont{Kovacs}}, \bibinfo {author}
  {\bibfnamefont{F.}~\bibnamefont{Paci}},\ and\ \bibinfo {author}
  {\bibfnamefont{I.}~\bibnamefont{Szapudi}}}%
   (\bibinfo {year} {2014}),\
  \Eprint{http://arxiv.org/abs/1405.1555}{arXiv:1405.1555 [astro-ph.CO]}%
  \bibAnnoteFile{NoStop}{Finelli:2014yha}%
\bibitem{Szapudi:2014zha}%
  \BibitemOpen
  \bibfield{author}{%
  \bibinfo {author} {\bibfnamefont{I.}~\bibnamefont{Szapudi}}, \bibinfo
  {author} {\bibfnamefont{A.}~\bibnamefont{Kovács}}, \bibinfo {author}
  {\bibfnamefont{B.~R.}\ \bibnamefont{Granett}}, \bibinfo {author}
  {\bibfnamefont{Z.}~\bibnamefont{Frei}}, \bibinfo {author}
  {\bibfnamefont{J.}~\bibnamefont{Silk}}, \emph{et~al.}}%
   (\bibinfo {year} {2014}),\
  \Eprint{http://arxiv.org/abs/1405.1566}{arXiv:1405.1566 [astro-ph.CO]}%
  \bibAnnoteFile{NoStop}{Szapudi:2014zha}%
\bibitem{2004Natur.427...45B}%
  \BibitemOpen
  \bibfield{author}{%
  \bibinfo {author} {\bibfnamefont{S.}~\bibnamefont{{Boughn}}}\ and\ \bibinfo
  {author} {\bibfnamefont{R.}~\bibnamefont{{Crittenden}}},\ }%
  \bibfield{journal}{%
  \Doi{10.1038/nature02139}{\bibinfo {journal} {\nat}}\ }%
  \textbf{\bibinfo {volume} {427}},\ \bibinfo {pages} {45} (\bibinfo {month}
  {Jan.}\ \bibinfo {year} {2004}),\
  \Eprint{http://arxiv.org/abs/astro-ph/0305001}{astro-ph/0305001}%
  \bibAnnoteFile{NoStop}{2004Natur.427...45B}%
\bibitem{2008PhRvD..78d3519H}%
  \BibitemOpen
  \bibfield{author}{%
  \bibinfo {author} {\bibfnamefont{S.}~\bibnamefont{{Ho}}}, \bibinfo {author}
  {\bibfnamefont{C.}~\bibnamefont{{Hirata}}}, \bibinfo {author}
  {\bibfnamefont{N.}~\bibnamefont{{Padmanabhan}}}, \bibinfo {author}
  {\bibfnamefont{U.}~\bibnamefont{{Seljak}}},\ and\ \bibinfo {author}
  {\bibfnamefont{N.}~\bibnamefont{{Bahcall}}},\ }%
  \bibfield{journal}{%
  \Doi{10.1103/PhysRevD.78.043519}{\bibinfo {journal} {\prd}}\ }%
  \textbf{\bibinfo {volume} {78}},\ \bibinfo {eid} {043519} (\bibinfo {month}
  {Aug.}\ \bibinfo {year} {2008}),\
  \Eprint{http://arxiv.org/abs/0801.0642}{arXiv:0801.0642}%
  \bibAnnoteFile{NoStop}{2008PhRvD..78d3519H}%
\bibitem{Giannantonio:2012aa}%
  \BibitemOpen
  \bibfield{author}{%
  \bibinfo {author} {\bibfnamefont{T.}~\bibnamefont{Giannantonio}}, \bibinfo
  {author} {\bibfnamefont{R.}~\bibnamefont{Crittenden}}, \bibinfo {author}
  {\bibfnamefont{R.}~\bibnamefont{Nichol}},\ and\ \bibinfo {author}
  {\bibfnamefont{A.~J.}\ \bibnamefont{Ross}},\ }%
  \bibfield{journal}{%
  \Doi{10.1111/j.1365-2966.2012.21896.x}{\bibinfo {journal}
  {Mon.Not.Roy.Astron.Soc.}}\ }%
  \textbf{\bibinfo {volume} {426}},\ \bibinfo {pages} {2581} (\bibinfo {year}
  {2012}),\ \Eprint{http://arxiv.org/abs/1209.2125}{arXiv:1209.2125
  [astro-ph.CO]}%
  \bibAnnoteFile{NoStop}{Giannantonio:2012aa}%
\bibitem{Francis:2009pt}%
  \BibitemOpen
  \bibfield{author}{%
  \bibinfo {author} {\bibfnamefont{C.}~\bibnamefont{Francis}}\ and\ \bibinfo
  {author} {\bibfnamefont{J.}~\bibnamefont{Peacock}},\ }%
  \bibfield{journal}{%
  \Doi{10.1111/j.1365-2966.2010.16866.x}{\bibinfo {journal}
  {Mon.Not.Roy.Astron.Soc.}}\ }%
  \textbf{\bibinfo {volume} {406}},\ \bibinfo {pages} {14} (\bibinfo {year}
  {2010}),\ \Eprint{http://arxiv.org/abs/0909.2495}{arXiv:0909.2495
  [astro-ph.CO]}%
  \bibAnnoteFile{NoStop}{Francis:2009pt}%
\bibitem{Francis:2009ps}%
  \BibitemOpen
  \bibfield{author}{%
  \bibinfo {author} {\bibfnamefont{C.}~\bibnamefont{Francis}}\ and\ \bibinfo
  {author} {\bibfnamefont{J.}~\bibnamefont{Peacock}},\ }%
  \bibfield{journal}{%
  \Doi{10.1111/j.1365-2966.2010.16278.x}{\bibinfo {journal}
  {Mon.Not.Roy.Astron.Soc.}}\ }%
  \textbf{\bibinfo {volume} {406}},\ \bibinfo {pages} {2} (\bibinfo {year}
  {2010}),\ \Eprint{http://arxiv.org/abs/0909.2494}{arXiv:0909.2494
  [astro-ph.CO]}%
  \bibAnnoteFile{NoStop}{Francis:2009ps}%
\bibitem{HernandezMonteagudo:2009fb}%
  \BibitemOpen
  \bibfield{author}{%
  \bibinfo {author} {\bibfnamefont{C.}~\bibnamefont{Hernandez-Monteagudo}}}%
   (\bibinfo {year} {2009}),\
  \Eprint{http://arxiv.org/abs/0909.4294}{arXiv:0909.4294 [astro-ph.CO]}%
  \bibAnnoteFile{NoStop}{HernandezMonteagudo:2009fb}%
\bibitem{Sawangwit:2009gd}%
  \BibitemOpen
  \bibfield{author}{%
  \bibinfo {author} {\bibfnamefont{U.}~\bibnamefont{Sawangwit}}, \bibinfo
  {author} {\bibfnamefont{T.}~\bibnamefont{Shanks}}, \bibinfo {author}
  {\bibfnamefont{R.}~\bibnamefont{Cannon}}, \bibinfo {author}
  {\bibfnamefont{S.}~\bibnamefont{Croom}}, \bibinfo {author}
  {\bibfnamefont{N.~P.}\ \bibnamefont{Ross}}, \emph{et~al.},\ }%
  \bibfield{journal}{%
  \Doi{10.1111/j.1365-2966.2009.16054.x}{\bibinfo {journal}
  {Mon.Not.Roy.Astron.Soc.}}\ }%
  \textbf{\bibinfo {volume} {402}},\ \bibinfo {pages} {2228} (\bibinfo {year}
  {2010}),\ \Eprint{http://arxiv.org/abs/0911.1352}{arXiv:0911.1352
  [astro-ph.CO]}%
  \bibAnnoteFile{NoStop}{Sawangwit:2009gd}%
\bibitem{LopezCorredoira:2010rr}%
  \BibitemOpen
  \bibfield{author}{%
  \bibinfo {author} {\bibfnamefont{M.}~\bibnamefont{Lopez-Corredoira}},
  \bibinfo {author} {\bibfnamefont{F.~S.}\ \bibnamefont{Labini}},\ and\
  \bibinfo {author} {\bibfnamefont{J.}~\bibnamefont{Betancort-Rijo}},\ }%
  \bibfield{journal}{%
  \Doi{10.1051/0004-6361/200912763}{\bibinfo {journal} {Astron.Astrophys.}}\ }%
  \textbf{\bibinfo {volume} {513}},\ \bibinfo {pages} {A3} (\bibinfo {year}
  {2010}),\ \Eprint{http://arxiv.org/abs/1001.4000}{arXiv:1001.4000
  [astro-ph.CO]}%
  \bibAnnoteFile{NoStop}{LopezCorredoira:2010rr}%
\bibitem{Efstathiou:2013via}%
  \BibitemOpen
  \bibfield{author}{%
  \bibinfo {author} {\bibfnamefont{G.}~\bibnamefont{Efstathiou}}}%
   (\bibinfo {year} {2013}),\
  \Eprint{http://arxiv.org/abs/1311.3461}{arXiv:1311.3461 [astro-ph.CO]}%
  \bibAnnoteFile{NoStop}{Efstathiou:2013via}%
\bibitem{Heymans:2013fya}%
  \BibitemOpen
  \bibfield{author}{%
  \bibinfo {author} {\bibfnamefont{C.}~\bibnamefont{Heymans}}, \bibinfo
  {author} {\bibfnamefont{E.}~\bibnamefont{Grocutt}}, \bibinfo {author}
  {\bibfnamefont{A.}~\bibnamefont{Heavens}}, \bibinfo {author}
  {\bibfnamefont{M.}~\bibnamefont{Kilbinger}}, \bibinfo {author}
  {\bibfnamefont{T.~D.}\ \bibnamefont{Kitching}}, \emph{et~al.}}%
   (\bibinfo {year} {2013}),\
  \Eprint{http://arxiv.org/abs/1303.1808}{arXiv:1303.1808 [astro-ph.CO]}%
  \bibAnnoteFile{NoStop}{Heymans:2013fya}%
\bibitem{Ade:2013lmv}%
  \BibitemOpen
  \bibfield{author}{%
  \bibinfo {author} {\bibfnamefont{P.}~\bibnamefont{Ade}} \emph{et~al.}
  (\bibinfo {collaboration} {Planck Collaboration})}%
   (\bibinfo {year} {2013}),\
  \Eprint{http://arxiv.org/abs/1303.5080}{arXiv:1303.5080 [astro-ph.CO]}%
  \bibAnnoteFile{NoStop}{Ade:2013lmv}%
\bibitem{Wyman:2013lza}%
  \BibitemOpen
  \bibfield{author}{%
  \bibinfo {author} {\bibfnamefont{M.}~\bibnamefont{Wyman}}, \bibinfo {author}
  {\bibfnamefont{D.~H.}\ \bibnamefont{Rudd}}, \bibinfo {author}
  {\bibfnamefont{R.~A.}\ \bibnamefont{Vanderveld}},\ and\ \bibinfo {author}
  {\bibfnamefont{W.}~\bibnamefont{Hu}},\ }%
  \bibfield{journal}{%
  \Doi{10.1103/PhysRevLett.112.051302}{\bibinfo {journal} {Phys.Rev.Lett.}}\ }%
  \textbf{\bibinfo {volume} {112}},\ \bibinfo {pages} {051302} (\bibinfo {year}
  {2014}),\ \Eprint{http://arxiv.org/abs/1307.7715}{arXiv:1307.7715
  [astro-ph.CO]}%
  \bibAnnoteFile{NoStop}{Wyman:2013lza}%
\bibitem{Battye:2013xqa}%
  \BibitemOpen
  \bibfield{author}{%
  \bibinfo {author} {\bibfnamefont{R.~A.}\ \bibnamefont{Battye}}\ and\ \bibinfo
  {author} {\bibfnamefont{A.}~\bibnamefont{Moss}},\ }%
  \bibfield{journal}{%
  \Doi{10.1103/PhysRevLett.112.051303}{\bibinfo {journal} {Phys.Rev.Lett.}}\ }%
  \textbf{\bibinfo {volume} {112}},\ \bibinfo {pages} {051303} (\bibinfo {year}
  {2014}),\ \Eprint{http://arxiv.org/abs/1308.5870}{arXiv:1308.5870
  [astro-ph.CO]}%
  \bibAnnoteFile{NoStop}{Battye:2013xqa}%
\bibitem{Drexlin:2013lha}%
  \BibitemOpen
  \bibfield{author}{%
  \bibinfo {author} {\bibfnamefont{G.}~\bibnamefont{Drexlin}}, \bibinfo
  {author} {\bibfnamefont{V.}~\bibnamefont{Hannen}}, \bibinfo {author}
  {\bibfnamefont{S.}~\bibnamefont{Mertens}},\ and\ \bibinfo {author}
  {\bibfnamefont{C.}~\bibnamefont{Weinheimer}},\ }%
  \bibfield{journal}{%
  \Doi{10.1155/2013/293986}{\bibinfo {journal} {Adv.High Energy Phys.}}\ }%
  \textbf{\bibinfo {volume} {2013}},\ \bibinfo {pages} {293986} (\bibinfo
  {year} {2013}),\ \Eprint{http://arxiv.org/abs/1307.0101}{arXiv:1307.0101
  [physics.ins-det]}%
  \bibAnnoteFile{NoStop}{Drexlin:2013lha}%
\bibitem{katrin}%
  \BibitemOpen
  \bibinfo {author} {\bibnamefont{http://www.katrin.kit.edu/}}%
  \bibAnnoteFile{NoStop}{katrin}%
\bibitem{Vergados:2012xy}%
  \BibitemOpen
\bibfield{author}{%
    }%
  \bibfield{author}{%
  \bibinfo {author} {\bibfnamefont{J.}~\bibnamefont{Vergados}}, \bibinfo
  {author} {\bibfnamefont{H.}~\bibnamefont{Ejiri}},\ and\ \bibinfo {author}
  {\bibfnamefont{F.}~\bibnamefont{Simkovic}},\ }%
  \bibfield{journal}{%
  \Doi{10.1088/0034-4885/75/10/106301}{\bibinfo {journal} {Rept.Prog.Phys.}}\
  }%
  \textbf{\bibinfo {volume} {75}},\ \bibinfo {pages} {106301} (\bibinfo {year}
  {2012}),\ \Eprint{http://arxiv.org/abs/1205.0649}{arXiv:1205.0649 [hep-ph]}%
  \bibAnnoteFile{NoStop}{Vergados:2012xy}%
\bibitem{2011MNRAS.410.2081J}%
  \BibitemOpen
  \bibfield{author}{%
  \bibinfo {author} {\bibfnamefont{E.}~\bibnamefont{{Jennings}}}, \bibinfo
  {author} {\bibfnamefont{C.~M.}\ \bibnamefont{{Baugh}}},\ and\ \bibinfo
  {author} {\bibfnamefont{S.}~\bibnamefont{{Pascoli}}},\ }%
  \bibfield{journal}{%
  \Doi{10.1111/j.1365-2966.2010.17581.x}{\bibinfo {journal} {MNRAS}}\ }%
  \textbf{\bibinfo {volume} {410}},\ \bibinfo {pages} {2081} (\bibinfo {month}
  {Jan.}\ \bibinfo {year} {2011}),\
  \Eprint{http://arxiv.org/abs/1003.4282}{arXiv:1003.4282 [astro-ph.CO]}%
  \bibAnnoteFile{NoStop}{2011MNRAS.410.2081J}%
\bibitem{Jennings:2012pt}%
  \BibitemOpen
  \bibfield{author}{%
  \bibinfo {author} {\bibfnamefont{E.}~\bibnamefont{Jennings}}, \bibinfo
  {author} {\bibfnamefont{C.~M.}\ \bibnamefont{Baugh}}, \bibinfo {author}
  {\bibfnamefont{B.}~\bibnamefont{Li}}, \bibinfo {author}
  {\bibfnamefont{G.-B.}\ \bibnamefont{Zhao}},\ and\ \bibinfo {author}
  {\bibfnamefont{K.}~\bibnamefont{Koyama}},\ }%
  \bibfield{journal}{%
  \Doi{10.1111/j.1365-2966.2012.21567.x}{\bibinfo {journal}
  {Mon.Not.Roy.Astron.Soc.}}\ }%
  \textbf{\bibinfo {volume} {425}},\ \bibinfo {pages} {2128} (\bibinfo {year}
  {2012}),\ \Eprint{http://arxiv.org/abs/1205.2698}{arXiv:1205.2698
  [astro-ph.CO]}%
  \bibAnnoteFile{NoStop}{Jennings:2012pt}%
\bibitem{Wyman:2013jaa}%
  \BibitemOpen
  \bibfield{author}{%
  \bibinfo {author} {\bibfnamefont{M.}~\bibnamefont{Wyman}}, \bibinfo {author}
  {\bibfnamefont{E.}~\bibnamefont{Jennings}},\ and\ \bibinfo {author}
  {\bibfnamefont{M.}~\bibnamefont{Lima}},\ }%
  \bibfield{journal}{%
  \Doi{10.1103/PhysRevD.88.084029}{\bibinfo {journal} {Phys.Rev.}}\ }%
  \textbf{\bibinfo {volume} {D88}},\ \bibinfo {pages} {084029} (\bibinfo {year}
  {2013}),\ \Eprint{http://arxiv.org/abs/1303.6630}{arXiv:1303.6630
  [astro-ph.CO]}%
  \bibAnnoteFile{NoStop}{Wyman:2013jaa}%
\bibitem{Williams:2004qba}%
  \BibitemOpen
  \bibfield{author}{%
  \bibinfo {author} {\bibfnamefont{J.~G.}\ \bibnamefont{Williams}}, \bibinfo
  {author} {\bibfnamefont{S.~G.}\ \bibnamefont{Turyshev}},\ and\ \bibinfo
  {author} {\bibfnamefont{D.~H.}\ \bibnamefont{Boggs}},\ }%
  \bibfield{journal}{%
  \Doi{10.1103/PhysRevLett.93.261101}{\bibinfo {journal} {Phys.Rev.Lett.}}\ }%
  \textbf{\bibinfo {volume} {93}},\ \bibinfo {pages} {261101} (\bibinfo {year}
  {2004}),\ \Eprint{http://arxiv.org/abs/gr-qc/0411113}{arXiv:gr-qc/0411113
  [gr-qc]}%
  \bibAnnoteFile{NoStop}{Williams:2004qba}%
\bibitem{Ade:2014xna}%
  \BibitemOpen
  \bibfield{author}{%
  \bibinfo {author} {\bibfnamefont{P.}~\bibnamefont{Ade}} \emph{et~al.}
  (\bibinfo {collaboration} {BICEP2 Collaboration}),\ }%
  \bibfield{journal}{%
  \Doi{10.1103/PhysRevLett.112.241101}{\bibinfo {journal} {Phys.Rev.Lett.}}\ }%
  \textbf{\bibinfo {volume} {112}},\ \bibinfo {pages} {241101} (\bibinfo {year}
  {2014}),\ \Eprint{http://arxiv.org/abs/1403.3985}{arXiv:1403.3985
  [astro-ph.CO]}%
  \bibAnnoteFile{NoStop}{Ade:2014xna}%
\end{thebibliography}%

\end{document}